\documentclass[fleqn,usenatbib]{mnras}

\usepackage{newtxtext,newtxmath}

\usepackage[T1]{fontenc}
\usepackage{ae,aecompl}
\usepackage[toc,page]{appendix}
\usepackage{xcolor}
\usepackage{soul}

\usepackage{graphicx}    
\usepackage{amsmath}    
\usepackage{float}
\usepackage{longtable}

\title[UCDs beyond the centre of the Fornax cluster]{Ultra-compact dwarfs beyond the centre of the Fornax galaxy cluster: Hints of UCD formation in low-density environments}

\author[T. Saifollahi et al.]
{Teymoor Saifollahi$^{1}$\thanks{E-mail: teymur.saif@gmail.com},
Joachim Janz$^{2,3}$,
Reynier F. Peletier$^{1}$,
Michele Cantiello$^{4}$,
\newauthor
Michael Hilker$^{5}$,
Steffen Mieske$^{6}$,
Edwin A. Valentijn$^{1}$,
Aku Venhola$^{3}$,
Gijs Verdoes Kleijn$^{1}$
\\\\
$^{1}$Kapteyn Astronomical Institute, University of Groningen, PO Box 800, 9700 AV Groningen, The Netherlands\\
$^{2}$Finnish Centre of Astronomy with ESO (FINCA), Vesilinnantie 5, FI-20014 University of Turku, Finland\\
$^{3}$Space Physics and Astronomy Research Unit, University of Oulu, P.O. Box 3000, FI-90014, Oulu, Finland\\
$^{4}$ INAF osservatorio astronomico d’Abruzzo, via Magrini snc, I-64100, Teramo, Italy \\
$^{5}$ European Southern Observatory, Karl-Schwarzschild-Str. 2, D-85748, Garching bei München, Germany \\
$^{6}$ European Southern Observatory, Alonso de Cordova 3107, Vitacura, Santiago, Chile \\
}

\date{Accepted XXX. Received YYY; in original form ZZZ}

\pubyear{2020}

\begin{document}
\label{firstpage}
\pagerange{\pageref{firstpage}--\pageref{lastpage}}
\maketitle

\begin{abstract}
Ultra-compact dwarf galaxies (UCDs) were serendipitously discovered by spectroscopic surveys in the Fornax cluster twenty years ago. Nowadays, it is commonly accepted that many bright UCDs are the nuclei of galaxies that have been stripped. However, this conclusion might be driven by biased samples of UCDs in high-density environments, on which most searches are based. With the deep optical images of the Fornax Deep Survey, combined with public near-infrared data, we revisit the UCD population of the Fornax cluster and search for UCD candidates, for the first time, systematically out to the virial radius of the galaxy cluster. Our search is complete down to magnitude m$_g$ = 21 mag or M$_g$ $\sim$ -10.5 mag at the distance of the Fornax cluster. The UCD candidates are identified and separated from foreground stars and background galaxies by their optical and near-infrared colours. This primarily utilizes the $u-i$/$i-Ks$ diagram and a machine learning technique is employed to incorporate other colour combinations to reduce the number of contaminants. The newly identified candidates (44) in addition to the spectroscopically confirmed UCDs (61), increases the number of known Fornax UCD considerably (105). Almost all of the new UCD candidates are located outside the Fornax cluster core (360 kpc), where all of the known UCDs were found. The distribution of UCDs within the Fornax cluster shows that a population of UCDs may form in low-density environments. This most likely challenges the current models of UCD formation.

\end{abstract}




\section{Introduction}

In the late 90s, through the spectroscopic surveys of the Fornax galaxy cluster, \citet{Hilker-1999} and \citet{Drinkwater-1999} independently reported the detection of very compact objects at the redshift of the cluster, brighter than globular clusters and fainter than compact dwarf galaxies. Since then, many studies have been carried out to investigate the origin of these so-called Ultra-Compact Dwarf Galaxies (UCDs, \citealp{phi2001}). UCDs are larger, brighter and more massive than typical globular clusters (GCs) with typical half-light radii of 10 $\leq$ r$_h$ $\leq$ 100 pc, luminosities between -13.0 $\leq$ M$_g$ $\leq$ -10.0 mag and masses in a range from 2$\times$10$^6$ M$_{\odot}$ to $<$ 10$^8$ M$_{\odot}$ (\citealp{Mieske-2008,Misgeld-2011}) with predominantly old stellar populations and a wide range of metallicities (\citealp{firth2009,janz2016,Zhang-2018,fahrion3,forbes2020}). Moreover, they have on average a dynamical to stellar mass ratio M$_{dyn}$/M$_*$ $>$ 1 (M$_{dyn}$/M$_*$ = 1.7$\pm$0.2 for massive UCDs with M $>$ 10$^7$ M$_\odot$), while for typical GCs, M$_{dyn}$/M$_*$ $\sim$ 1 (\citealp{Mieske-2013}). 

As a result of the studies in the past two decades, two main formation scenarios for UCDs are suggested. In the first scenario, UCDs are the remnant nuclei of tidally disrupted galaxies (\citealp{Bekki-2003}). In this scenario, when a nucleated galaxy with a distinct and high-density nuclear star cluster (NSC) undergoes tidal stripping, it loses most of its stars except the central ones, where the gravitational potential is strong enough to confine stars. Additionally, gravitational interactions induce gas in-fall into the centre of the galaxy and initiate star formation which can change the stellar populations of the nucleus (\citealp{mark,yasna2018,johnston}). In response, the stellar populations and star formation history of the future UCD deviates from the original populations of the nucleus. The detection of tidal features (\citealp{Voggel2016}, \citealp{Schweizer-2018}), extended halos (\citealp{Evstigneeva-2008}, \citealp{Liu-2020}) and asymmetries in the morphology (\citealp{wittmann-2016}) in addition to an extended star formation history (\citealp{Norris-2015}) and multiple stellar populations (\citealp{Mieske-2008,DaRocha-2011}) that have been observed in some UCDs, support the stripped nuclei scenario. Moreover, the progenitor's central black hole remains unchanged by the stripping, leading to a higher M$_{dyn}$/M$_*$. A SMBH with a mass of 15\% on the average can explain the elevated M$_{dyn}$/M$_*$ (\citealp{Mieske-2013}). However, there are cases for which this does not apply or it is not the whole answer since the SMBH is unlikely to be massive enough to explain the observed dynamical to stellar
mass ratio (\citealp{janz2015}). High-resolution observations (spatial and spectral) of the brightest UCDs found supermassive black holes ($>$ 10$^6$ M$_{\odot}$) in a few cases (\citealp{Seth-2014,ahn2017,ahn2018,afanasiev})\footnote{At the time of writing this paper, SMBH is confirmed in 5 UCDs: 4 UCDs in the Virgo cluster (M60-UCD1, M59-UCD3, VUCD3, M59cO) and 1 in the Fornax cluster (Fornax-UCD3)}. Alternatives such as variations in the initial mass function (IMF) are proposed to explain the elevated M$_{dyn}$/M$_*$ (\citealp{forbes2014,alexa,kroupa,haghi}).

In the second scenario, UCDs are the outcome of star cluster formation processes, either as the brightest and most massive examples of GCs at the bright end of the GCs luminosity function (GCLF) (\citealp{mieske2002}) or as the result of merging stellar super clusters (\citealp{Fellhauer-2002}). In this case, UCDs are the extension of GCs and share most of their properties. Detailed studies of Milky Way GCs have not yet found any evidence for the presence of a dark matter halo (\citealp{mash2005,conroy2011,iba2013}) or a massive black hole in the centre of GCs (\citealp{Baum2003b}) and their M$_{dyn}$/M$_*$ is around unity. However, alternatives to the SMBH, like variations in the IMF as mentioned above, or a very concentrated cluster of massive stellar remnants (\citealp{mahani}) have been proposed to explain the observed elevated M$_{dyn}$/M$_*$. While the former can not explain the detected supermassive black hole in some UCDs, the latter predicts an observationally identified central kinematical peak which can be interpreted as a SMBH.

\begin{figure*}
\centering
        \includegraphics[width=\linewidth]{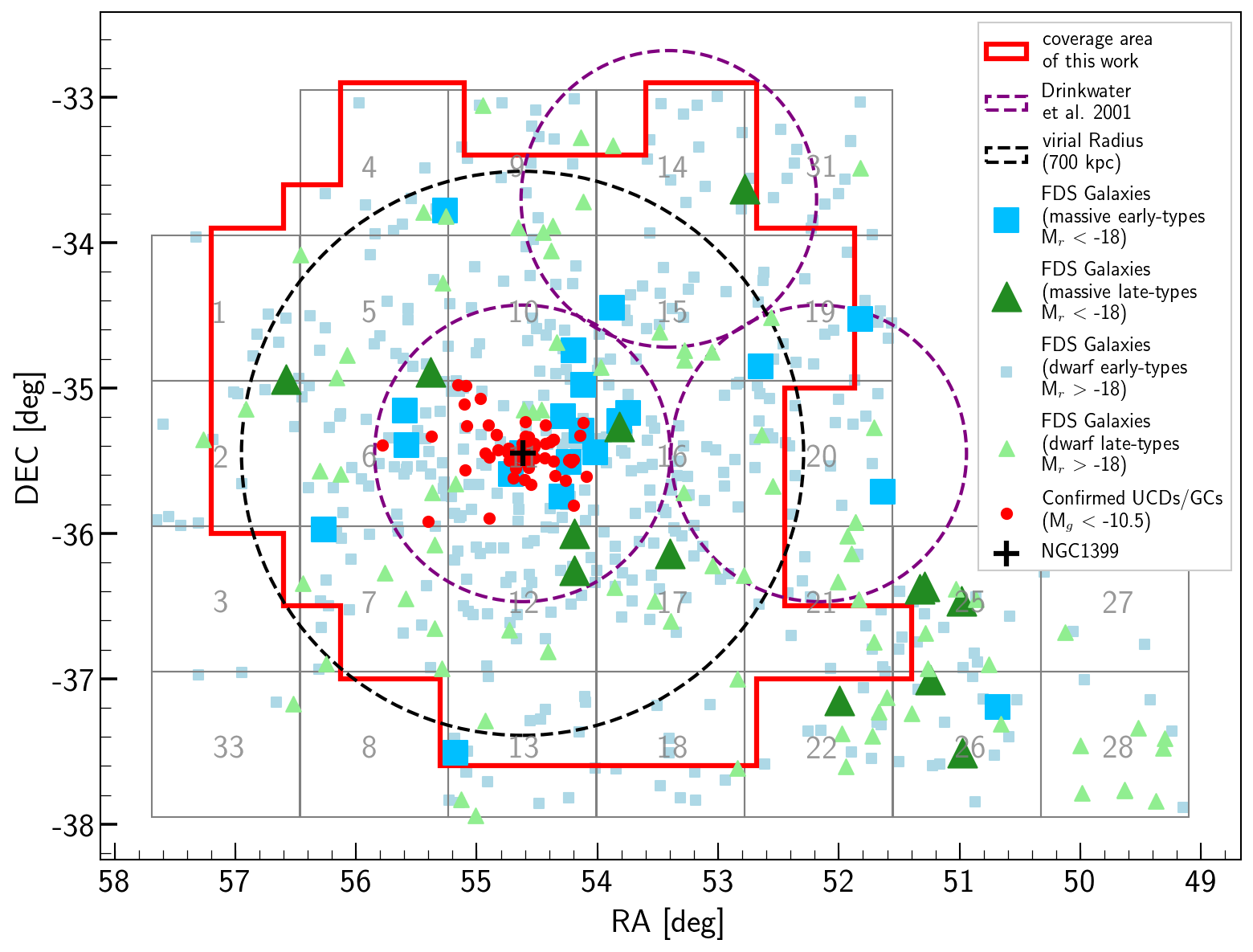}
\caption{The coverage of the combined optical and near-infrared data in this work. The dataset covers the Fornax cluster within its virial radius in 6 filters ($u$, $g$, $r$, $i$, $J$ and $Ks$) and allows us to search for the compact sources in the cluster outskirts. Most of the previous surveys of compact sources targeted the central 1 square degree. As a result, all of the known UCDs (red points) are located in the core of the cluster, within half the virial radius from NGC1399 in the centre. \citet{drinkwater2000} covered a larger fraction of the cluster (purple circles) compared to the other surveys. Grey grid and numbers represent the Fornax Deep Survey (FDS) observed fields.}
\label{coverage}
\end{figure*}

The wide range of properties of the known UCDs and their similarity to both globular clusters and nuclei of dwarf galaxies suggest that they are formed through multiple and co-existing formation channels (\citealp{Francis-2012,Norris-2015}) and the contribution of the stripped nuclei of dwarf galaxies increases with mass and brightness (\citealp{Brodie-2011,Pfeffer-2016,mayes}). However, our current knowledge of UCDs is limited by the selection function of the surveys within which UCDs were searched for. The known UCDs were found through spectroscopic surveys around massive galaxies (\citealp{Norris-2014}) or in the cores of galaxy clusters/groups. Therefore, there is not much known about UCDs (or similar objects) in the outskirts of galaxy clusters. In the Fornax cluster, all of the confirmed UCDs are located in the inner 360 kpc (about half of the cluster virial radius). This is because almost all of the previous spectroscopic surveys of compact sources in the Fornax cluster, covered the core of the cluster (\citealp{Hilker-1999,Mieske2004,bergond2007,firth2007,firth2008,gregg2009,schuberth2010,Pota-2018}). \citet{drinkwater2000}\footnote{This survey led to the discovery of the UCDs and identifying such sources was not planned.} covered a larger fraction of the cluster. However, the data reached m$_g$ = 20 mag (80\% complete) and were therefore limited to the brightest and most massive UCDs. This magnitude corresponds to M$_g$ = -11.5 mag and stellar mass of $10^7 M_{\odot}$ at the distance of Fornax. \citet{drinkwater2000} did not find any UCD outside of the cluster's core (Fig. \ref{coverage}).

Because of the small size of the majority of the UCD/GCs, identification of these objects in images is challenging. To spatially resolve them at low redshifts, ground-based observations in excellent seeing conditions or space-based observations are needed (\citealp{richtler2005,mark,jordan2015,voggel-2016}). Otherwise, the majority of UCD/GCs appear as point-sources which makes them indistinguishable from foreground stars (Milky Way stars) or distant background galaxies. Traditionally, optical photometry is used to select UCD/GC candidates around galaxies and in galaxy clusters (\citealp{taylor2016,mlgc,prole2019,Cantiello2020}). Adding information from infrared bands can substantially improve this selection. \citet{Munoz-2014} showed that the compact sources (UCD/GCs) in the Virgo cluster follow a well-defined sequence in the optical/near-infrared colour-colour diagram which can be employed as a tool for identifying them (\citealp{Liu-2015,Powalka-2017,Cantiello2018,gonz2019,Liu-2020}). 

Using the optical data from the Fornax Deep Survey (FDS, \citealp{Iodice-2016,venhola2018}) combined with the near-infrared observations of the Vista Hemisphere Survey (VHS) and ESO/VISTA archival data, we aim to identify the possible population of UCDs in the outskirts of the Fornax galaxy cluster within its virial radius (700 kpc, \citealp{drinkwater2001}). In \citet{Cantiello2018}, we derived a list of GC candidates in the cluster using the optical data of the FDS. While having the near-infrared data helps to reduce the contamination, the shallower observations in the near-infrared than the optical limits this study to the UCDs and the brightest GCs. 

In the next sections, the data, methods, and results are presented. The paper is structured as follows: Section \ref{sec2} describes the data and data reduction. Section \ref{sec3} gives a detailed description of the data analysis, which includes source extraction, photometry, creation of the multi-wavelength catalogues of the observed sources and measurement of sizes of the detected sources. Section \ref{sec4} describes our methodology to identify UCD candidates and the applied machine learning technique. In section \ref{sec5}, we inspect the performance of our methodology using the ACSFCS catalogue of UCD/GCs (\citealp{jordan2015}) and the stars from Gaia DR2 (\citealp{gaia}). We also present the catalogue of UCD candidates in the Fornax cluster, including the spectroscopically confirmed UCDs and the most likely UCD candidates from our analysis. This catalogue supersedes the UCD catalogue presented in \citealp{Cantiello2020}. We discuss the results in section \ref{sec6} and conclude our findings in section \ref{sec7}. 

In this paper, we refer to UCDs as objects brighter than m$_g$ $=$ 21 mag or M$_g$ = -10.5 mag at the Fornax cluster's distance. This definition is adopted from \citealp{Cantiello2020} (see references therein). This magnitude corresponds to the stellar mass of $4 \times 10^6 M_{\odot}$ \footnote{Using the photometric transformations in \citet{jester2005} and considering the average colour g-r $\sim$ 0.7 mag, as is observed for the confirmed UCD/GCs, the magnitude M$_g$ = -10.5 mag corresponds to M$_V$ = -10.9 mag. Assuming an average (M$_*$/L$_V$) $\simeq$ 2 (\citealp{kruij2008}), this magnitude corresponds to the stellar mass 4 $\times$ 10$^6$ $M_{\odot}$.}, about the same order of magnitude as $\omega$ Centauri (NGC5139, \citealp{dsouza,Baumgardt2018}). Note that we do not adopt any lower limit for the sizes of UCDs. As was discussed here, nowadays it is believed that the observed UCDs are objects of different origins (\citealp{Hilker2006}). However, the term "UCD" is still used\footnote{The term ultra-compact dwarf (UCD) was introduced in the early years of the 21st century since they look like the compact dwarf galaxies, but fainter and more compact.}. Therefore, our definition based on magnitude (brightness) may include different types of compact objects and does not mean that they necessarily are dwarf galaxies. 

Throughout this paper, the term "core" and "outskirts" of the Fornax cluster refers to the area within the virial radius of the cluster, inside and outside the central 1 degree (360 kpc, $\sim$half the virial radius) from NGC1399. Optical and near-infrared magnitudes are expressed in the AB and Vega magnitude systems, respectively. We assume a distance modulus m-M = 31.50 mag (a distance of $\sim$ 20 Mpc) for the Fornax cluster (\citealp{jerjen2003, Blakeslee2009}) and the following cosmological parameters $\Omega_{M}$ = 0.3, $\Omega_{\Lambda}$ = 0.7 and H$_{0}$ = 70 km s$^{-1}$ Mpc$^{-1}$. 

\section{Observations}
\label{sec2}
The optical and near-infrared images stem from three different data sets and cover 16 square degrees out to the virial radius of the Fornax cluster (Fig. \ref{coverage}) in 6 filters: the optical data of the Fornax Deep Survey (FDS) in $u$, $g$, $r$, $i$ and near-infrared data from the VISTA Hemisphere Survey (VHS) in $J$ and ESO/VISTA archival data in $Ks$.

\subsection{Optical \textit{ugri} data: Fornax Deep Survey (FDS)}
The Fornax Deep Survey (FDS, \citealp{fds}) is a deep imaging survey using OmegaCAM camera at the ESO VLT survey telescope (VST) and it is used to study a wide range of topics, including intra-cluster light (\citealp{Iodice-2016,Iodice-2017,spavone}), galaxy assembly history (\citealp{raj}), dwarf galaxies (\citealp{venhola2018} and \citealp{venhola-2019}), ultra-diffuse galaxies (\citealp{Venhola-2017}) and globular clusters (\citealp{Cantiello2020}). This survey covers 26 square degrees of the Fornax cluster in 4 optical bands, $u$, $g$, $r$, and $i$ with an average FWHM of 1.2, 1.1, 1.0, and 1.1 arcsec and a 5$\sigma$ limiting magnitude (point-source detection with S/N = 5) of 24.1, 25.4, 24.9 and 24.0 mag. The depth of this survey reaches the turn-over of the globular clusters luminosity function at m$_g$ $\sim$ 24 mag (\citealp{jordan2007}). The optical data is described in more detail in \citet{venhola2018}. 

\begin{figure}
\centering
        \includegraphics[width=\linewidth]{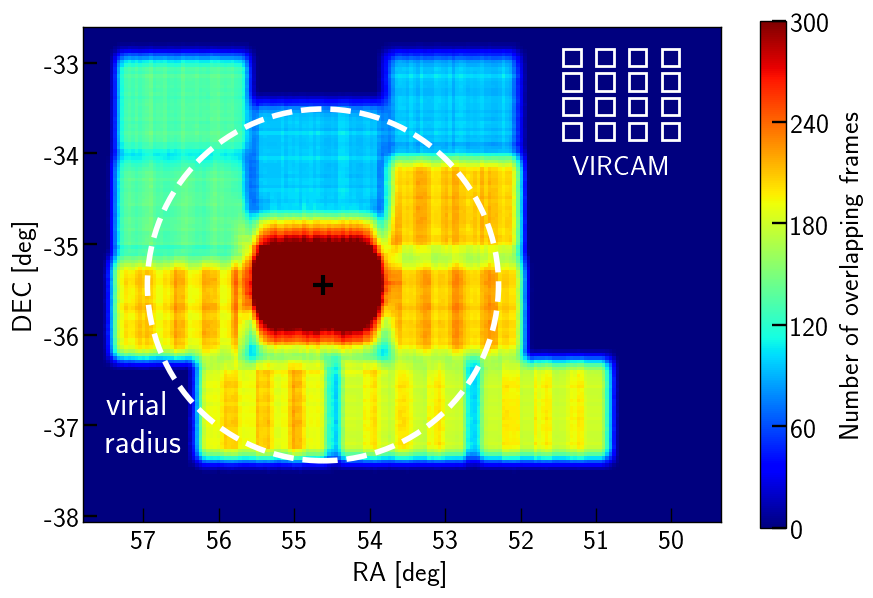}
    \caption{Footprint of the $Ks$ data. The number of overlapping exposures changes across the data, with $\sim$ 300 exposures in the centre (FDS field \#11) and $\sim$ 100-200 exposures outside of the centre. The black cross in the centre and white dashed circle indicate NGC1399 and the virial radius of the cluster. The relative sizes and positions of VIRCAM chips (16 chips) projected on the sky are shown in the top-right of the figure. The total area covered by pixels (16 chips) in one exposure (paw) is 0.6 deg$^2$ and 6 exposures cover a tile of 1.5 deg$^2$. The offsets between chips along the x-axis and y-axis are 47.5\% and 95\% of the size of one chip.}
\label{footprint}
\end{figure}

\begin{figure*}
\centering
        \includegraphics[width=\linewidth]{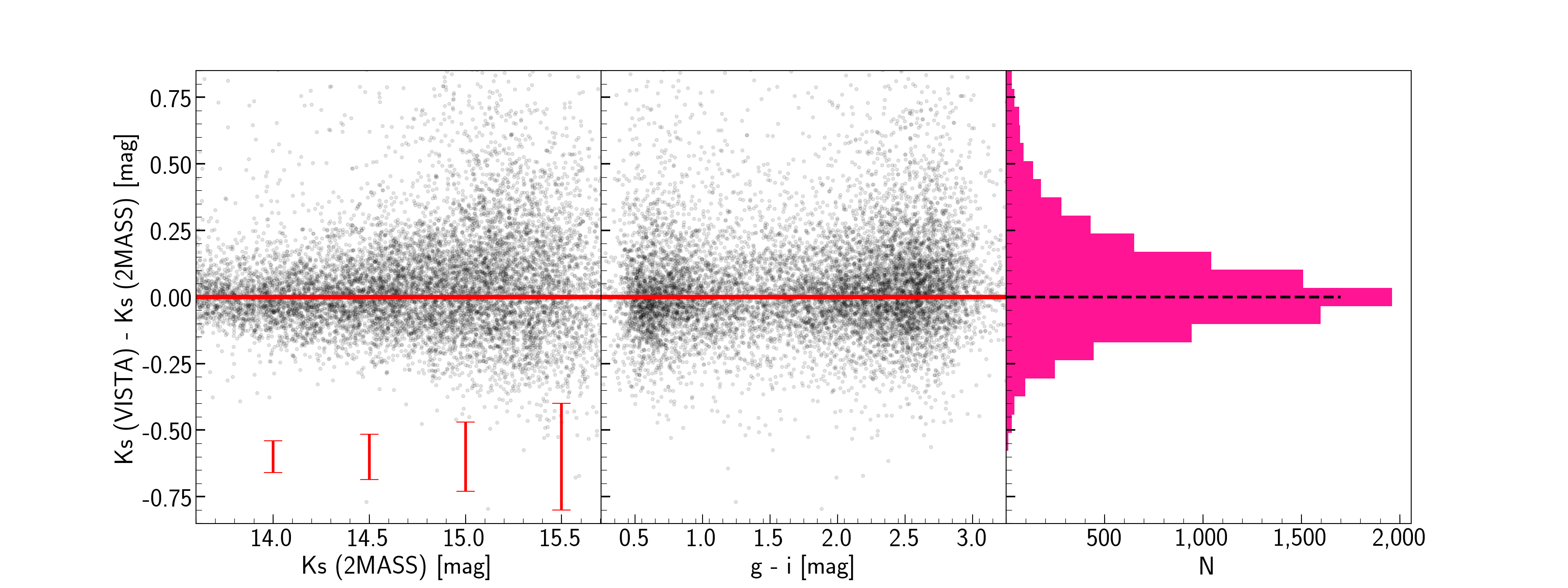}
\caption{Comparison between the $Ks$-magnitudes of point-sources in the $Ks$ final (co-added) frame and 2MASS. For this comparison, point-sources within the interval 13.5 $>$ m$_{Ks}$ $>$ 15.5 mag and 0.6 $<$ FWHM $<$ 0.9 arcsec were selected. The scatter in the magnitude residuals is mainly dominated by the uncertainties in the 2MASS photometry. The typical 2MASS uncertainties are indicated by red error bars in the left panel.}
\label{kmagcomparison}
\end{figure*}

\subsection{Near-infrared $J$ data: Vista Hemisphere Survey (VHS)}

The VISTA Hemisphere Survey (VHS, \citealp{vhs}) is a large survey that covers almost the whole southern hemisphere ($\sim$20,000 square degrees) in the near-infrared using the VISTA telescope. This survey is the near-infrared companion of the Dark Energy Survey (DES) and the VST ATLAS survey. VHS provides observations 30 times deeper than 2MASS in $J$ and $Ks$. We retrieved the reduced frames of the VHS in $J$ (and not $Ks$) from the ESO science portal\footnote{\url{http://archive.eso.org/scienceportal/}}. The final $J$-band frames have an average FWHM of 1.1 arcsec and a 5$\sigma$ limiting magnitude of 20.7 mag. 

\subsection{Near-infrared $Ks$ data: ESO/VISTA archival data}

In addition to $J$ band data of the VHS, we used a deeper set of $Ks$ band observations of the Fornax cluster (ESO programme ID 090.B-0.477(A), 092.B-0.622(A), 094.B-0.536(A), 096.B-0.532(A), PI: Mu\~noz) that were carried out using VIRCAM, the near-infrared camera of the VISTA telescope from October 2012 to March 2016. VIRCAM consists of 16 chips of 2048 $\times$ 2048 pixels and pixel scale of 0.334 arcsec pixel$^{-1}$. However, because of the physical distances between the chips, six exposures are needed to cover the telescope's field of view uniformly. This dither pattern covers each piece of sky at least twice (at least two pixels of the camera) except for the edges. 

The dataset consists of 6,000 exposures with exposure times of 24 or 30 seconds and covers the Fornax cluster and partly the Fornax A group in the south-west. The footprint of the data is shown in Fig. \ref{footprint}. When combined, as we show later, these data provide a deeper dataset than the VHS $Ks$ data ($\sim$0.7 mag deeper). The same data is used by \citet{yasna2018}, who reduced the data independently for the central region of the Fornax cluster (1 degree around NGC1399).

The $Ks$ data reduction was performed to remove the instrumental signatures, contaminant sources, and atmospheric effects. Subsequently, the data were calibrated astrometrically and photometrically. The final mosaic was made by stacking all the reduced/calibrated frames. For the data reduction, we used Astro-WISE (OmegaCEN, \citealp{aw2,McFarland-2013}), an environment developed to analyze wide-field astronomical imaging data in the optical. At first, master-flats were made by median stacking of all the flat-fields taken during each year. Bad-pixel maps were made using the master flats for each chip (16 chips in total). VIRCAM is known to have many bad pixels and bad patches across its chips\footnote{ \url{http://casu.ast.cam.ac.uk/surveys-projects/vista/technical/known-issues}}. Therefore, making bad pixel maps is necessary to flag these regions through the data reduction. Following this, background frames were made for each science frame. For this purpose, we followed the procedure that has been used in \citealp{venhola2018}. After flagging and masking bad pixels and big/bright objects, science frames were normalized. Then the six science frames that have been taken before and after a science frame within $\pm$600 seconds were median stacked to make a background frame. At the end, the background frames were subtracted from the science frames.

The astrometric and photometric calibration of the data was done using the Two Micron All-Sky Survey (2MASS, \citealp{2mass}) as reference catalogue. Astrometric solutions were calculated for individual chips (16 chips), and zero-points were measured on a nightly basis. The photometric calibration on a nightly basis was adopted to increase the number of reference sources and to increase the S/N of the detected sources and achieve a reliable photometry. To do that, all the science frames (reduced and astrometrically calibrated) that were observed during one night were corrected for the atmospheric extinction and then co-added using \textsc{SWarp} (\citealp{swarp}). 95\% of the exposures in $Ks$ have airmass values in a range between 1.0 and 1.3. Considering the $Ks$ extinction coefficient of 0.08 mag airmass$^{-1}$ (VIRCAM user manual\footnote{\url{https://www.eso.org/sci/facilities/paranal/instruments/vircam/doc/}}), corrections are $<$0.025 mag.

For each nightly co-added frame, point-sources were extracted and magnitudes were measured using \textsc{SExtractor} (\citealp{sex}). Next, to scale the flux values of all the co-added frames, the \textsc{SExtractor} MAG$\_$AUTO values were compared to the 2MASS $Ks$-magnitude, and relative zero-points were calculated. In this step, saturated objects (m$_{Ks}$ $<$ 13.5) and faint stars with large uncertainties in their 2MASS photometry (m$_{Ks}$ $>$ 15.5) were excluded from the analysis. For the \textsc{SExtractor} MAG$\_$AUTO, a KRON$\_$FACTOR = 2.5 was used. Using this value of KRON$\_$FACTOR, we expect to capture 94\% of the total light (\citealp{kron}). In that case, 6\% of the light is lost, which leads to under-estimating the zero-points. A larger KRON$\_$FACTOR leads to larger apertures, lower signal-to-noise ratio and larger uncertainty in photometry. No correction for the missing light and zero-points was applied in this step. This correction was done after making the final mosaics (co-addition) in which a more accurate approach was used for photometry of the point sources.

The measured relative zero-points were used for the frame by frame (nightly frames) relative calibration to scale flux values in different frames before the final co-addition. Next, the nightly co-added frames and the measured relative zero-points were used to make the final mosaic (using \textsc{SWarp}). We used the bad pixel maps for the final co-addition to create weight maps with 0 weight for bad pixels and equal weight for all valid pixels. Next, aperture-corrected photometry was done, and the absolute zero-points were calculated. The aperture-corrected photometry is described in section \ref{sec3}.

The uncertainties of the photometric reference catalogue (2MASS) for objects fainter than the saturation limit of the $Ks$ observations, makes the photometric calibration challenging. Fig. \ref{kmagcomparison} compares the measured $Ks$ magnitudes of the point-sources, extracted from the data, compared to the corresponding 2MASS values. For this comparison, sources were detected and their magnitudes were measured with \textsc{SExtractor} using BACK$\_$SIZE = 64 pixel and BACK$\_$FILTERSIZE = 3 pixel. Then, bright and not-saturated point-sources within 13.5 $>$ m$_{Ks}$ $>$ 15.5 mag (to avoid too large uncertainties of the 2MASS photometry and to avoid saturated sources in our $Ks$ data) and 0.6 $<$ FWHM $<$ 0.9 arcsec were selected. Note that in this magnitude range, the 2MASS magnitudes have uncertainties larger than 0.1 mag, which is consistent with the scatter in Fig. \ref{kmagcomparison}. In this figure, we also explore the colour dependence of the acquired photometric solution. No significant colour dependency is seen in the data. Because of the in-homogeneity of the observed regions, the signal-to-noise ratio varies across the field. The signal-to-noise ratio can drop significantly in some regions, which introduces some low-signal-to-noise regions in the final co-added frame, which cover around 3\% of the data.

The final co-added frames\footnote{The final reduced frames can be provided for interested parties upon request.} have an average FWHM $\sim$0.7 arcsec and a 5$\sigma$ limiting magnitude of 18.4 mag ($\sim$1.0 arcsec and 17.7 mag for the $Ks$ data of VHS). The values of FWHM, depth and limiting magnitude of the final $Ks$ frame are listed in Table \ref{fwhmanddepth} for each FDS field.

\begin{table}
\centering
\caption{FWHM values, FWHM standard deviations and 5$\sigma$ limiting magnitudes of the $Ks$ data for each FDS field.}
\begin{tabular}{ cccc } \hline  
FDS Field & FWHM & $\sigma$\textsubscript{FWHM} & limiting magnitude\\
- & arcsec & arcsec & mag\\ 
\hline
$2$ & $0.75$ & $0.03$ & $18.5$ \\ 
$4$ &  $0.73$ & $0.05$ &  $18.1$ \\ 
$5$ &  $0.72$ & $0.04$ &  $18.2$ \\ 
$6$ &  $0.71$ & $0.03$ &  $18.5$ \\ 
$7$ &  $0.72$ & $0.03$ &  $18.5$ \\ 
$9$ &  $0.70$ & $0.05$ & $18.1$ \\ 
$10$ &  $0.67$ & $0.04$ & $18.3$ \\  
$11$ & $0.68$ & $0.01$ & $19.1$ \\  
$12$ &  $0.71$ & $0.03$ &  $18.6$ \\  
$13$ &  $0.75$ & $0.03$ &  $18.4$ \\  
$14$ &  $0.70$ & $0.05$ & $18.1$ \\  
$15$ &  $0.73$ & $0.05$ & $18.4$ \\ 
$16$ &  $0.72$ & $0.03$ &  $18.7$ \\  
$17$ &  $0.71$ & $0.04$ &  $18.4$ \\  
$18$ & $0.71$ & $0.04$ & $18.3$ \\  
$19$ &  $0.74$ & $0.04$ & $19.0$ \\  
$20$ &  $0.75$ & $0.03$ &  $18.5$ \\ 
$21$ & $0.72$ & $0.04$ &  $17.4$ \\
\hline 
\end{tabular}
\label{fwhmanddepth}
\end{table}

\section{Data processing}
\label{sec3}

Because of the small angular sizes of the compact stellar objects (UCD/GCs), identification of this type of object based on size and morphology is challenging. Using the space-based observations of the ACS/HST, in the absence of atmospheric turbulence and using the small pixel size of the ACS camera (0.049 arcsec/pixel), several thousands of resolved UCDs and GCs have been identified around nearby galaxies and galaxy clusters (\citealp{jordan2004,sharina2005,mark,jordan2015}). However, ground-based observations of compact objects are limited by the seeing. For a typical UCD with half-light radius of $\sim$20 pc, assuming a Gaussian profile for the object, the FWHM at the distance of Fornax cluster would be 0.4 arcsec (FWHM = 2 $\times$ r$_h$). Convolving this with the typical PSF (Gaussian) FWHM of 1.1 arcsec (for FDS g-band) leads to a final FWHM of 1.17 arcsec. This difference is 1/3 of the pixel-size of the OmegaCAM (0.21 arcsec/pixel) and is of the same order as the 1$\sigma$ uncertainties in g-band FWHM measurements ($\sigma$ = 0.07 arcsec or 0.33 pixel). In the case of the $Ks$ data which provides better image quality than the optical data (FWHM = 0.7 arcsec), the expected FWHM for a typical UCD is 0.1 arcsec ($\sim$0.3 pixel) larger than the PSF (atmospheric point spread function) FWHM. Note that the near-infrared data has lower signal-to-noise ratio than the optical which makes FWHM measurements less certain. As a result, many of the UCDs and all the GCs appear as point-sources, which makes them indistinguishable in shape from foreground stars in the Milky Way or distant galaxies. However, even when they are large enough to be separated from foreground stars, it is still a challenge to distinguish them from slightly resolved background galaxies. 

Traditionally, in addition to the sizes of sources, optical colours are used to find UCD/GC candidates around galaxies. However, separating compact sources (UCD/GCs) from the foreground stars solely using the optical colours is very challenging and optically selected candidates can be highly contaminated by foreground stars. \citet{Munoz-2014} showed that the GCs and UCDs of the Virgo galaxy cluster follow a well-defined sequence in the $u-i$/$i-Ks$ colour-colour diagram (the compact object sequence). This sequence provides a way to identify UCD/GCs and separate them from the foreground stars and background galaxies. However, because most of the time the near-infrared data is shallower than the optical, we can only use this selection for the brightest sources. 

\subsection{Source extraction}

Catalogues of the detected point-sources in each filter (6 filters in total: $u$, $g$, $r$, $i$, $J$ and $Ks$) were made using \textsc{SExtractor}. Because of the differences in the depth, FWHM and pixel-scales in different filters, different sets of \textsc{SExtractor} parameters were used for the source extraction and photometry. These parameters are presented in Table \ref{sexparams}. The PHOT\_APERTURES parameter in the table indicates the sizes (diameter) of the apertures used for aperture-corrected photometry (in pixels) and it is described later in the text.

When running \textsc{SExtractor}, we used a separate frame for detection. The detection frames were made by subtracting the median filtered science frames using a 16 $\times$ 16 pixels filter. This removes the extended light of the galaxies and increases the efficiency of the source detection around the bright objects except of near their centres (angular distances smaller than 30 arcsec or 3 kpc).

\begin{table}
\caption{\textsc{SExtractor} parameters that were used for source extraction and photometry expressed for each filter.}
\begin{tabular}{ ccccc } \hline  
 & $u$ & $g,r,i$ & $J$ & $Ks$ \\
\hline
DETECT$\_$MINAREA & 5 & 8 & 5 & 5 \\
DETECT$\_$THRESH & 1.5 & 2.0 & 1.5  & 1.5 \\
ANALYSIS$\_$THRESH & 1.5 & 2.0 & 1.5  & 1.5  \\
DEBLEND$\_$NTHRESH & $64$ & $64$ & $64$ & $64$ \\
DEBLEND$\_$MINCONT & $0.005$ & $0.005$ & $0.005$ & $0.005$ \\
PHOT$\_$APERTURES & 8,75  & 8,60 & 6,30 & 6,30 \\
BACKPHOTO$\_$TYPE & LOCAL & LOCAL & LOCAL  & LOCAL \\
BACKPHOTO$\_$THICK & $20$ & $20$ & $20$ & $20$\\
\hline
\end{tabular}
\label{sexparams}
\end{table}

To clean the \textsc{SExtractor} output catalogues and remove bad (false) detections, such as bad pixels, noise and cosmic rays, the extracted sources with FLAGS\footnote{The FLAGS parameter in \textsc{SExtractor} output is an integer which indicates the confidence level and the quality of the source extraction/photometry for a detection.} $>$ 3 were removed. In this way, we only include good detections and blended sources. Moreover, detected sources with FWHM$\_$IMAGE $<$ 1.0 pixel were excluded from the catalogues since they are too small to be real and most likely are bad detections or cosmic rays.

Once the single-band catalogues are made and cleaned, they were cross-matched within 1.0 arcsec uncertainty in their positions to create a multi-wavelength catalogue. Through this paper, we refer to this catalogue as the master catalogue. The master catalogue includes the multi-wavelength photometric information (magnitude, colour, FWHM, ellipticity, etc.) of $\sim$ 1,000,000 objects in at least 3 filters ($g$, $r$ and $i$) and $\sim$ 120,000 objects in all the 6 filters ($u$, $g$, $r$, $i$, $J$ and $Ks$). For analysis, we only consider the subset with full photometric coverage in 6 filters and throughout the paper we refer to it as the main catalogue. Among all the filters, the completeness of sources in the main catalogue is mostly limited by (in order) $Ks$, $u$ and $J$. In total, 88\% and 68\% of the extracted sources in \textit{g} data brighter than m$_{g}$ = 21 mag and m$_{g}$ = 22 mag, respectively, are also identified in $u$, $J$ and $Ks$ and included in the main catalogue. By assuming that the $g$ detections are complete up to m$_{g}$ = 22 mag, this means that the sources in the main catalogue are 88\% and 68\% complete to m$_{g}$ = 21 mag and m$_{g}$ = 22 mag.

\begin{figure*}
\centering
        \includegraphics[width=0.48\linewidth]{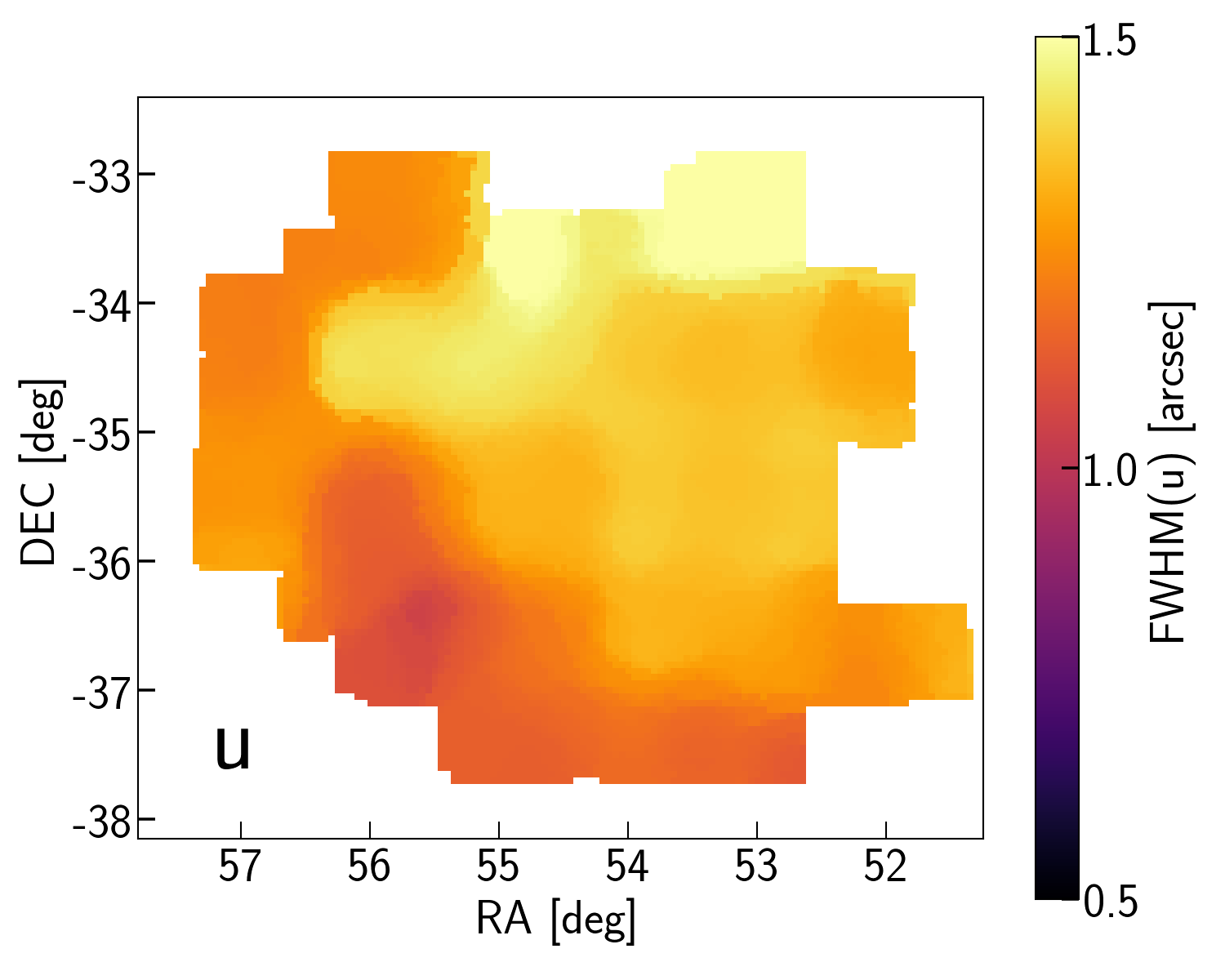}
        \includegraphics[width=0.48\linewidth]{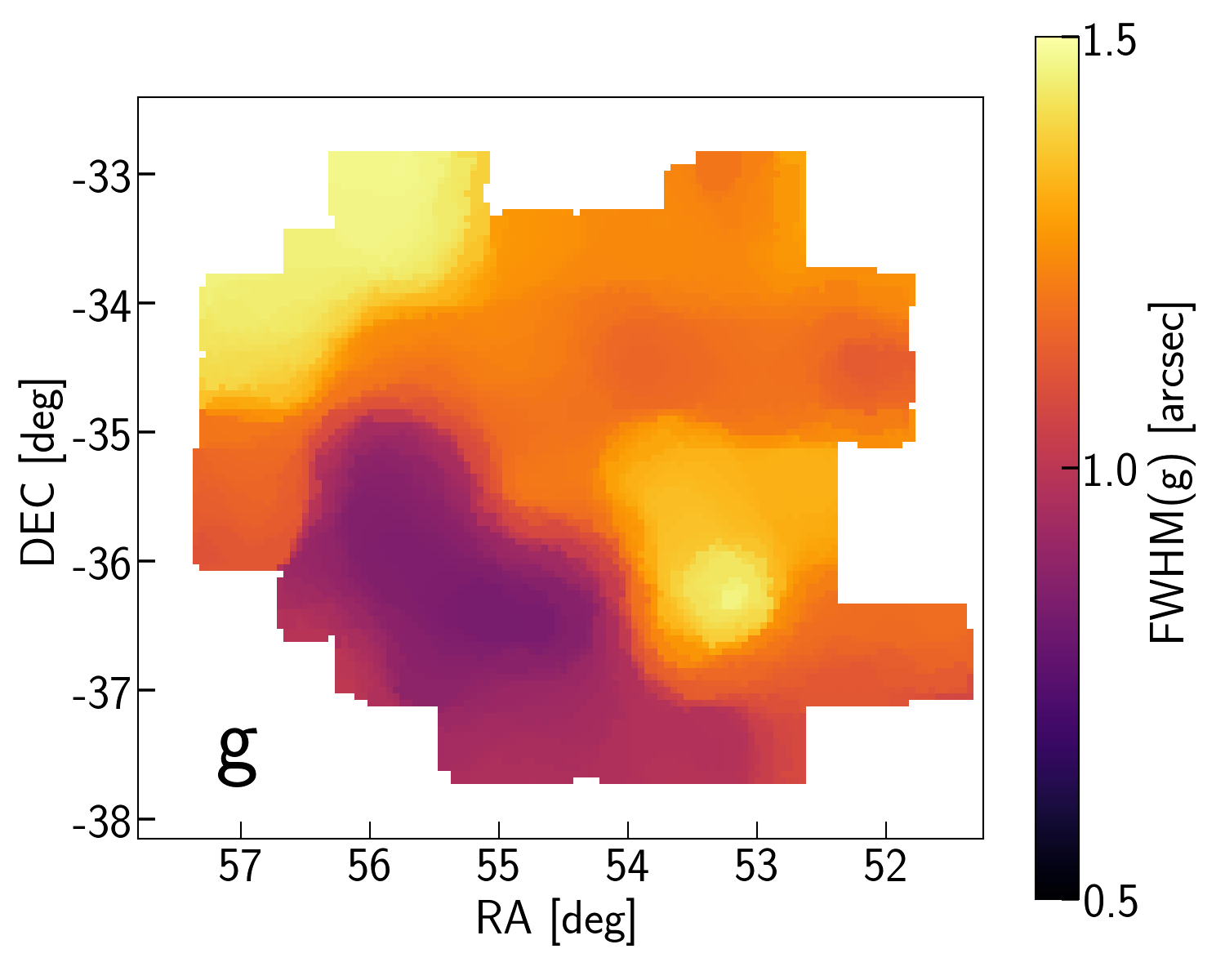}
        \includegraphics[width=0.48\linewidth]{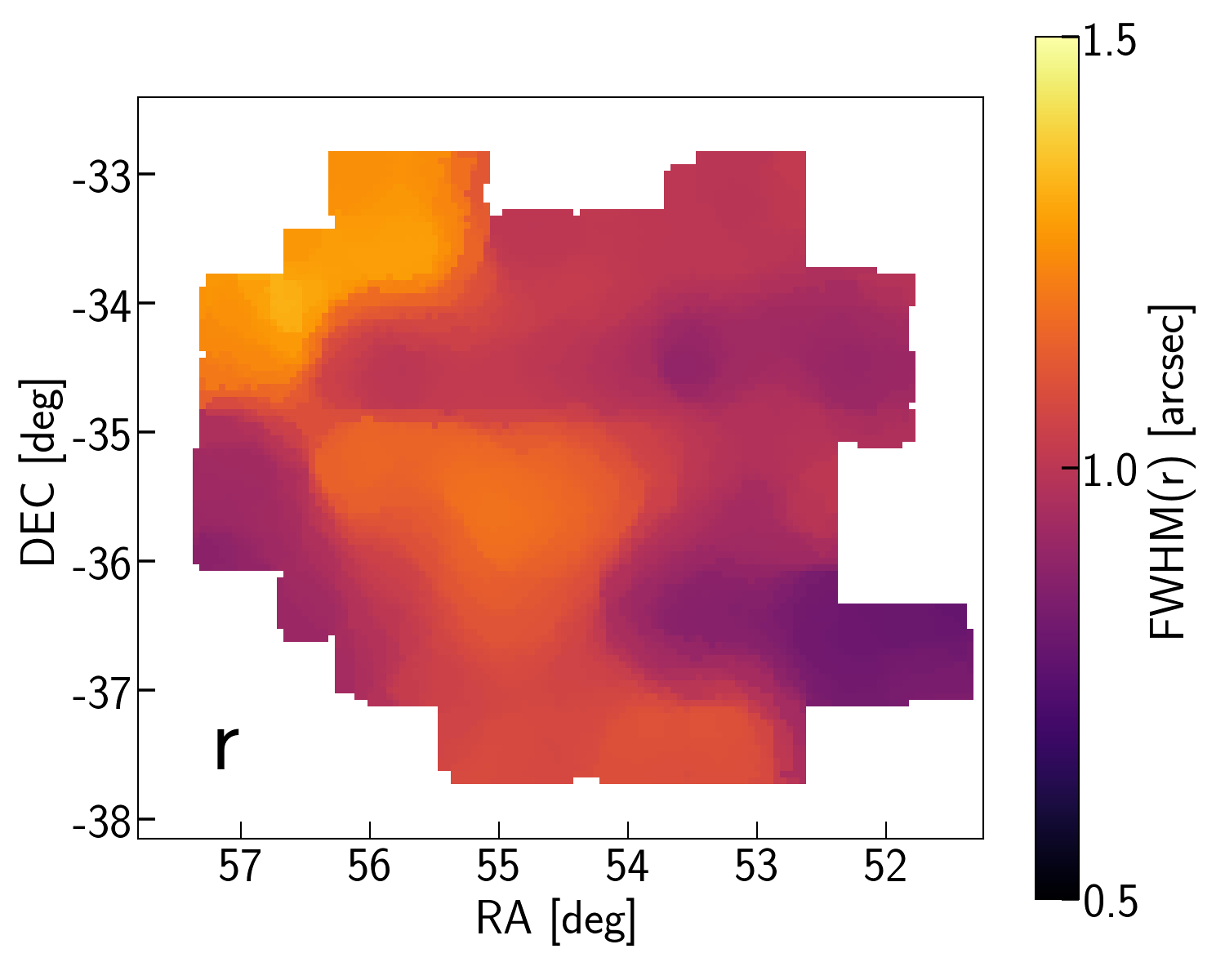}
        \includegraphics[width=0.48\linewidth]{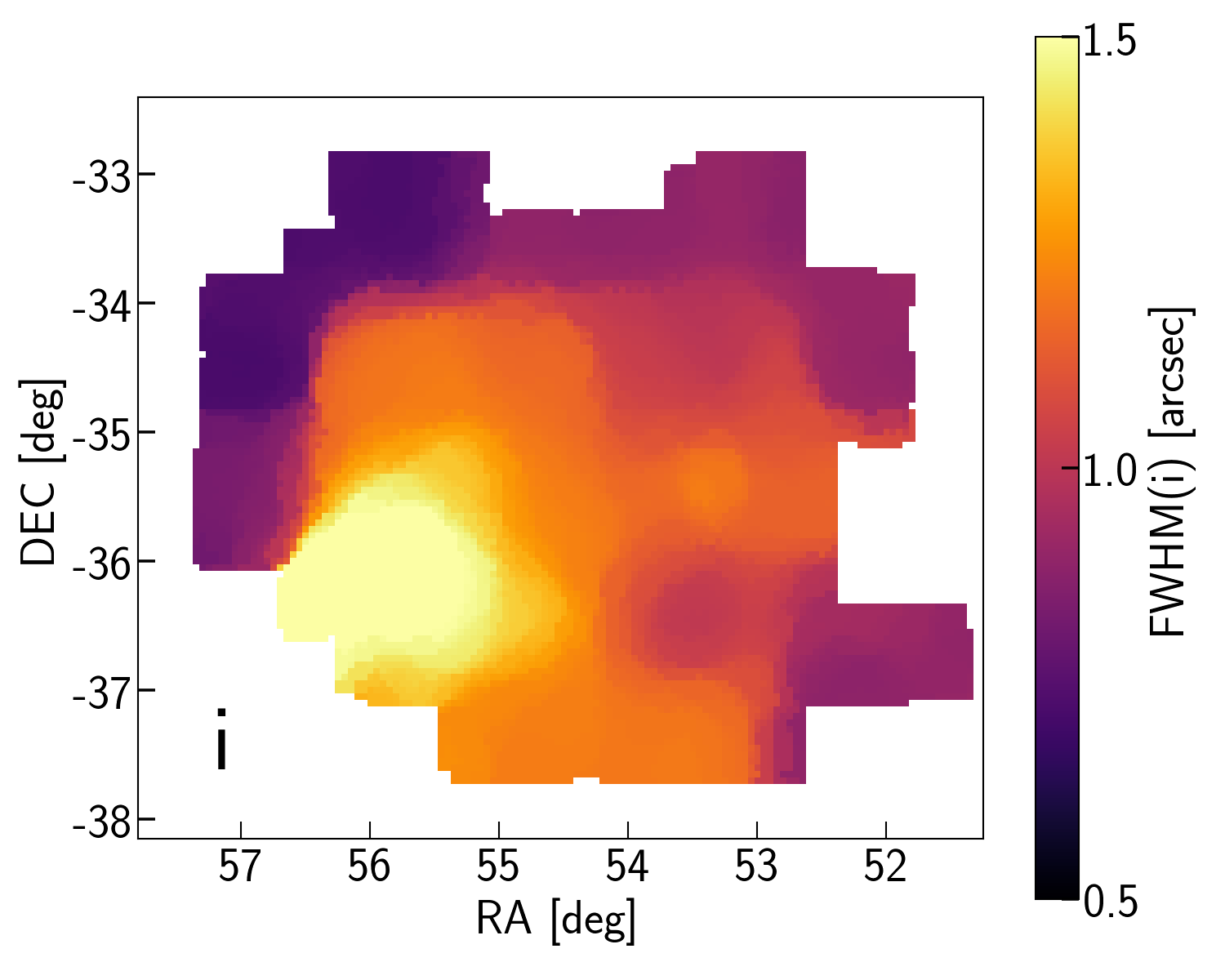}
        \includegraphics[width=0.48\linewidth]{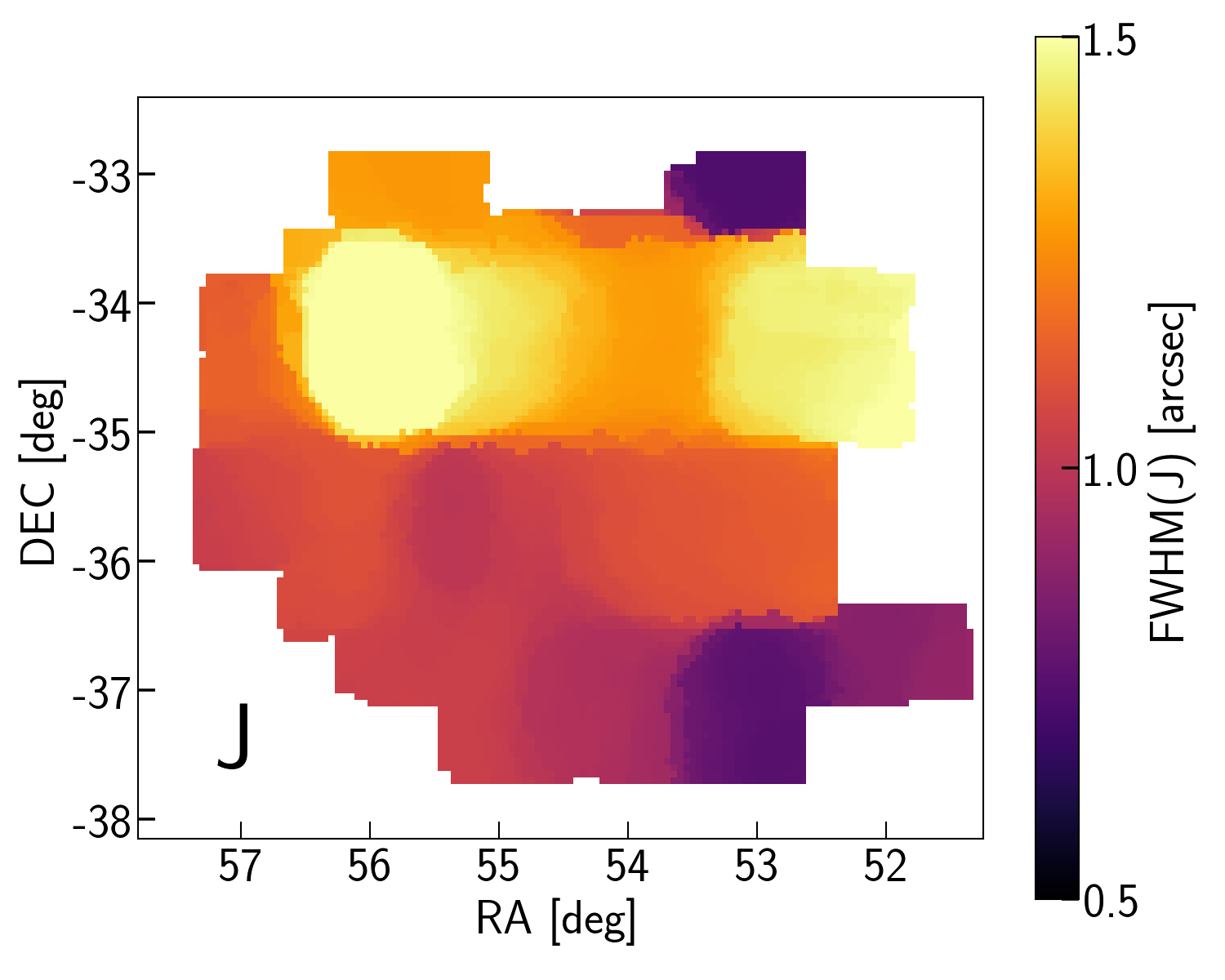}
        \includegraphics[width=0.48\linewidth]{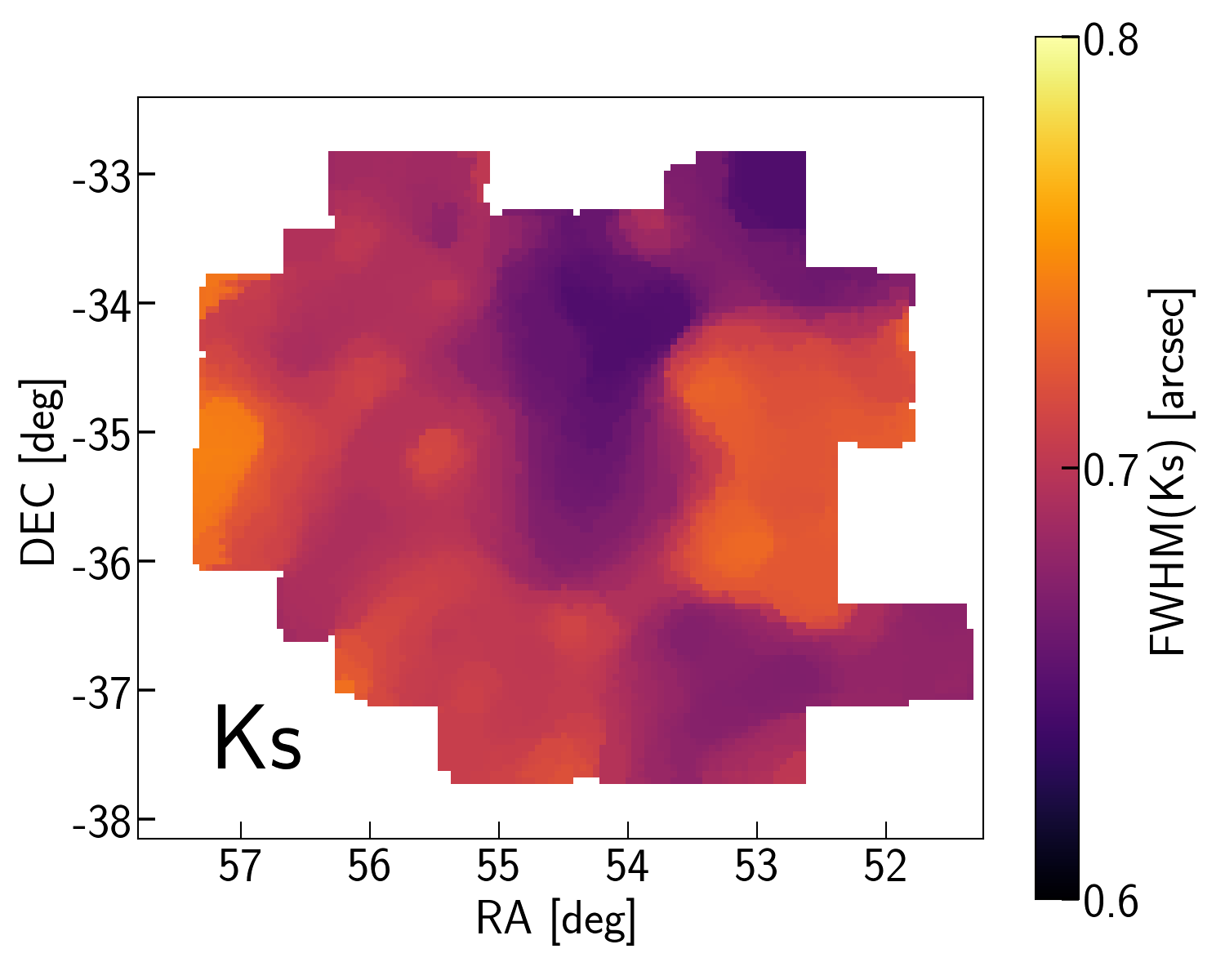}
    \caption{FWHM across the observed field in $u$, $g$, $r$, $i$, $J$ and $Ks$. Among these filters, $Ks$ has the best and the most stable FWHM. Note that the colourbar limits for the $Ks$ are different than those for the other filters.}
\label{fwhmmaps}
\end{figure*}

\begin{figure*}
\centering
        \includegraphics[width=\linewidth]{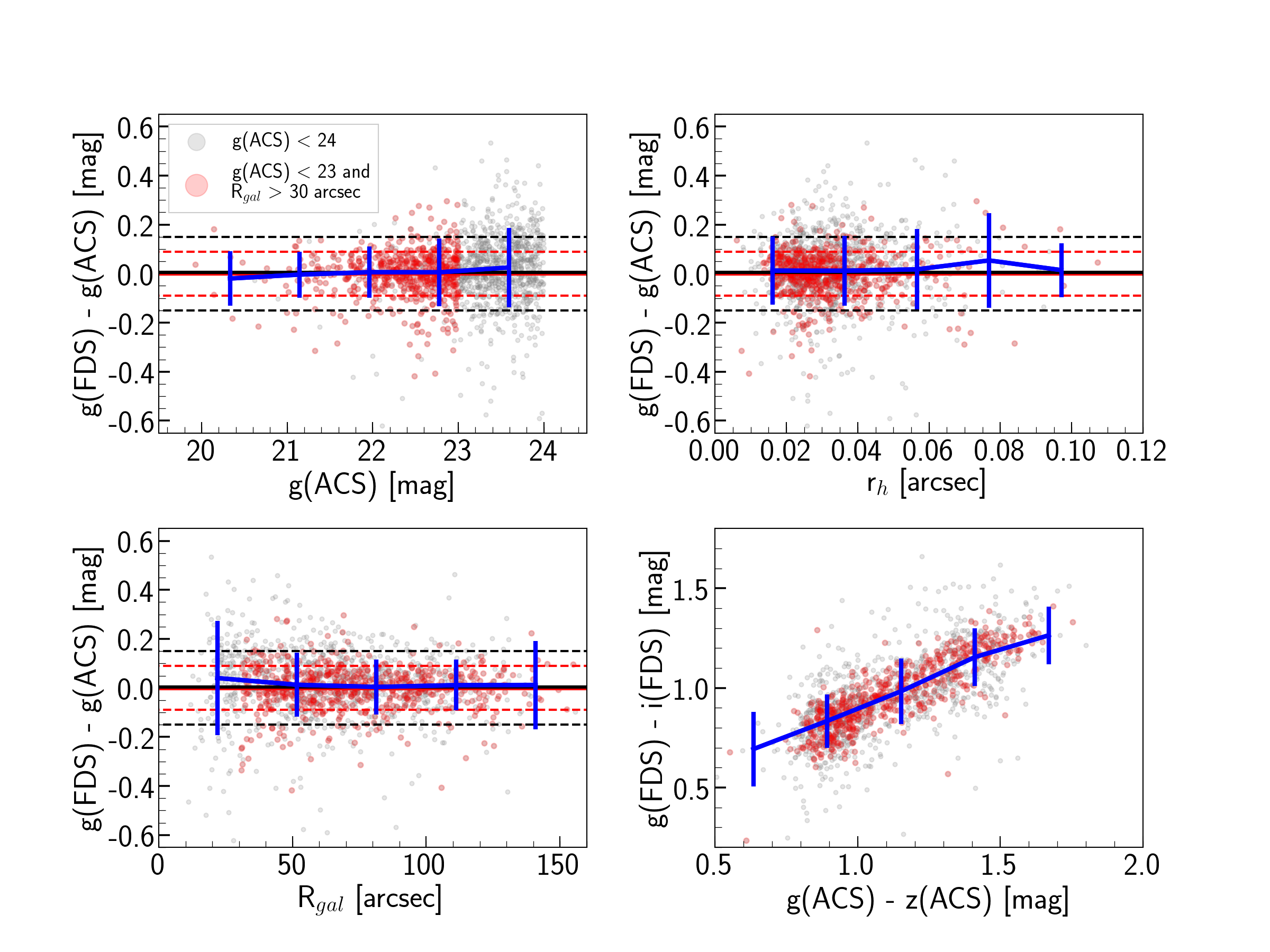}
\caption{Comparison between our photometry and the ACS survey of the Fornax cluster (ACSFCS, \citealp{jordan2015}). For the comparison objects with high GC probability $p_{gc}$ $>$ 0.75 are selected (from ACSFC). Grey points show sources brighter than m$_g$ $=$ 24 mag. Red points represent sources brighter than m$_g$ $=$ 23 mag and with R$_{gal}$ $>$ 30 arcsec (further than 30 arcsec or 3 kpc from their host galaxy). The solid and dashed lines represent the average and 1$\sigma$ deviation from average for the corresponding data points of the same colour. For both cases, the average is consistent with zero, and r.m.s. are 0.15 mag and 0.09 mag for grey and red points, respectively. The blue lines and error bars indicate the medians and standard deviations within different bins. The upper right panel shows the correlation between the magnitude residuals and size of the sources (King radius from ACSFCS). It is clear that the photometry is consistent for the whole range of sizes.}
\label{fdsvsacs}
\end{figure*}

\begin{figure*}
\centering
        \includegraphics[width=\linewidth]{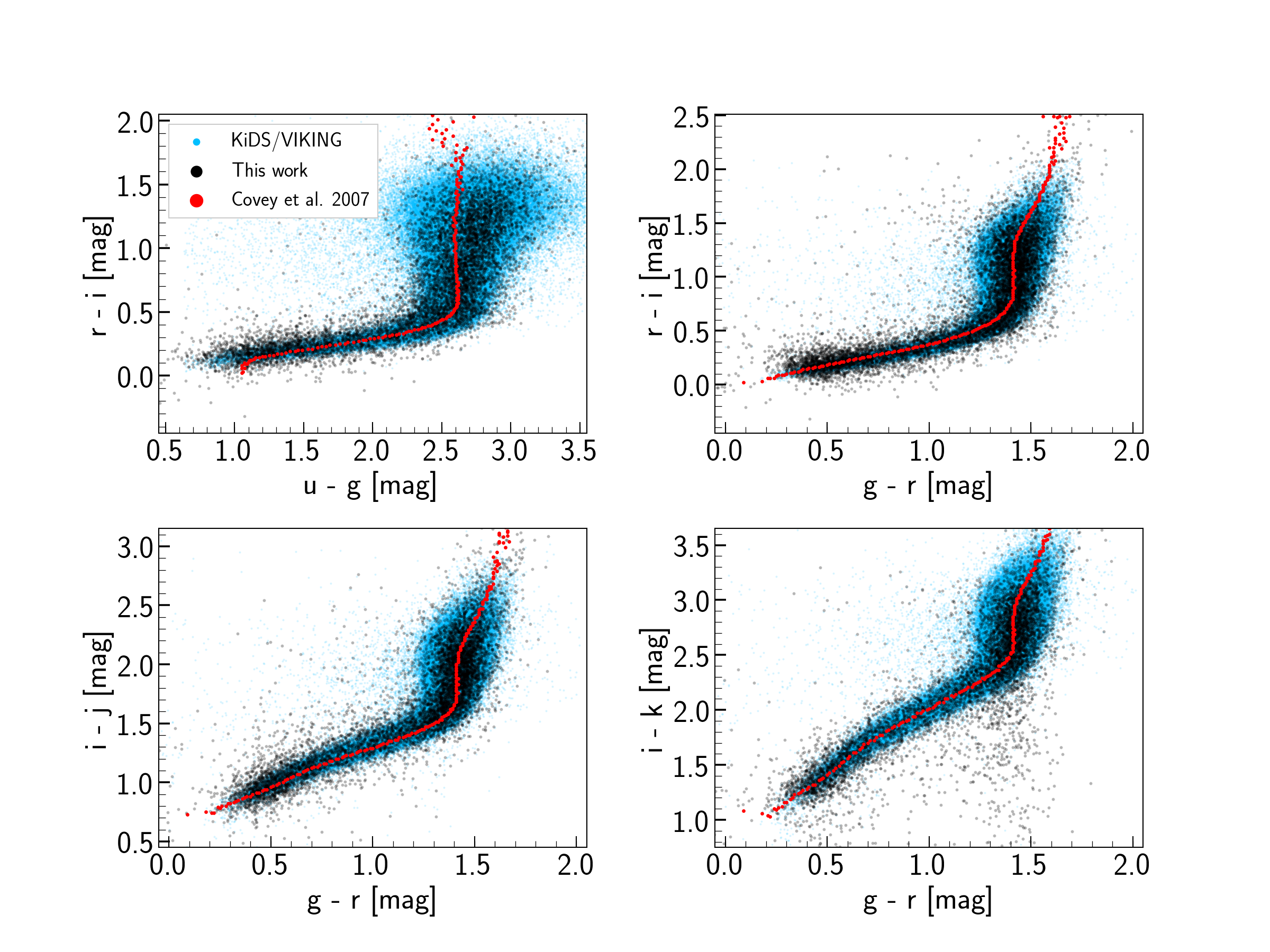}
\caption{Stellar locii and photometric comparison between our photometry (black dots) and KiDS DR4 (blue dots, \citealp{kids-dr4}). On top, the stellar locii from \citet{covey} based on the SDSS/2MASS observations of the bright stars are shown in red.}
\label{fdsvskids}
\end{figure*}

\subsection{Photometry}

Good photometry for the faint point sources can be achieved using aperture photometry. However, because of the spatial variation of the FWHM (Fig. \ref{fwhmmaps}), aperture photometry with a constant aperture size does not provide \textit{solid results}. In that case, the aperture sizes should be adjusted to FWHM and filter. A solution is to perform photometry within a constant aperture (using \textsc{SExtractor}) for each filter and correct the photometry for the applied aperture-size as the following: At first, magnitudes are measured using a smaller aperture (initial aperture) and then, they are corrected for the fraction of missing light. This fraction is estimated by doing aperture photometry for the nearest bright and isolated stars (stars in the same region as the object which have the same FWHM) using two different aperture sizes; the size of the initial aperture and a larger aperture to capture the total light of the bright stars. The difference between these two magnitudes is an estimation for the fraction of the missing light and should be added to the aperture magnitude (using initial-aperture size) of all the objects.

We use the \textsc{SExtractor} output (aperture magnitude: MAG\_APER) to perform aperture corrected photometry for all the sources. At first, we made a star catalogue (real point-sources) based on the second Gaia data release (Gaia DR2, \citealp{gaia}) by selecting the objects with with accurate distance measurements, given by $ \frac{\text{parallax}}{\text{parallax$\_$error}} $ $>$ 5.0. Afterwards, we used aperture sizes as presented in Table \ref{sexparams}. We measured the magnitude difference between the small and large apertures for the stars in the Gaia-based star catalogue and calculated the median of this difference. The calculated value was added to the measured aperture magnitude (with smaller aperture size) of all the sources. To avoid uncertainties caused by FWHM variations, we selected the nearest bright (and not-saturated) stars (closer than 15 arcmin) for each source. The aperture correction values for each pointing are calculated based on 50-100 stars and have an average r.m.s of $\sim$0.02 mag.

Once the magnitudes were estimated (using aperture corrected photometry), we re-calibrated the zero-points of the optical data using the overlapping regions between fields as follows. For a given filter, we identified stars in the overlap region between pairs of fields. Next, we measured the average magnitude residuals between the fields using the brightest (non-saturated) over-lapped stars. Then, we calculated the absolute value of the zero-point corrections for a given field. The uncertainties associated to the zero-points of the FDS as is estimated by \citet{venhola2018} are 0.03 mag in $g$, $r$, $i$ and 0.04 mag in $u$. These values are calculated using the stellar locus test (\citealp{ivezic}). Therefore, since the global zero-points are precise enough, to calculate the (relative) zero-point re-calibration values, we assumed that the average re-calibration value is zero. In other words, the global re-calibration value of the whole data is zero. Table \ref{zpcorr} presents the values of zero-point re-calibration corresponding to the FDS fields. All the optical magnitudes from aperture-corrected photometry were corrected accordingly. Next, we address the consistency of the measured magnitudes and colours by direct comparison with the sources in the ACS survey of the Fornax cluster (ACSFCS, \citealp{jordan2015}) and KiDS DR4 (\citealp{kids-dr4}). 

\begin{table}
\centering
\caption{Zero-point re-calibration values of the FDS fields. The list of fields here represents all the fields with available $u$, $g$, $r$ and $i$ data. Note that the $u$-band data do not cover fields 22, 25, 26, 27, 28. Moreover, the $u$-band data have the largest zero-point re-calibration values. Values $\leq$0.02 are ignored.}
\begin{tabular}{ ccccc } \hline  
FDS Field & $\Delta$ZP$_u$ & $\Delta$ZP$_g$ & $\Delta$ZP$_r$ & $\Delta$ZP$_i$\\
- & mag & mag & mag & mag \\ 
\hline
1 & 0.27 & --- & --- & -0.03 \\
2 & 0.29 & --- & --- & ---\\
4 & 0.15 & --- & --- & ---\\
5 & -0.1 & --- & 0.03 & ---\\
6 & -0.07 & 0.07 & --- & 0.06 \\
7 & -0.03 & --- & 0.04 & ---\\
9 & 0.07 & --- & --- & ---\\
10 & -0.09 & --- & --- & ---\\
11 & 0.06 & 0.04 & --- & 0.03 \\
12 & --- & 0.03 & 0.04 & ---\\
13 & 0.37 & --- & 0.04 & ---\\
14 & --- & --- & --- & ---\\
15 & --- & --- & --- & ---\\
16 & -0.17 & 0.04 & --- & 0.03 \\
17 & 0.11 & --- & --- & ---\\
18 & 0.31 & --- & --- & ---\\
19 & -0.05 & --- & --- & ---\\
20 & -0.05 & --- & --- & ---\\
21 & 0.15 & --- & --- & ---\\
22 & --- & --- & --- & ---\\
25 & --- & --- & --- & ---\\
26 & --- & 0.06 & --- & 0.05 \\
27 & --- & --- & --- & ---\\
28 & --- & --- & --- & ---\\
31 & --- & --- & --- & ---\\
\hline 
\end{tabular}
\label{zpcorr}
\end{table}

A sample of GCs from the ACSFCS survey (\citealp{jordan2015}) is selected based on the high-resolution imaging of the ACS camera of the Hubble space telescope (HST), which resolves the GCs larger than 0.01 arcsec ($\sim$0.2 ACS pixels) or 0.96 pc at the distance of the Fornax cluster. This catalogue provides accurate magnitudes and sizes of sources in two filters g$_{475}$ and z$_{850}$ and GC probabilities $p_{gc}$, which is the probability that an object is a globular cluster. This catalogue also provides a very accurate (relative) astrometry for the sources. In \citet{jordan2007}, the authors stated that for 35 galaxies out of 43 target galaxies of the ACSFCS, they could find enough astrometric reference sources to derive a reliable astrometry. The galaxies for which they could not carry out the correction were usually the brighter ones\footnote{Jord{\'a}n, private conversation}. We also found an offset between the coordinates in our data and ACSFCS catalogue mainly around the brightest galaxies. In the most extreme cases, for the objects around NGC1399 (FCC213) and NGC1316 (FCC21), while the relative astrometry is good, an $>$ 1 arcsec offset in absolute astrometry can be seen. Therefore, before any direct comparison, we corrected the absolute astrometry of the sources. The correction values in RA ($\Delta$RA) and DEC ($\Delta$DEC) are presented in Table \ref{acscorr}. $\Delta$RA and $\Delta$DEC are the average offset values between our data and \citet{jordan2015}: $\Delta$RA = RA$_{FDS}$ - RA$_{ACS}$ and
$\Delta$DEC = DEC$_{FDS}$ - DEC$_{ACS}$.
\\

\begin{table}
\centering
\caption{Astrometric offsets of the ACSFCS catalogue of GCs (\citealp{jordan2015})}
\begin{tabular}{ c c c } \hline  
Host galaxy & $\Delta$RA & $\Delta$DEC \\
FCC number & arcsec & arcsec  \\ 
\hline
19  & 0.53  & -0.25 \\
21  & 0.08  & 1.30 \\
43  & -0.08  & -0.23 \\
47  & 0.17  & -0.36 \\
55  & 0.11  & -0.23 \\
83  & 0.27  & -0.37 \\
90  & 0.11  & -0.00 \\
95  & 0.06  & -0.20 \\
100  & 0.23  & -0.22 \\
106  & 0.25  & -0.22 \\
119  & -0.13  & -0.18 \\
136  & -0.19  & -0.09 \\
143  & 0.22  & -0.24 \\
147  & 0.49  & -0.84 \\
148  & 1.11  & 0.22 \\
153  & 0.02  & -0.03 \\
167  & 0.09  & -0.08 \\
170  & -0.05  & -0.27 \\
177  & -0.13  & -0.29 \\
182  & -0.08  & 0.03 \\
184  & -0.10  & -0.19 \\
190  & -0.02  & -0.44 \\
193  & 0.04  & -0.32 \\
202  & 0.10  & -0.14 \\
203  & -0.14  & -0.23 \\
204  & 0.01  & -0.22 \\
213  & 1.77  & -0.10 \\
219  & 0.39  & -0.09 \\
249  & 0.14  & -0.17 \\
255  & 0.10  & -0.22 \\
276  & 0.05  & -0.20 \\
277  & 0.04  & -0.02 \\
288  & 0.01  & -0.17 \\
301  & 0.04  & -0.27 \\
303  & 0.06  & -0.31 \\
310  & 0.27  & -0.13 \\
335  & 0.25  & -0.32 \\
\hline 
\end{tabular}
\label{acscorr}
\end{table}

We compare the measured $g$-band magnitudes with the g$_{475}$ magnitudes of the ACSFCS in Fig. \ref{fdsvsacs}. For the comparison, objects with high GC probability $p_{gc}$ $>$ 0.75 are selected. The average value for objects brighter than m$_g$ $=$ 24 mag is consistent with zero and r.m.s. is $\sigma_{FDS-ACS}$ = 0.15 mag. For objects brighter than m$_g$ $=$ 23 mag with a distance from their host (the nearest bright galaxy) larger than R$_{gal}$ $=$ 30 arcsec (3 kpc) the scatter drops to $\sigma_{FDS-ACS}$ = 0.09 mag. 

Aperture photometry is a reliable method for point-source photometry. However, for extended objects, it misses some of the light, which leads to a larger (fainter) magnitude. Since UCDs can appear slightly extended in the FDS images, we investigated the correlation between the $g$-band magnitude residuals with the objects' sizes (from ACSFCS). As Fig. \ref{fdsvsacs} shows there is no clear correlation between magnitude difference and size of the objects smaller than r$_h$ $\sim$ 10 pc. Fig. \ref{fdsvsacs} also compares the \textit{g-z} colours of the ACS with the $g-i$ colours from our measurements. Moreover, the stellar loci and colours are consistent with KiDS DR4 (\citealp{kids-dr4}) and \citet{covey}. Fig. \ref{fdsvskids} compares the stellar locus in the optical/near-infrared colours space. Stars in our data and KiDS are selected based on the Gaia DR2 parallaxes.

We also compare our magnitudes with the magnitudes of UCDs in \citet{voggel-2016}. These authors, using better seeing conditions ($\sim$0.5 arcsec) and smaller pixel-size of the instrument (FORS2/VLT, 0.126 arcsec/pixel) reported the $V$-band magnitude and half-light radii of 13 UCDs in the Fornax cluster and in the halo of NGC1399 of which, 12 out of 13 UCDs are identified in our data. These sizes and magnitudes are measured by profile-fitting and take into account the differences in light profiles of UCDs. Fig. \ref{size} compares the $V$-band magnitudes from our work with those of the full UCD sample from \citet{voggel-2016}, also including UCDs without a size but magnitude measurement. $V$-band magnitudes m$_V$ (this work) are derived from the photometric transformations in \citet{jester2005} using m$_V$ = m$_g$ - 0.58 $\times$ (m$_g$ - m$_r$) - 0.01. The consistency of the magnitudes from both works implies that for UCDs with sizes between 20 to 50 pc, the aperture-corrected photometry is reliable.

\subsection{Measuring sizes}

As discussed above, measuring angular sizes of UCDs at the distance of the Fornax cluster is challenging. Accurate size measurement of compact-sources demands high-resolution space-based observations. However, for the largest UCDs, ground-based observations in an excellent seeing condition ($<$0.5 arcsec) are sufficient to measure sizes (half-light radii) as small as $\sim$10 pc in the nearby galaxy clusters (\citealp{richtler2005,voggel-2016,Liu-2020}).

\citet{andres} showed that fitting a profile to extra-galactic star clusters, such as a King profile (\citealp{king}) needs high angular resolution and high signal-to-noise ratio data.
In case of lower S/N data, the authors suggest adopting a simple Gaussian distribution. Following the suggestion in \citet{andres}, we assume a Gaussian light profile for the UCD/GCs. In this case, the observed light profile of the objects is the result of a Gaussian distribution (intrinsic distribution) convolved with another Gaussian distribution (atmospheric+instrumental PSF). Therefore, the value of the true (intrinsic) FWHM in the absence of atmospheric seeing (or as we call it FWHM$^*$) can be measured using:\\\\
$FWHM* = \sqrt[]{{FWHM }^2 - {FWHM_{PSF}}^2}$
\\

Here FWHM is the measured $g$-band FWHM of the object (using \textsc{SExtractor}) and FWHM$_{PSF}$ represents the FWHM of the point spread function (PSF). FWHM$_{PSF}$ is measured as the mean value of the FWHM of point-sources (stars from the Gaia DR2 star catalogue) within 15 arcmin from the object. For objects with FWHM $<$ FWHM$_{PSF}$, FWHM$^*$ is set to zero. The measured FWHM$^*$ value is converted to the half-light radius r$_h$ (also called effective radius $R_e$) using FWHM = 2 $\times$ r$_h$ (value for a Gaussian distribution).

\begin{figure}
        \includegraphics[width=\linewidth]{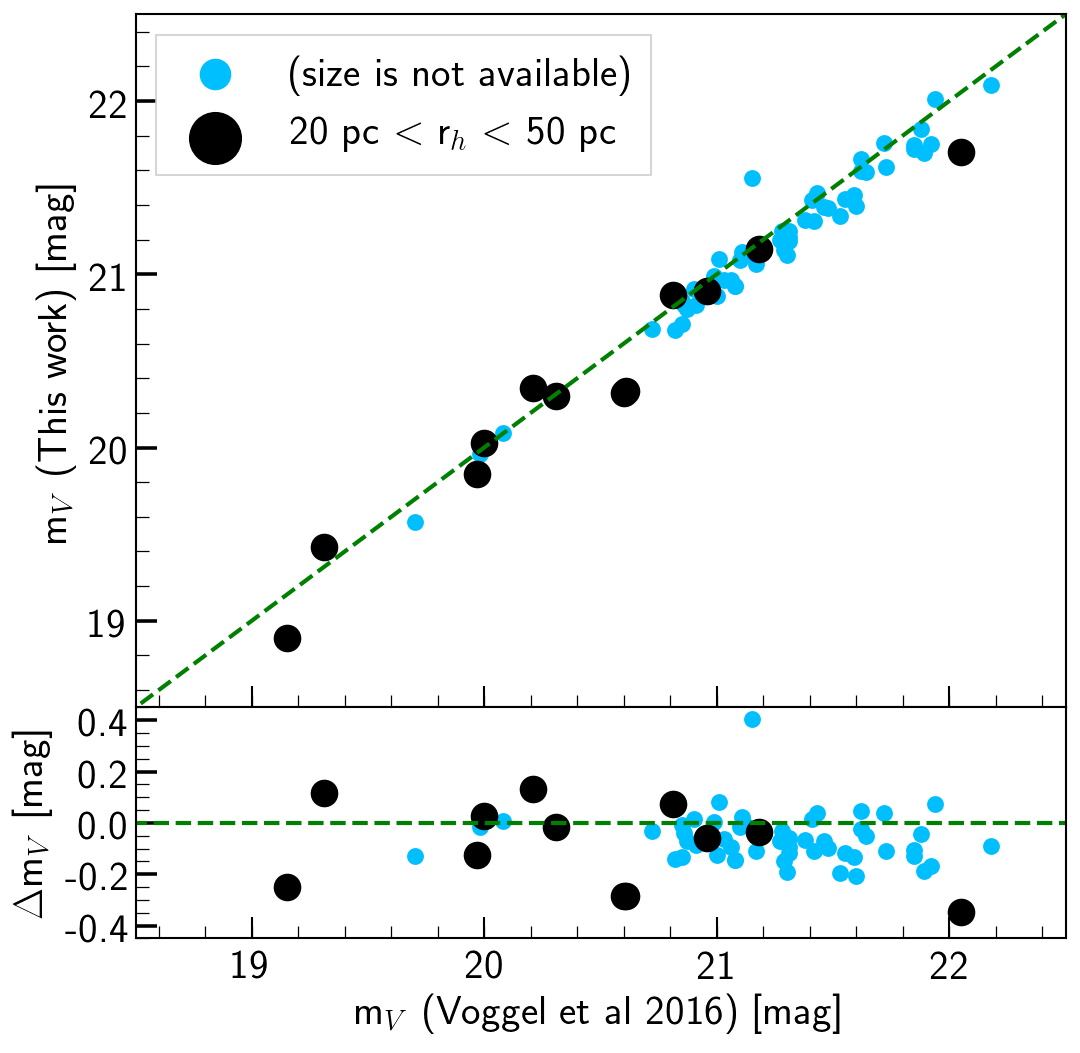}
        \includegraphics[width=\linewidth]{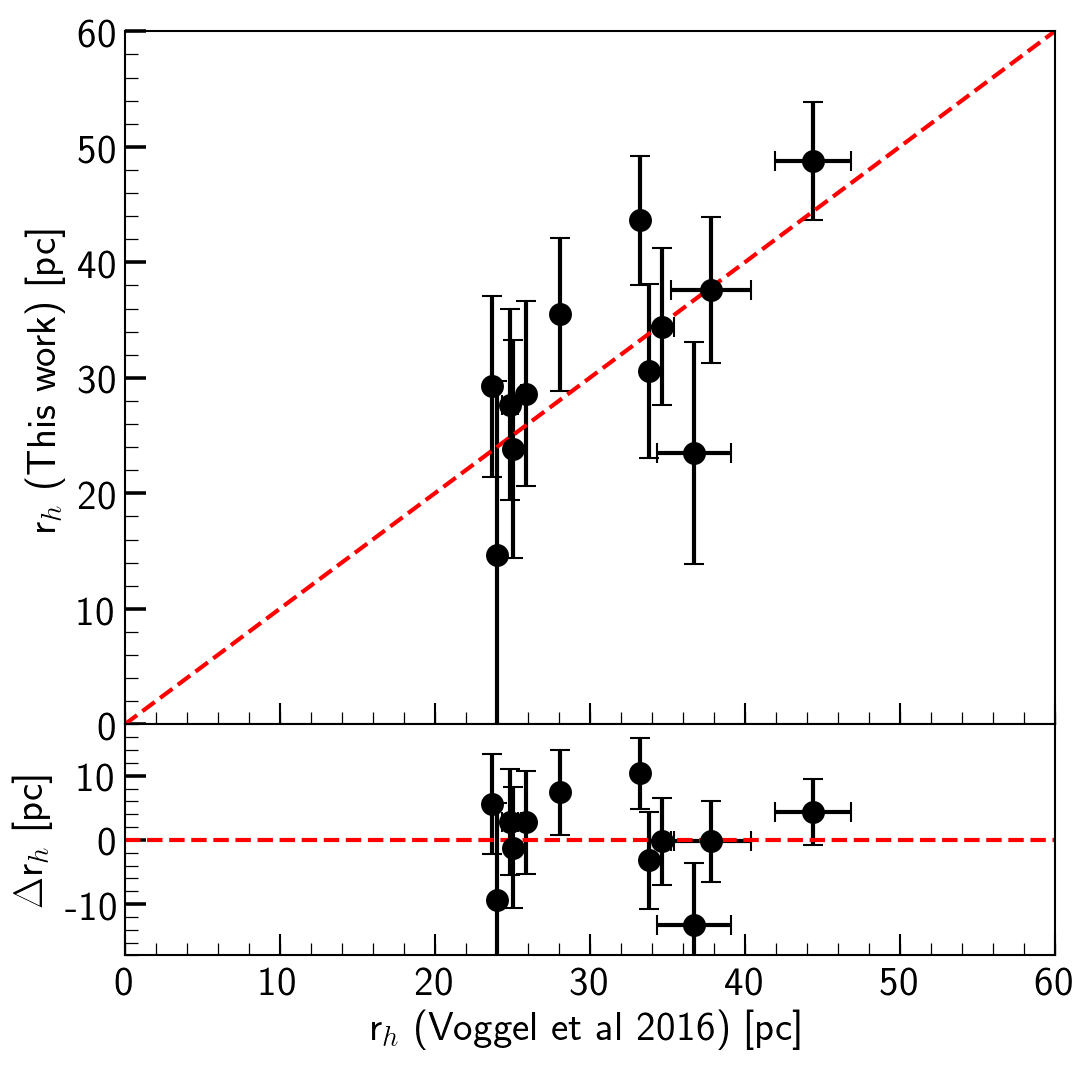}
\caption{Comparison between the $V$-band magnitudes (top) and sizes (bottom) of this work and \citet{voggel-2016} for 12 UCDs (black points). These UCDs are located in the halo of NGC1399 in the centre of the cluster and have sizes between 20 to 50 pc. Blue points represent the rest of the initial UCD sample of \citet{voggel-2016} without size estimates since they are too small for model-fitting even in an excellent seeing condition. The uncertainties in magnitudes are smaller than the drawn data points and are not shown in the figure. $\Delta$m$_V$  and $\Delta$r$_h$ represent the magnitude and size difference between our work and \citet{voggel-2016}.}
\label{size}
\end{figure}

Using the FDS data with average $g$-band seeing of 1.1 arcsec, we are able to measure FWHM$^*$ within 0.4 arcsec uncertainties which implies that, for a Gaussian distribution, the uncertainties in measuring the half-light radius of sources is 0.2 arcsec (FWHM = 2 $\times$ r$_h$) or 20 pc at the distance of the Fornax cluster (1 arcsec = 100 pc). In Fig. \ref{size} we make a comparison between our measured sizes and \citet{voggel-2016}. Given the better seeing condition and smaller pixel-size of the instrument of their observations, their reported values are expected to be more precise than ours which provides a benchmark to inspect the accuracy of our measurements. As is seen in Fig \ref{size} (bottom), our measured values, within the given uncertainties are consistent with those of \citet{voggel-2016}.

For objects smaller than 20 pc, we do not expect to measure reliable sizes. This limit is almost twice as large as the largest UCD/GCs in ACSFCS. We use the FWHM$^*$ values in the first step of the UCD identification to exclude the extended sources and we do not define a lower limit on FWHM$^*$ (sizes) of objects. The sizes of the identified UCDs are presented later in the result section.

\section{UCD identification}
\label{sec4}
We use the main catalogue to discover new UCD candidates. We do that in three steps. Here, for clarity,  we first shortly summarize these steps.

\textit{Step i.} Make the \textit{KNOWN} and \textit{UNKNOWN} catalogues: First we divide our sources in the main catalogue into two different catalogues, the \textit{KNOWN} and \textit{UNKNOWN} catalogues containing sources with and without available spectroscopic data (radial velocity) in three reference catalogues: \citet{wittmann-2016,Pota-2018,Maddox-2019} and references therein. In the \textit{KNOWN} catalogue, sources are divided into three classes based on their radial velocities. These three classes are: foreground stars, confirmed UCD/GCs and background galaxies.

\textit{Step ii.} Pre-selection: Based on the observed magnitude, size and $u-i$/$i-Ks$ colours of the confirmed UCD/GCs in the \textit{KNOWN} catalogue, we determine the ranges of their magnitudes, sizes and colours to settle limits on these parameters. In turn  we apply these limits to all the objects in the \textit{KNOWN} and \textit{UNKNOWN} catalogues, to exclude likely non-UCDs in the samples. These criteria exclude $>$ 95\% of the foreground stars and background galaxies (non-UCD/GCs) of the \textit{KNOWN} catalogue while keeping almost all the confirmed UCD/GCs. Therefore, it is also expected that the applied criteria remove the majority ($>$ 95\%) of the non-UCD/GCs from the \textit{UNKNOWN} catalogue as well.

\textit{Step iii.} Selection: we use 5 independent colours, namely $u-g$, $g-r$, $r-i$, $i-J$ and $J-Ks$ and perform a supervised classification to identify UCD/GCs among the pre-selected \textit{UNKNOWN} sources. It means that we train a machine learning model using the pre-selected objects in the \textit{KNOWN} catalogue (as the training-set) and apply the model to classify the pre-selected sources of the \textit{UNKNOWN} catalogue. By classifying objects, we assign a class (label) to each object in the \textit{UNKNOWN} catalogue. These classes are similar to the three classes of objects in the \textit{KNOWN} catalogue (foreground stars, confirmed UCD/GCs and background galaxies). Eventually, we select compact sources based on the outcome of the classification as the objects that are classified as UCD/GC. As a result, we have two catalogues: a catalogue of compact sources with m$_g$ $<$ 21 mag (UCD candidates in our definition) and one with 21 $<$ m$_g$ $<$ 24.5 mag (GC candidates).

\subsection{Catalogue of \textit{KNOWN} and \textit{UNKNOWN} sources (step i)}

\begin{figure}
\centering
        \includegraphics[width=\columnwidth]{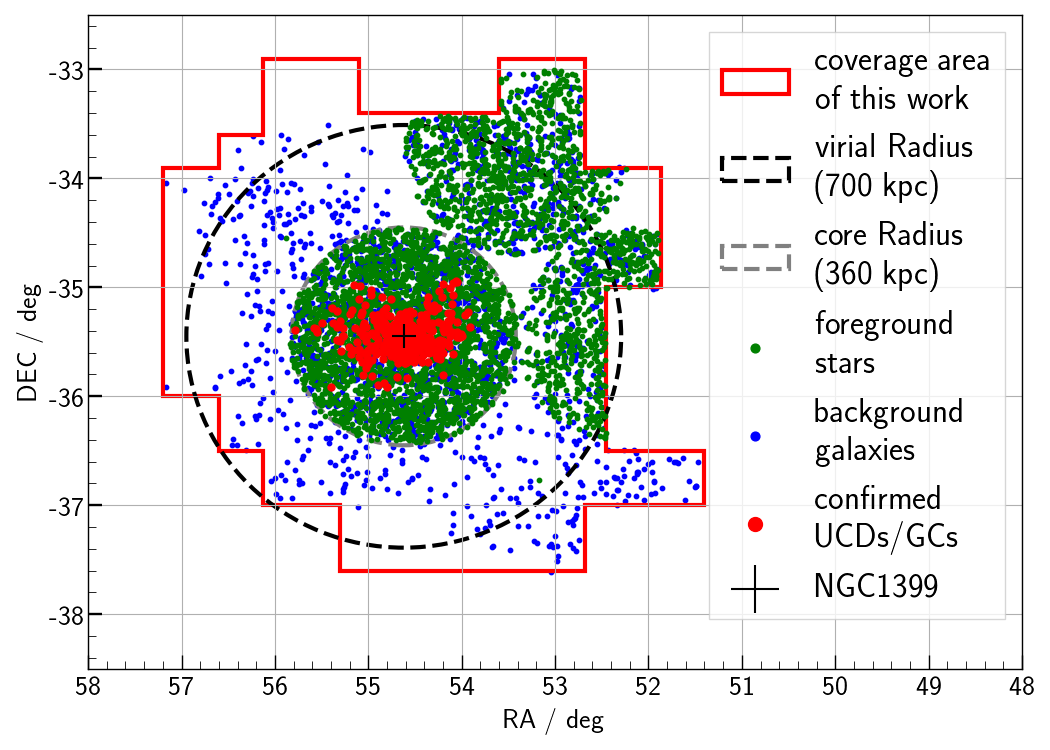}
\caption{The projected distribution of the sources in the \textit{KNOWN} catalogue. The circular shape of the data-points in the centre is a signature of the 2dF survey of the Fornax cluster (\citealp{drinkwater2000}). The foreground stars (green) and background galaxies (blue) are mainly from two surveys: \citet{drinkwater2000} (almost all the forgeround stars and 70\% of the background galaxies) and \citet{Maddox-2019} (25\% of the background galaxies). Within the covered area, these two surveys are about 80\% complete down to m$_g$ = 20 mag. The confirmed UCD/GCs (red) are a compilation of many other deeper surveys in the central region of the cluster (\citealp{wittmann-2016,Pota-2018} and references therein)}.
\label{knowndist}
\end{figure} 

\begin{figure*}
        \includegraphics[width=\linewidth]{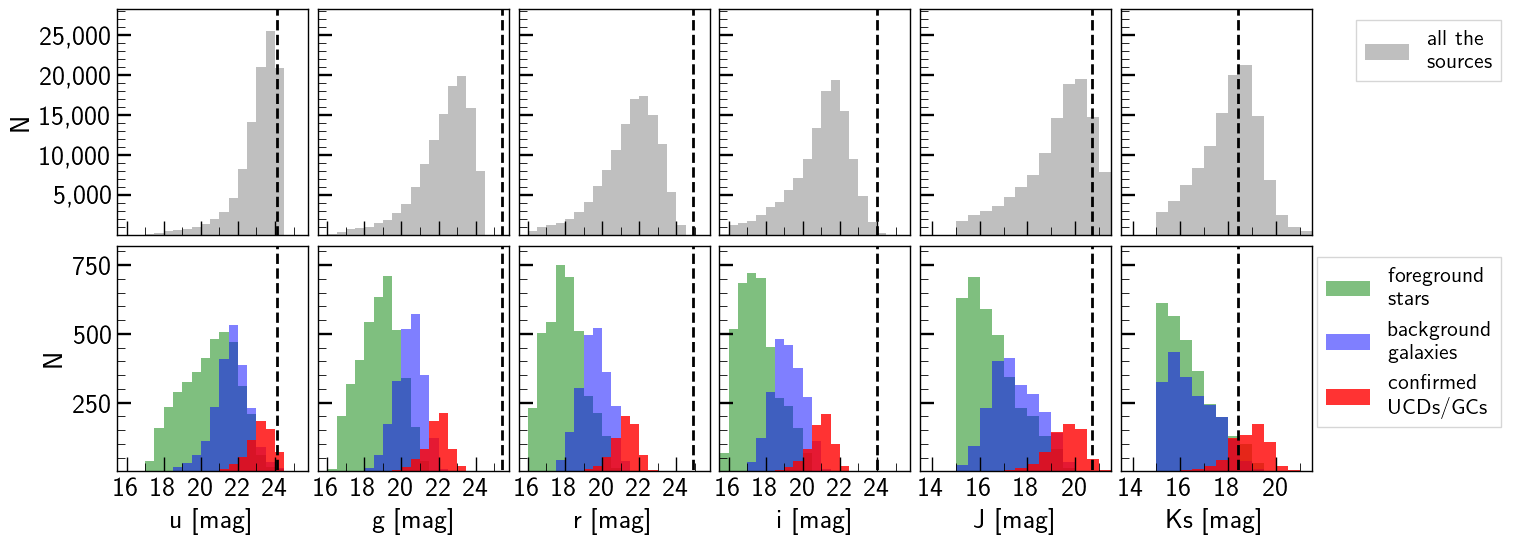}
\caption{Magnitude distribution of the sources in different filters in the \textit{UNKNOWN} catalogue (top) and the \textit{KNOWN} catalogue (bottom). Foreground stars, confirmed UCD/GCs and background galaxies are shown in green, red and blue. The 5$\sigma$ limiting magnitude in each band is indicated by the dashed black line. }
\label{hist-mag}
\end{figure*}

\begin{figure}
\centering
        \includegraphics[width=\columnwidth]{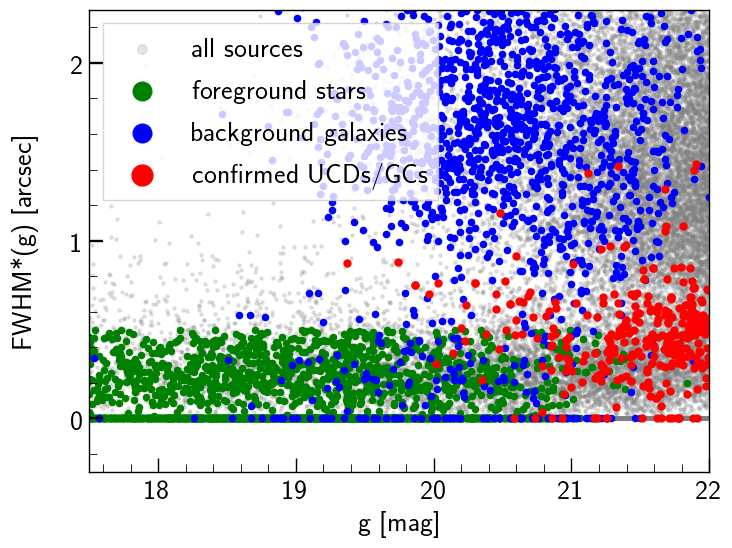}
        \includegraphics[width=\columnwidth]{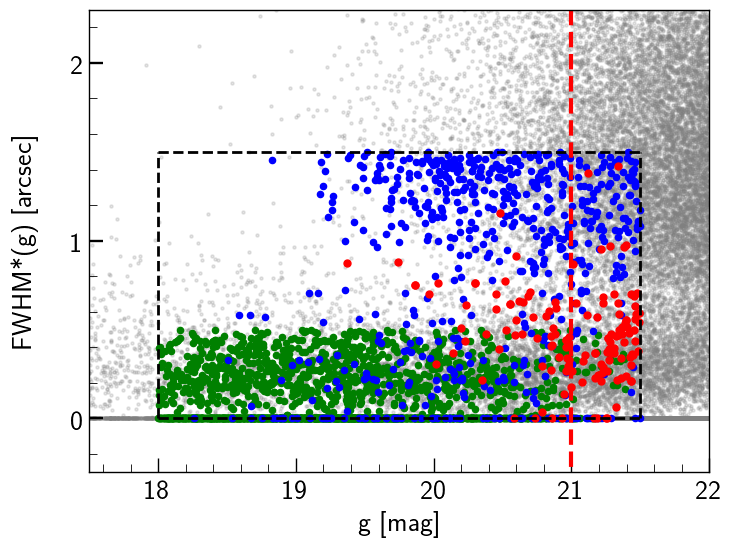}
\caption{Top: FWHM$^*$-magnitude diagrams (in $g$-band) for sources in the whole dataset (grey points), foreground stars (green points), confirmed UCD/GCs (red points) and background galaxies (blue points). The FWHM$^*$ parameter is used as a proxy for the angular sizes of the sources. Bottom: The selection boundaries in magnitude and size are shown with black dashed box. The red dashed line indicates our adopted magnitude limit for defining UCDs.  \citep{drinkwater2000}.}
\label{fwhmmag}
\end{figure} 

The main catalogue has been split into a sample with measured radial velocities in the spectroscopic references (\textit{KNOWN} catalogue) and the rest (\textit{UNKNOWN} catalogue). In the following, these catalogues are described.

\subsubsection{The \textit{KNOWN} catalogue}
Sources in the \textit{KNOWN} catalogue (sources with available radial velocities) are categorized into three classes: foreground stars, confirmed UCD/GC and background galaxies. UCD/GCs are from the spectroscopically (kinematically) confirmed UCD/GCs in the Fornax cluster (\citealp{wittmann-2016,Pota-2018} and references therein). These UCD/GCs reference catalogues contain 1285 sources in total of which 639 have full photometric coverage and are added to the \textit{KNOWN} catalogue. Table \ref{specucdgccat} presents the optical and near-infrared magnitudes of these UCD/GCs. The remaining 646 sources lack either the $u$-band or $Ks$-band photometry and are not included in the \textit{KNOWN} catalogue. 

Foreground stars and background galaxies were selected based on the spectroscopic data of \citet{Maddox-2019} (and references therein). Objects with V\textsubscript{rad} $<$ 300 km s$^{-1}$ and V\textsubscript{rad} $>$ 3,000 km s$^{-1}$ are selected as foreground stars and background galaxies, respectively. Fornax cluster galaxies have an average radial velocity of 1,442 km s$^{-1}$ and a velocity dispersion of 318 km s$^{-1}$ (\citealp{Maddox-2019}). \citet{Maddox-2019} presented a compilation of spectroscopic redshifts in the Fornax cluster (\citealp{Hilker-1999,drinkwater2000,Mieske2004,bergond2007,firth2007,firth2008,gregg2009,schuberth2010}), also using the extended galaxies of the Fornax cluster Dwarf Galaxy Catalogue (FDSDC, \citealp{venhola2018}) to cross-identify objects. Table \ref{fore} and Table \ref{back} gives an overview of the catalogues of foreground stars and background galaxies.

Fig. \ref{knowndist} shows the projected distribution of the \textit{KNOWN} sources on the sky. For convenience, throughout this section of the paper, the spectroscopically confirmed foreground stars, UCD/GCs and background galaxies (the three classes of objects) in the figures are shown in green, red and blue, respectively. Almost all the foreground stars in the \textit{KNOWN} catalogue are from the 2dF survey of the Fornax cluster (\citealp{drinkwater2000}); 70\% and 25\% of the background galaxies are from the Fornax cluster surveys by \citet{drinkwater2000} and \citet{Maddox-2019}. Both surveys are about 80\% complete down to m$_g$ = 20 mag. However, the UCD/GCs in the centre of the cluster are a compilation of many other, deeper, surveys. Note that for the \textit{KNOWN} catalogue, we only use sources with available radial velocities in the literature (spectroscopically confirmed). We do not include more stars into the \textit{KNOWN} catalogue from the Gaia DR2 since the spectroscopic sample of foreground stars is rich enough and contains sources over the whole colour range. However, in section \ref{sec5}, when the UCD candidates are identified, we investigate if any of Gaia DR2 stars are incorrectly identified as UCD.

The \textit{KNOWN} catalogue contains 6,670 objects of which 3,880, 639 and 2,151 are foreground stars, UCD/GCs and background galaxies, respectively. This is the first optical/near-infrared dataset of the Fornax cluster UCD/GCs and is a valuable source for studying stellar populations of UCD/GCs. This is, however beyond the scope of this paper and we postpone it to a future publication.

The bottom panel of Fig \ref{hist-mag} shows the magnitude distribution of the sources in the \textit{KNOWN} catalogue. We already discussed that the main catalogue is 88\% complete down to magnitude m$_{g}$= 21 mag. This magnitude corresponds to the adopted magnitude for defining UCDs. We also inspect the completeness of the main catalogue for the red and blue UCD/GCs detected in $u$ and $Ks$ data respectively. Considering objects with magnitudes m$_g$ $<$ 21 mag, 21 $<$ m$_g$ $<$ 21.5 mag and 21.5 $<$ m$_g$ $<$ 22 mag, 100\%, 96\%, 83\% of the red UCD/GCs ($g-i$ $>$ 1.0 mag) are detected in $u$ and 100\%, 98\%, 86\% of the blue UCD/GCs ($g-i$ $<$ 1.0 mag) are detected in $Ks$.

\subsubsection{The \textit{UNKNOWN} catalogue}
The \textit{UNKNOWN} catalogue contains $\sim$110,000 objects brighter than m$_g$ $=$ 24.5 mag. This magnitude corresponds to the turn over of the globular clusters luminosity function (GCLF) at the distance of the Fornax cluster (\citealp{Cantiello2020}). The UCD candidates will be selected from the sources in this catalogue. Table \ref{unk} presents an overview of the \textit{UNKNOWN} catalogue. The top panel of Fig. \ref{hist-mag} presents the magnitude distribution of sources in the \textit{UNKNOWN} catalogue in each filter. 

\subsection{Pre-selection of the UCD/GC candidates (step ii)}

\begin{figure}
\centering
\includegraphics[width=\columnwidth]{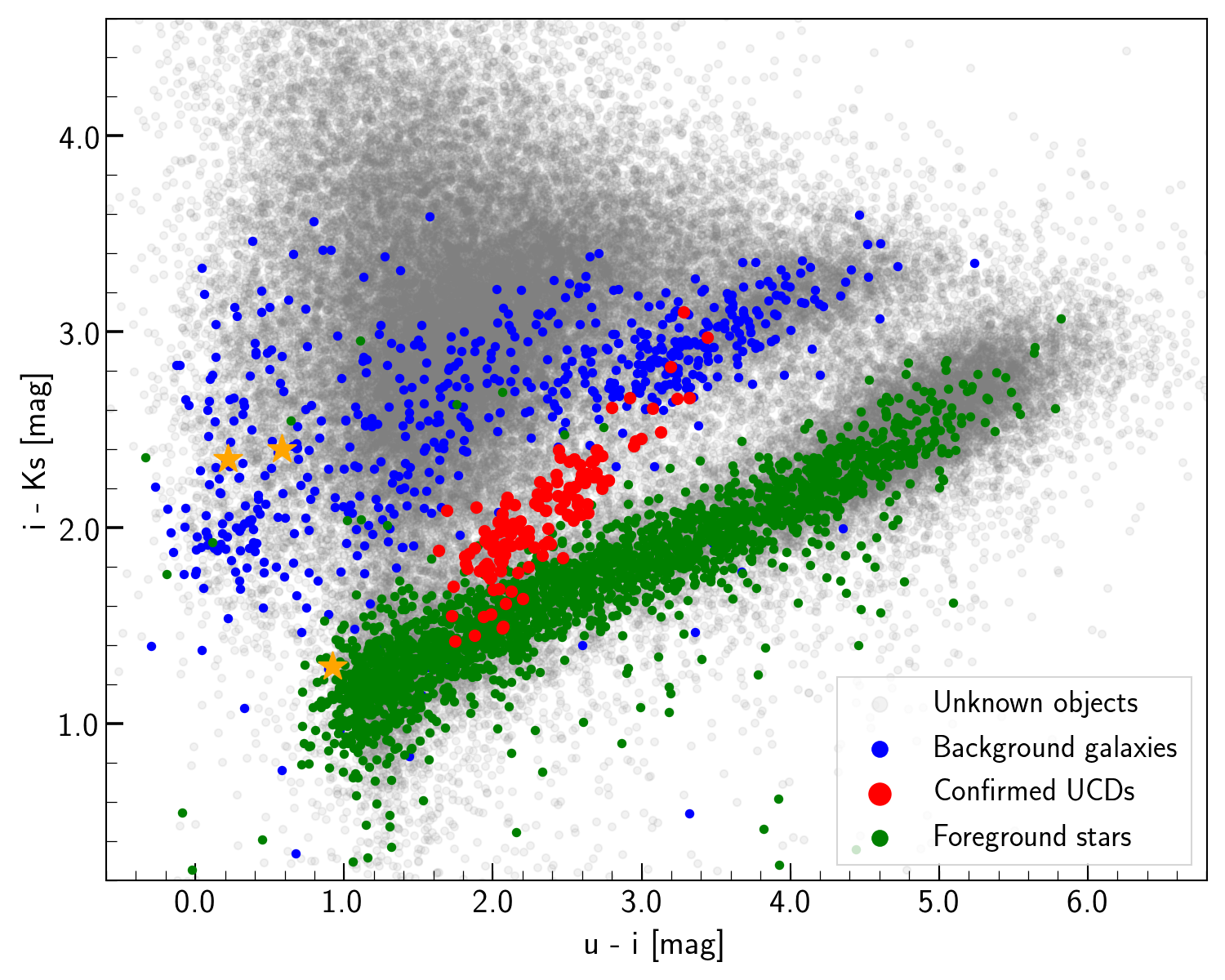}
\includegraphics[width=\columnwidth]{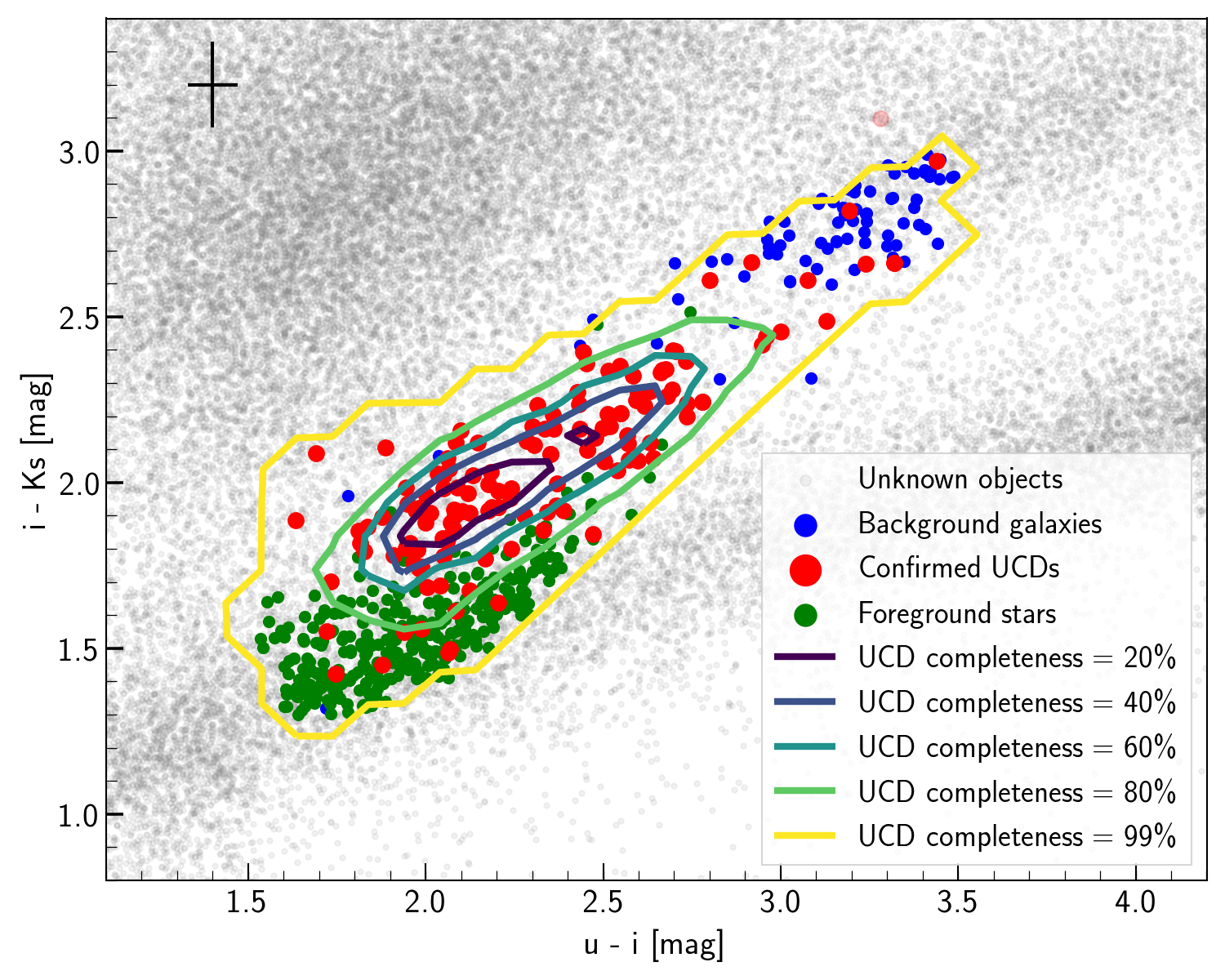}
\caption{top: The $u-i$/$i-Ks$ colour-colour diagram of the \textit{UNKNOWN} sources (grey points), foreground stars (green points), confirmed UCD/GCs (brighter than m$_g$ $=$ 21.5, red points) and background galaxies (blue points). The three orange stars indicate the three outliers of the UCD/GC sequence ([BAL2007] gc152.1, [BAL2007] gc290.6 and [BAL2007] gc235.7). These objects were excluded from the analysis. bottom: Pre-selection of UCD/GC candidates using the UCD/GC sequence in $u-i$/$i-Ks$ colour-colour diagram. After the size-magnitude pre-selection, all the \textit{UNKNOWN} sources (grey points) within the 99\% UCD/GC contour have been selected. Because of the uncertainties in the measurements, the pre-selected sample itself can be largely contaminated with foreground stars (green points) and background galaxies (blue points). The light-red point outside of the UCD/GC sequence indicates a UCD/GC that was excluded after applying the 99\% contour. The error bar on top left corner indicates the average uncertainties in colours of a UCD with m$_g$ = 21 mag.}
\label{preselection}
\end{figure}

\begin{table*}
\centering
\caption{Number of selected objects in \textit{step i} and \textit{step ii}. }
\begin{tabular}{ ccccc } \hline  
Sample & master catalogue ($gri$) & main catalogue ($ugriJKs$) & size-magnitude (pre-)selected & $u-i$/$i-Ks$ (pre-)selected \\
\hline
\textit{UNKNOWN} & $\sim$ 1,000,000 &  $\sim$110,000 & $\sim$60,000 & 4,079\\
\textit{KNOWN} &  10,258 & 6,670 & 2,913 & 547\\
foreground stars & 5,700 & 3,880 & 2,172 & 331\\
confirmed UCD/GCs & 1,285 & 639 & 138 & 137\\
background galaxies & 3,273 & 2,151 & 603 & 79\\ 
\hline 
\end{tabular}
\label{numbers}
\end{table*}

In this step, using the observed properties of the confirmed UCDs in the \textit{KNOWN} catalogue, namely $g$-band magnitude, size and the $u-i$/$i-Ks$ colours, we define some criteria to pre-select UCD candidates. The top panel of Fig. \ref{fwhmmag} shows the FWHM$^*$-magnitude diagram of the sources in the \textit{KNOWN} and \textit{UNKNOWN} catalogues. All the
UCDs (red points brighter than m$_{g}$= 21 mag) are fainter than m$_{g}$ $=$ 19 and have FWHM$^*$ $<$ 1.2 arcsec. By extending these limits in magnitude and size, we adopted a magnitude-size criteria for UCDs as sources with 18 $<$ m$_g$ $<$ 21.5 mag and 0 $\leq$ FWHM$^*$ $<$ 1.5 arcsec (0 $\leq$ r$_h$ $<$ 75 pc). The bottom panel of Fig. \ref{fwhmmag} demonstrates these criteria in the FWHM$^*$ vs. magnitude diagram. The lower limit on magnitude is 0.5 magnitude fainter than the magnitude limit of defining UCDs (m$_g$ = 21 mag). This decision was made to increase the number of the selected confirmed compact sources which is important for what comes next in the analysis ($u-i$/$i-Ks$ colour-colour pre-selection and machine learning). The number of spectroscopically confirmed UCDs in the Fornax cluster brighter than m$_g$ $=$ 21 mag is 61. By including fainter compact sources (GCs in our definition) in the magnitude range 21 $<$ m$_g$ $<$ 21.5 mag, we increase the number of sources by more than a factor of 2 to 138. We note that the adopted upper limit on FWHM$^*$ does not include Fornax-UCD3 (\citealp{drinkwater2000}), the brightest UCD in the Fornax cluster. This UCD has FWHM$^*$ = 2.78 arcsec (r$_h$ = 139.0 pc), which is about 3 times larger than the second brightest UCD in our sample (r$_h$ = 43.6 pc). Also, since a few UCDs have FWHM$^*$ = 0 arcsec (indistinguishable from 0), we did not define a lower limit on the size. However, the majority of the UCDs (about 90\%) have FWHM$^*$ $>$ 0 arcsec. 

Next, we applied the size-magnitude criteria to the \textit{KNOWN} catalogue and \textit{KNOWN} objects that did not satisfy these criteria were excluded from the rest of the analysis. The total number of size-magnitude selected sources in the \textit{KNOWN} catalogue is 2,913 of which 2,172, 138 and 603 are foreground stars, UCD/GCs and background galaxies respectively. For the \textit{UNKNOWN} catalogue, we only use the size criteria (and not the magnitude criteria). This is done for two reasons: first, to inspect the quality of our method in rejecting sources brighter than m$_g$ = 18.0 as UCDs (which are very likely foreground stars) and second, to extend our search to GCs fainter than m$_g$ = 21.5 mag. With these criteria, $\sim$60,000 objects in the \textit{UNKNOWN} catalogue were selected.

We continued the pre-selection using the $u-i$ and $i-Ks$ colours of the sources and the UCD/GC sequence (\citealp{Munoz-2014}). Fig. \ref{preselection} (top) shows the $u-i$/$i-Ks$ colour-colour diagram of the size-magnitude selected sources of the \textit{KNOWN} catalogue. In this figure, UCD/GCs (red points) occupy a region in the colour-colour diagram, which makes them distinguishable from foreground stars (green points) and background galaxies (blue points). The red tail of the sequence is as red as $u-i$ $\sim$ 3.3 mag and $i-Ks$ $\sim$ 3.0 mag. According to single stellar population models (MILES, \citealp{vazdekis,vazdekis2016}) objects with an old age (10-14 Gyr) and a super-solar metallicity ([M/H]$\sim$0.5) can be as red as $u-i$ $\sim$ 3.3 mag and $i-Ks$ $\sim$ 2.5 mag which, within uncertainties are consistent with the observed colours of the red \textit{KNOWN} UCD/GCs in our sample.

The location of the UCD/GC sequence provides a tool to identify UCD/GCs photometrically. To define the UCD/GC sequence in Fig. \ref{preselection}, we used the colours of the confirmed UCD/GCs. The $u-i$/$i-Ks$ colour-colour space was divided into grids with a grid size of 0.1 mag and UCD/GCs' density was measured within the grids. Then, the isodensity contour enclosing 99\% of the UCD/GCs was drawn and UCD/GCs outside this contour were removed from the analysis. As the result, only one object was removed from the initial set of UCD/GCs. We repeated this procedure for the selected UCD/GCs from the first run. No UCD/GC was removed in the second run. Fig. \ref{preselection} (bottom) shows the corresponding isodensity contours that enclose 20\%, 40\%, 60\%, 80\% and 99\% of the spectroscopically confirmed UCD/GCs which indicates the UCD/GC completeness of the enclosed colour-colour space. Note that the 99\% contour is based on compact sources brighter than m$_g$ = 21.5 mag. This contour encloses all the confirmed UCDs (brighter than m$_g$ = 21 mag). Once the 99\% contour is drawn, objects in the \textit{KNOWN} catalogue (size-magnitude selected) and the \textit{UNKNOWN} catalogue (size selected) within this contour were selected and the pre-selection step was completed. 

The pre-selection step selects $\sim$4,079 and 547 objects from \textit{UNKNOWN} and \textit{KNOWN} catalogues (331 foreground stars, 137 UCD/GCs and 79 background galaxies). Table \ref{numbers} summarizes the numbers of the selected sources in this step. The pre-selected sample, as it is implied from the pre-selected \textit{KNOWN} sources, can be largely contaminated by foreground stars (60\%). 

The pre-selection based on magnitude, size and $u-i$/$i-Ks$ colours was done to clean the \textit{KNOWN} and \textit{UNKNOWN} catalogues from the obvious non-UCD/GCs (95\% of the sources) while it keeps almost all (99\%) of the confirmed UCD/GCs. Therefore, the resulting samples are expected to represent a complete sample of all the UCD/GCs while the majority of the foreground stars and background galaxies are removed. The remaining contamination is mainly due to the scatter in the $i-Ks$ colours of the UCD/GCs that originates from the uncertainties in the near-infrared photometry (mainly $Ks$). Narrowing down the selection area to the densest part of the UCD/GC sequence in the $u-i$/$i-Ks$ diagram (for example 80\% completeness contour), while it keeps the majority of the UCD/GCs (80\%), it reduces the contamination (85\% of foreground stars and 94\% of the background galaxies will be removed). However, in this case, a fraction of possible UCDs, in particular the most metal-rich ones, are missed (20\%) and therefore, the resulting sample has a lower completeness. Additionally, for the next step (\textit{step iii}) of the UCD selection (supervised machine learning), the pre-selected \textit{KNOWN} catalogue will be used for the training of the model and it should contain more or less the same number of foreground stars, confirmed UCD/GCs and background galaxies. Otherwise, the outcome will be biased toward the class with the majority of objects. Our choice with the 99\% contour will lead to an equal sample of all three classes. In other cases (for example using 90\% or 95\% completeness contours), only a few background galaxies will be included and biases the supervised classification. In the next step, we use all the available colours in the main catalogue (5 independent colours: $u-g$, $g-r$, $r-i$, $i-J$ and $J-Ks$) and try to clean the pre-selected samples.

\subsubsection{Outliers from the $u-i$/$i-Ks$ UCD/GC sequence }
While inspecting the $u-i$/$i-Ks$ colour-colour diagram of the confirmed UCDs, we found that 3 spectroscopically confirmed UCDs (brighter than m$_g$ $=$ 21) with Simbad ID [BAL2007] gc152.1 (V\textsubscript{rad} = 947 km s$^{-1}$), [BAL2007] gc290.6 (V\textsubscript{rad} = 1901 km s$^{-1}$) and [BAL2007] gc235.7 (V\textsubscript{rad} = 1310 km s$^{-1}$) are located outside of the UCD/GC sequence. Two of these objects ([BAL2007] gc152.1 and [BAL2007] gc290.6) are located in the same region as the quasars and high-redshift objects and one ([BAL2007] gc235.7) is on the stellar sequence. These objects are indicated by orange stars in Fig. \ref{preselection} (top panel). As discussed in \citealp{bergond2007}, the measured radial velocities of these objects are not very certain (flagged "B" in their catalogue). The recent spectroscopic survey of \citet{Maddox-2019} updated the radial velocity of [BAL2007] gc152.1 ($\sim$ 3.8 $\times$ 10$^5$ km s$^{-1}$). Therefore this object is likely a background high-redshift galaxy, as it is expected from its $u-i$/$i-Ks$ colours. Therefore, we conclude that the colours of these three objects do not give enough credibility to them to be a member UCD and they have been excluded from the sample of confirmed UCD/GCs.

\subsection{Selection of the UCD/GC candidates (step iii)}

\begin{figure}
\centering
  \includegraphics[width=\linewidth]{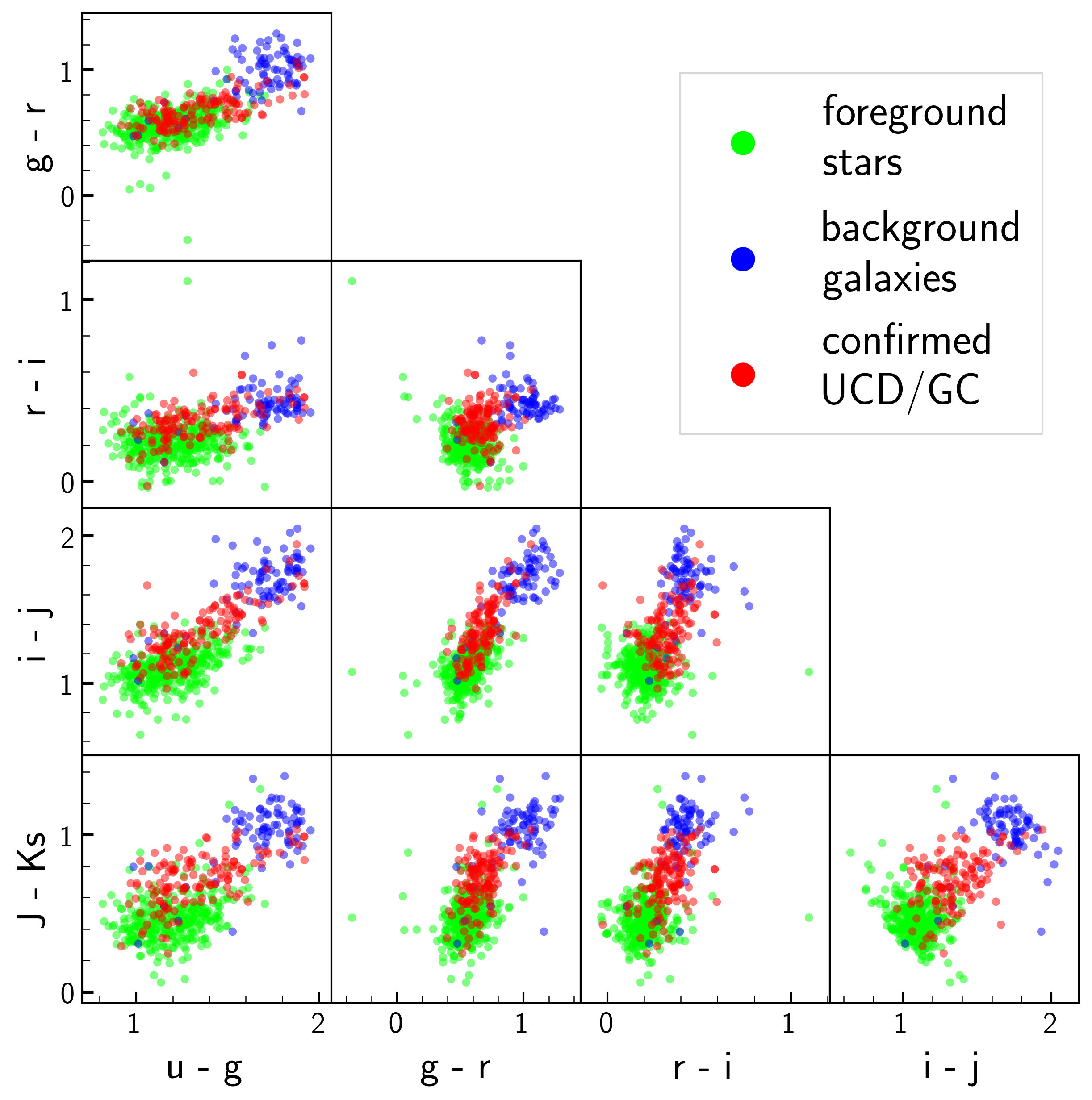}
  \caption{Different projections of the 5-D colour-colour diagram of the pre-selected \textit{KNOWN} sources. Foregrounds stars, UCD/GCs and background galaxies are shown in green, red and blue.}
\label{5dcc}
\end{figure}

In the third step, we use the pre-selected \textit{KNOWN} sources, train a machine learning model using the K-nearest neighbours technique (KNN) with an adjustment to the original technique (described later in \ref{knn100}) and perform a supervised classification to classify \textit{UNKNOWN} sources into three classes: foreground star, UCD/GC, background galaxy. Our methodology is described in detail in the rest of this section. For the classification, we use 5 features (parameters): 5 independent colours namely $u-g$, $g-r$, $r-i$, $i-J$ and $J-Ks$. The combination of these colours defines a 5-D colour-colour diagram. Fig. \ref{5dcc} shows the different projections of this colour-colour diagram for the pre-selected \textit{KNOWN} sources. Note, we do not use the measured sizes and magnitudes for the classification in this step. Our size measurements are not accurate enough to be used in the classification. Moreover, since the depth of the \textit{KNOWN} catalogues is different for different type of objects (due to different selection criteria and depth in the respective surveys that detected them), including the measured magnitudes in the parameter space biases the classification; for brighter and fainter objects, the classification will be biased toward the foreground stars (which are generally brighter) and UCD/GCs (which are generally fainter). 

\subsubsection{K-nearest neighbours (KNN)}

We aim to find objects that have similar properties as the already known UCDs in the \textit{KNOWN} catalogue. On the other hand many of the objects are foreground stars or background galaxies which have properties slightly different from those of UCDs. Using machine learning, we try to identify whether the properties of a given object resemble more those of known foreground stars, background galaxies and UCDs and label them (classify) accordingly. We use the K-Nearest Neighbours (KNN, \citealp{knn}) method, a supervised machine learning technique for classification. Note that our methodology for this step is based on KNN with an adjustment. Here, we briefly explain how KNN works and later, the adopted adjustment to KNN is described.

For each unlabelled object, the KNN algorithm measures its euclidean distance\footnote{In KNN, it is possible to weigh the importance of the neighbours based on their distance in the N-dimensional parameter space, here 5-dimensional colour-colour space. During the optimization of the algorithm, we tried weighting and did not find the outcome satisfying. Therefore, for the classification, we did not assign any weight to the neighbours (this can be interpreted as assigning equal weight to all the neighbours).} to other data-points and searches for the K-nearest labelled neighbours (called reference-set or training-set) in the parameter space. Based on the classes of the nearest labelled neighbours (foreground stars, UCD/GCs or background galaxies), KNN classifies (labels) the unlabelled object as the most common class in the neighbourhood. In KNN, the value of K (the number of nearest neighbours) is not arbitrary and it must be optimized based on the efficiency of the classifier using different K values. In any stage of this process, KNN does not apply any normalization to the data. However, a sort of normalization before running KNN is crucial. Finding the best K value and normalization of the data are described further in section \ref{knn100}.

In KNN, the labelled dataset is used to make both training-set and validation-set. Normally, the majority of the labelled data is used for the training-set and a smaller fraction is used for the validation-set. Once the classification with the given training-set is done, the validation-set is used to evaluate the accuracy of the classification using the given training-set.

The accuracy of the classifier in classifying a subset (for example the validation-set) can be expressed by two parameters called precision and recall. The precision describes the reliability of the classifier when it classifies an object as UCD/GC and is defined as the ratio between the number of correctly classified objects as UCD/GCs (true positives) and the total number of objects classified as UCD/GCs (true positives and false positives): \\\\
Precision = $\frac{\text{True Positive}}{\text{True Positive + False Positive}}$ \\\\
Recall is the fraction of the confirmed UCD/GCs that are classified correctly (recovered) by the classifier\footnote{In multi-class classification problems, precision and recall can be measured for the combination of all the classes. However, for simplicity, we measure precision and recall which corresponds to the UCD/GC class} and can be measured as the ratio between objects that are (correctly) identified as UCD/GC (true positives) to the total number of the real UCD/GCs (true positives + false negatives):\\\\
Recall = $\frac{\text{True Positive}}{\text{True Positive + False Negatives}}$
\\\\
Precision and recall measure different aspects of a classifier. Another parameter known as F1-score (\citealp{f1}), can also be used to expresses the accuracy of a classifier using one parameter. The F1-score is defined as follows. \\\\
$F_{1}=\frac{\text{Precision} \: \times \: \text{Recall}}{\text{Precision}\: + \: \text{Recall}}$ \; \;
\\\\

Normally, the training and validation are done N times using $\frac{1}{N}$ of the whole dataset as validation-set and the remaining $\frac{N-1}{N}$ as training-set. At the end, the average and r.m.s. of the precision and recall values for the N runs are the final value and uncertainty of precision and recall of the classifier. This approach is called "N-fold cross-validation"\footnote{This method is known as "K-fold cross-validation". However, in this paper, we use "N-fold cross-validation" to avoid confusions with the K value in the K-nearest neighbours method.}. The value of K for KNN can be optimized by running the N-fold cross-validation and select the K value which leads to the highest F1-score (the best K value).

\subsubsection{KNN+100 and classification}
\label{knn100}

\begin{figure}
\centering
\includegraphics[width=\linewidth]{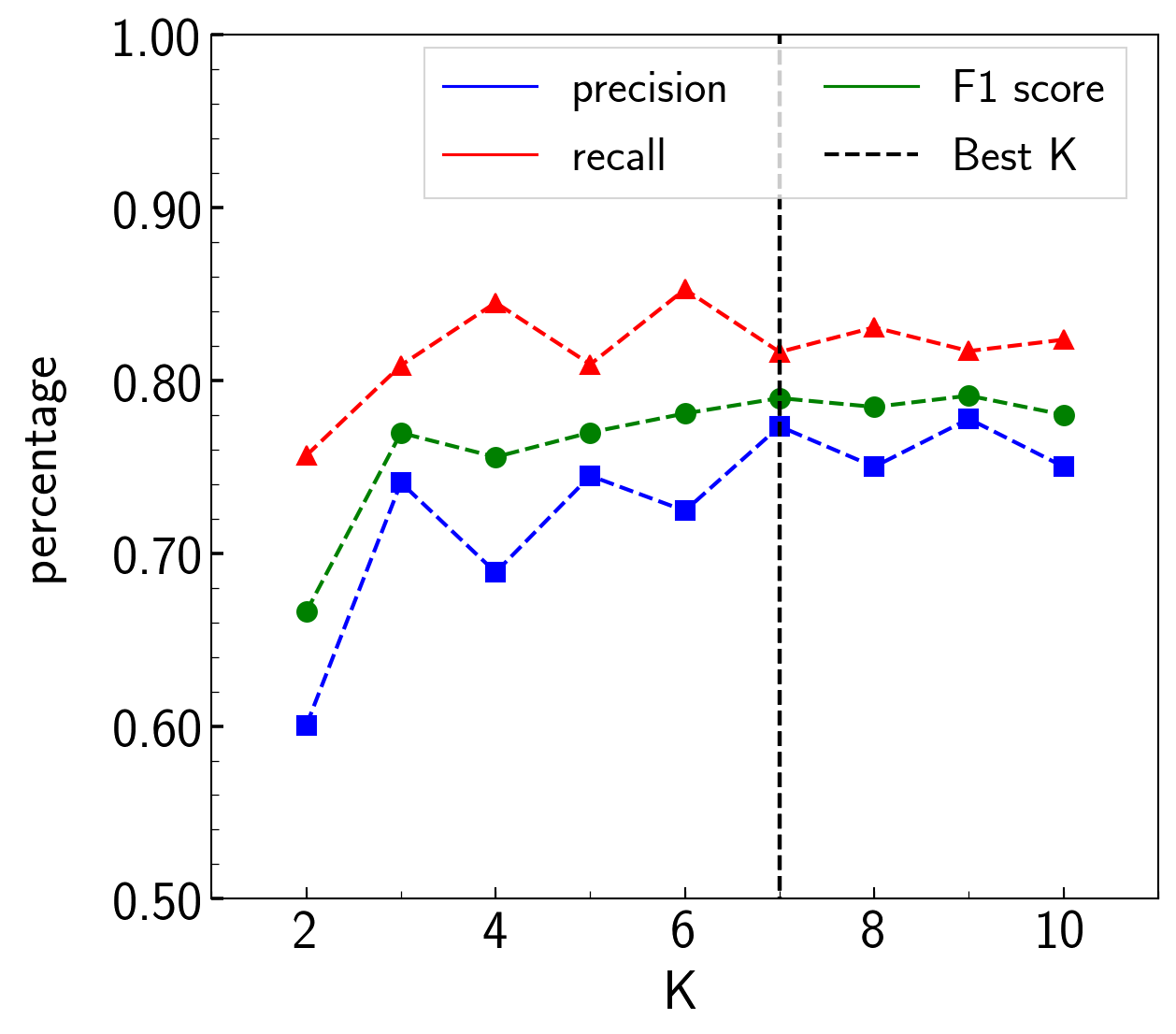}
\caption{Selecting the best value of K using the estimated F1-score, precision and recall of different values of K. This result led to the best K value K = 7 for the classification. The values of precision, recall and F1-score are the average of 10 different runs using 10 different training and validation sets.}
\label{bestk}
\end{figure}

We used the pre-selected \textit{KNOWN} and \textit{UNKNOWN} catalogues as the labelled and unlabelled datasets. Moreover, for our purpose, we adopted a modification to the KNN. For a given training-set, we run KNN for 100 times and each time we randomly select and use 90\% of the training-set. Therefore we classify objects 100 times with a random subset of the training-set. The main motivation behind this modification is to take into account the various outcomes of the classification when a slightly different training-set is used. At the end when classification is done, we count the number of runs that an object is classified as a UCD (UCD$\_$SCORE), foreground star (STAR$\_$SCORE), or background galaxy (GAL$\_$SCORE). These score values are an integer between 0 and 100, and for a given object, the summation of all the scores is 100. The UCD/GC candidates are objects with UCD$\_$SCORE $>$ STAR$\_$SCORE and UCD$\_$SCORE $>$ GAL$\_$SCORE. Clearly sources with higher UCD$\_$SCORE are more likely to be a UCD. To distinguish the modified version of KNN from the original KNN, we refer to our methodology as KNN+100. Note that in each run, the training-set should contain the same number (same order) of objects of each class. Otherwise, classification is biased toward the class with the majority of the data-points. 

Before performing the classification, by visual inspection, we checked that all the confirmed UCD/GCs are in the area outside of the low-signal-to-noise regions of the $Ks$ data. Afterwards, features (i.e., colours) were z-score normalized. z-scores were calculated using $z = (x-\mu)/\sigma$ where $\mu$ and $\sigma$ are the mean and standard deviation of the data-points for a given $x$. Next, the best K value was derived using N-fold cross-validation with N = 10 (10-fold cross-validation) and results are shown in Fig. \ref{bestk}. We chose K = 7 as the best K value which leads to precision 77$\pm$12\%, recall 81$\pm$8\% and F1-score 79$\pm$8\%. In practice, there is not much difference between values of K between K = 5 to K = 10. We continued the classification using KNN+100 with K = 7 to classify all the sources in the \textit{UNKNOWN} catalogue. The outcome is presented in section \ref{sec5}.

\begin{figure}
\centering
        \includegraphics[width=\linewidth]{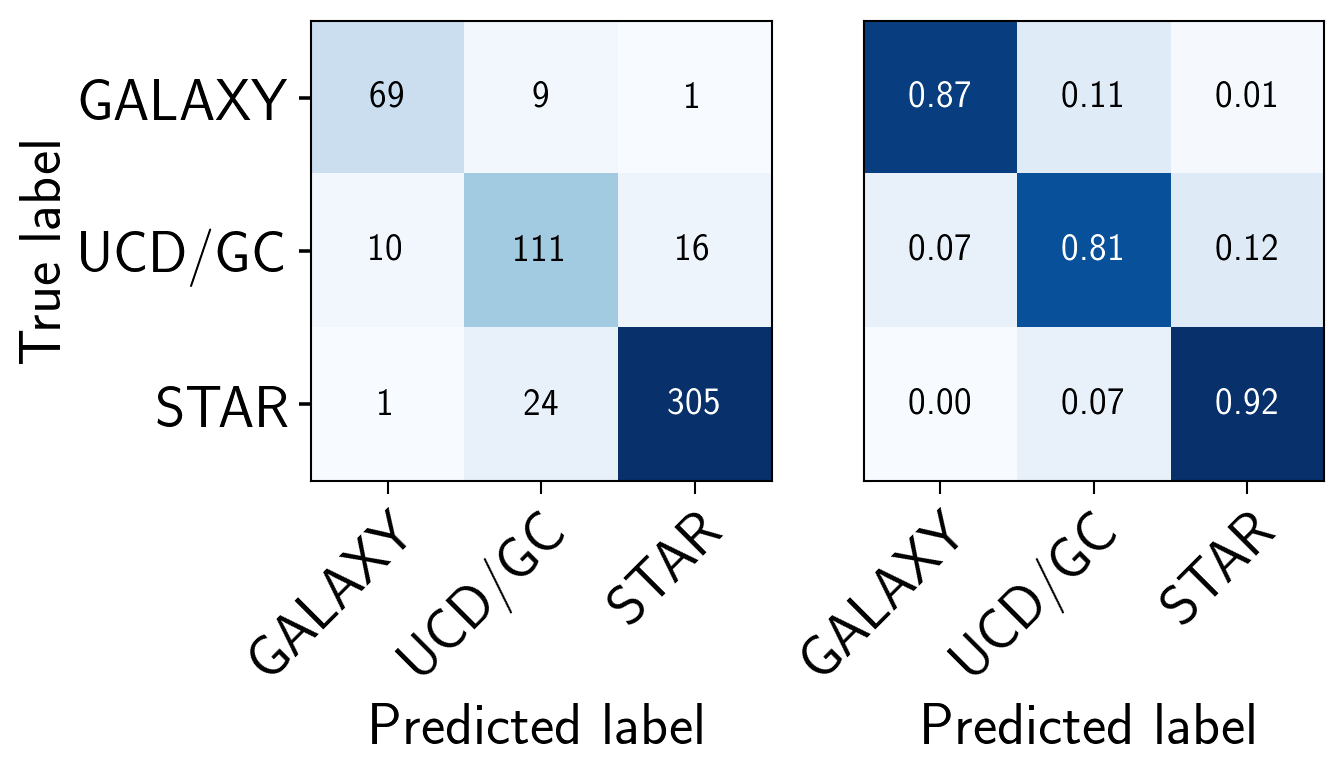}\\
\caption{The confusion matrix of the classification for the 10 validation-sets in absolute numbers (left) and percentage (right). The values in each validation-set can vary by $\sim$10\%.}
\label{matrix}
\end{figure}

\begin{figure*}
        \includegraphics[width=\linewidth]{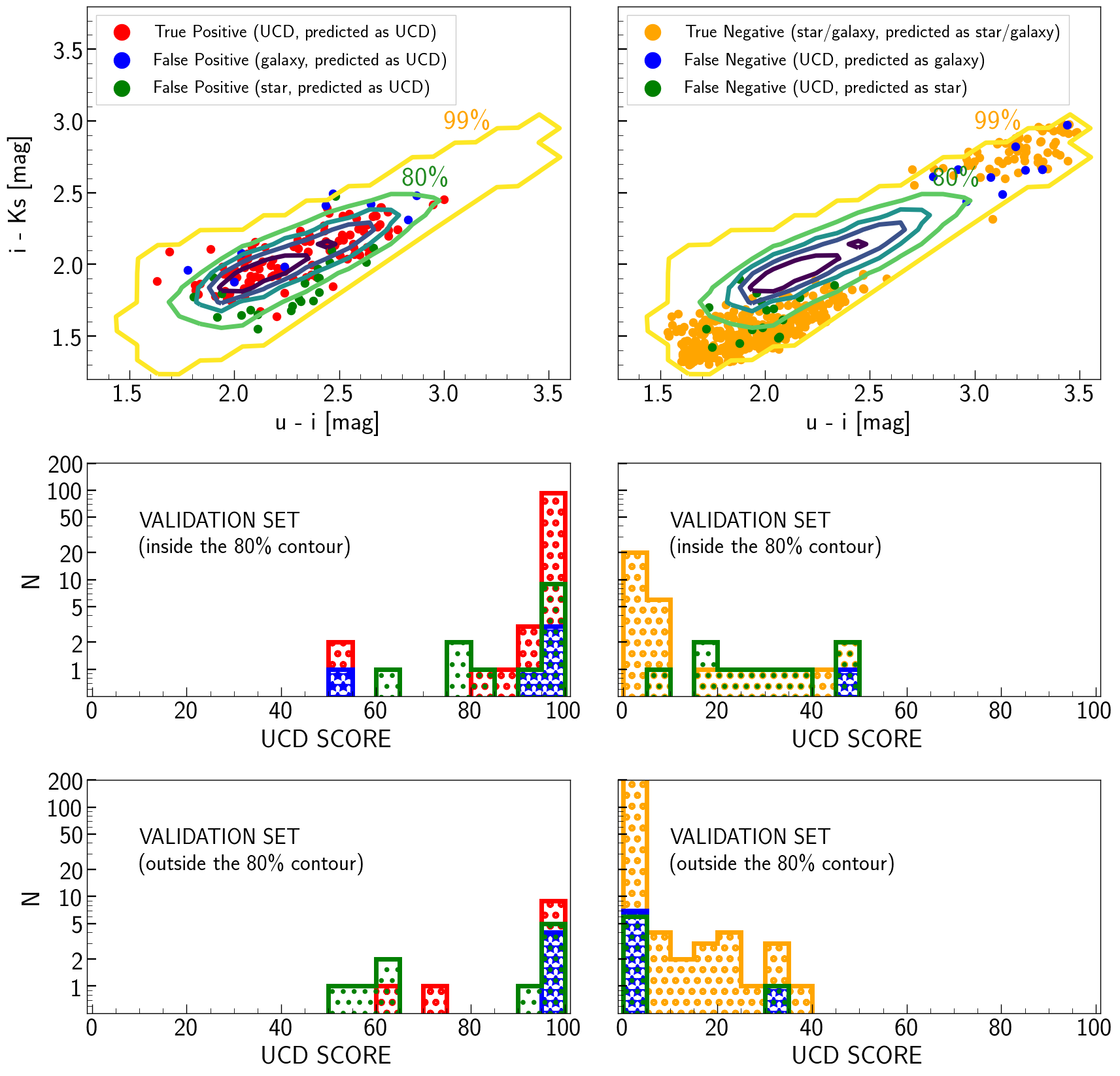}
\caption{The colour-colour diagram and UCD\_SCORE values of the true-positives (TP), true-negatives (TN), false-positives (FP) and false-negatives (FN) in the 10 validation-sets used for 10-fold cross-validation. }
\label{fpfn}
\end{figure*}

\subsubsection{Performance of the KNN+100 classifier}


Fig. \ref{matrix} presents the result of this classification during 10-fold cross-validation in a confusion matrix. This NxN matrix shows the three different classes and the number of objects identified in each class. The main-diagonal entries of the confusion matrix show the number (or percentage) of objects classified correctly (true positives) and off-main-diagonal entries show the number (or percentage) of objects classified incorrectly (false positives and false negatives). As is shown in Fig. \ref{matrix}, our classifier can predict (recover) 81\% of the UCD/GCs (recall = 0.81). The remaining 19\% are classified (incorrectly) as foreground stars (12\%) or background galaxies (7\%). Moreover, 77\% of the predicted UCD/GCs are true UCD/GC (precision = 0.77), 17\% are foreground stars and 6\% background galaxies. For the bright sources with m$_g$ $<$ 20.5, while precision drops to 42\%, recall increases to 90\%. For the faint objects with m$_g$ $>$ 20.5, precision and recall are 92\% and 79\%. Precision and recall for both bright and faint sources improve when objects inside the 80\% UCD completeness contour (in $u-i$/$i-Ks$ diagram) are considered. In this case, for bright sources, precision and recall are 55\% and 100\%. For faint sources, these rates are 94\% and 89\%. Therefore, to have a cleaner sample of UCDs, one might restrict the candidate selection to a lower UCD completeness contour, for example the 80\% completeness contour. The lower rate of precision of the bright sources compared to the faint sources is mainly because of the smaller number of UCDs at brighter magnitudes while the number of total sources (and therefore the expected false positives) are the same.

Fig. \ref{fpfn} shows the $u-i$/$i-Ks$colour-colour diagrams (top) and UCD$\_$SCORE values (bottom) of the true-positives and false-positives (TP and FP, on the \text{left}) and true negatives and false-negatives (TN and FN, on the \text{right}) in all the 10 validation-sets. The performance of our classifier is bounded to the data points in the training-set and for this reason, except of a few objects, it misses the majority of the confirmed UCD/GCs outside of the 80\% UCD completeness contour (false negatives in the figure on right, 10\% of the total UCD/GCs). The 80\% contour, while it encloses 80\% (completeness) of the UCD/GCs, selects 13.5\% of the non-UCD/GCs (15\% and 6\% for the foreground stars and background galaxies) and has the highest rate of selecting UCDs over non-UCDs. The middle panels of Fig. \ref{fpfn} present the UCD\_SCORE of the sources inside the 80\% UCD completeness contour. Except of two objects, all the confirmed UCD/GCs have UCD\_SCORE $>$ 80. Only 2 out of 119 UCDs within the 80\% contour have UCD\_SCORE $<$ 80 which corresponds to 1.6\% of the sample. Also, 14 foreground stars and 5 background galaxies within the 80\% contour (19 in total) have UCD\_SCORE $>$ 80 which corresponds to 14\% of the total sources identified as UCDs.

Normally, in addition to constraints on size and magnitude, the $u-i$/$i-Ks$ colour-colour diagram is used to select UCD/GCs (as we did in \textit{step ii} of the UCD selection). However, as we showed, this criterion alone leads to a contaminated sample (the pre-selected sample) by foreground stars and background galaxies. This is the case especially when the near-infrared data is not as deep as the optical. This is the motivation behind \textit{step iii} in which we cleaned the pre-selected sample from the possible contaminants using all the available colours (5 independent colours) in our dataset. This leads to the removal of 70\% of the contaminants (64\% for the sources inside the 80\% contour). However, for objects outside the 80\% contour, our methodology is not reliable. This is because our UCD/GC sample does not provide enough data points outside the 80\% contour. We also discussed that this contour has the highest rate of selecting UCDs over non-UCDs. To summarize, when objects inside the 80\% contour are considered, our methodology is more reliable. Moreover, we showed that among the UCDs inside the 80\% contour, almost all the UCDs have UCD\_SCORE $>$ 80. In the next section and after presenting the result of classification of the \textit{UNKNOWN} sources, we select the most likely UCD candidates as objects inside the 80\% contour and with UCD\_SCORE $>$ 80. We call these candidates as the "BEST" UCD candidates. 

\section{Results}
\label{sec5}

We initially identified sources with full photometric coverage in 6 filters (main catalogue) and divided it into a \textit{KNOWN} and an \textit{UNKNOWN} catalogue, respectively for the objects with and without available radial velocity data in our reference literature (\citealp{wittmann-2016,Pota-2018,Maddox-2019}. Then, a sample of UCD/GC candidates was pre-selected using size-magnitude criteria and the $u-i$/$i-Ks$ colour-colour diagram. Subsequently, and by applying a supervised machine learning technique trained on the pre-selected sources in the \textit{KNOWN} catalogue, we classified all the pre-selected sources in the \textit{UNKNOWN} catalogue into three classes: foreground stars, UCD/GCs and background galaxies. In this section, the results of the classification and the final catalogue of the UCDs are presented.
\begin{figure*}
        \includegraphics[width=0.32\linewidth]{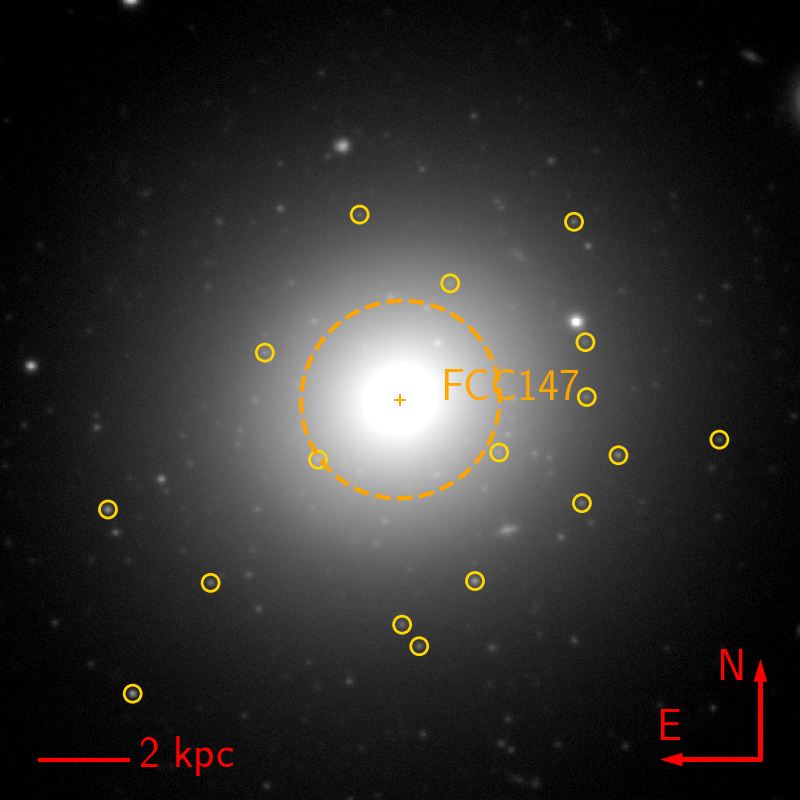}
        \includegraphics[width=0.32\linewidth]{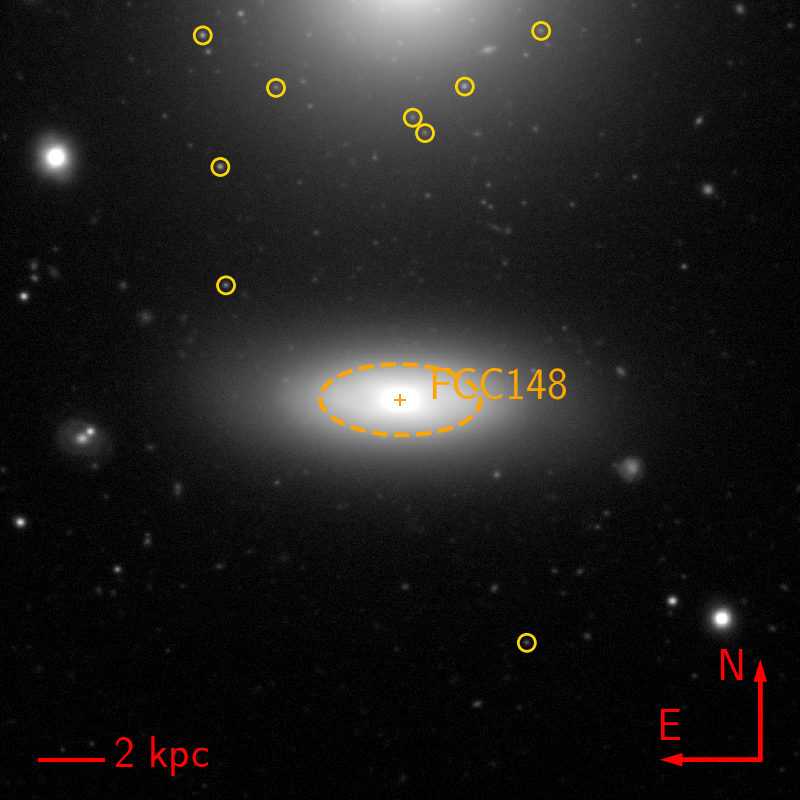}
        \includegraphics[width=0.32\linewidth]{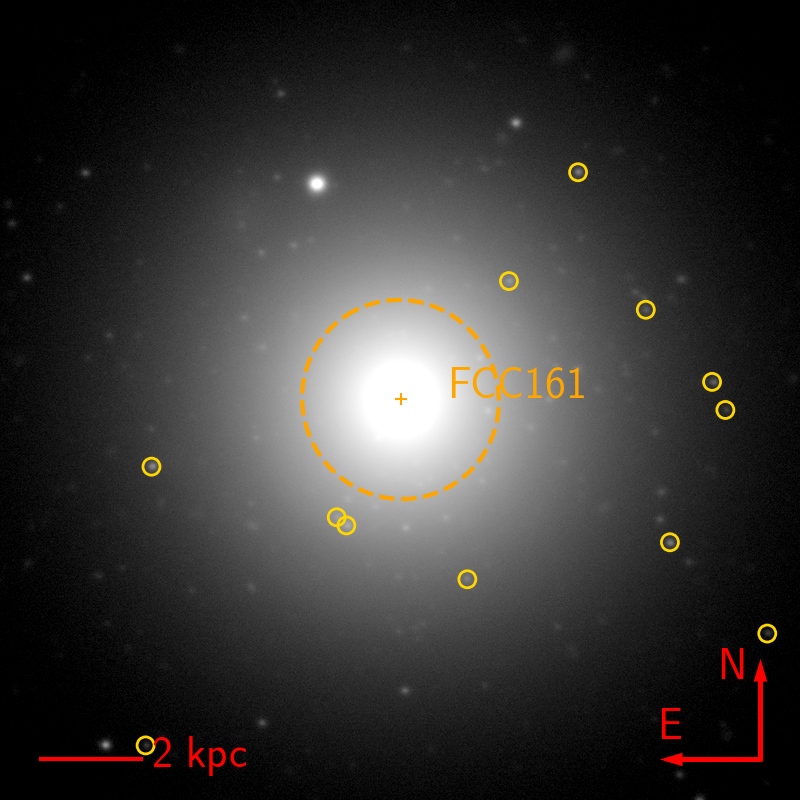}
        \\
        \includegraphics[width=0.32\linewidth]{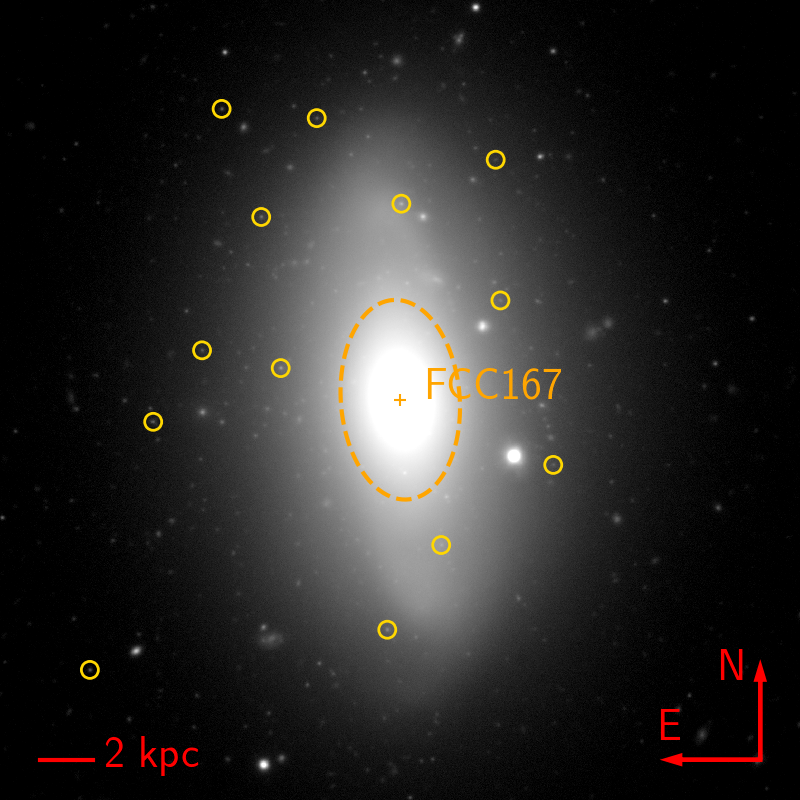}
        \includegraphics[width=0.32\linewidth]{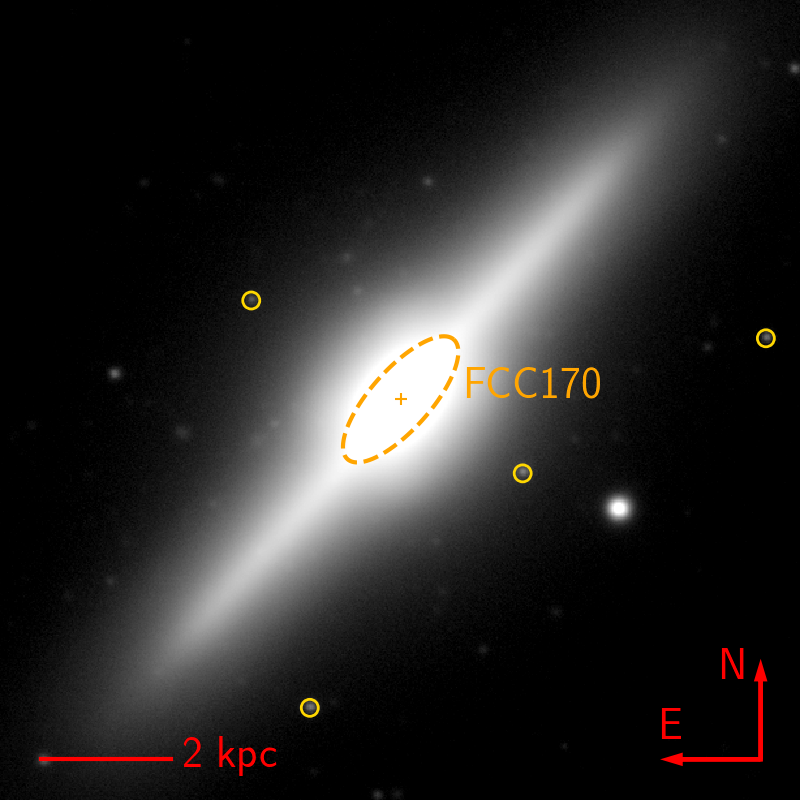}
        \includegraphics[width=0.32\linewidth]{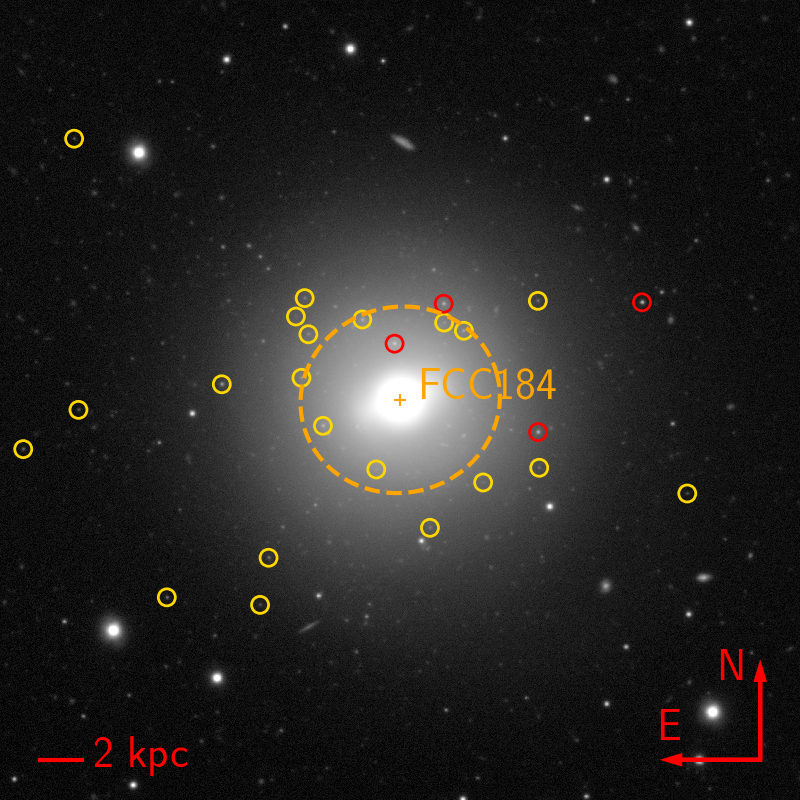}
        \\
        \includegraphics[width=0.32\linewidth]{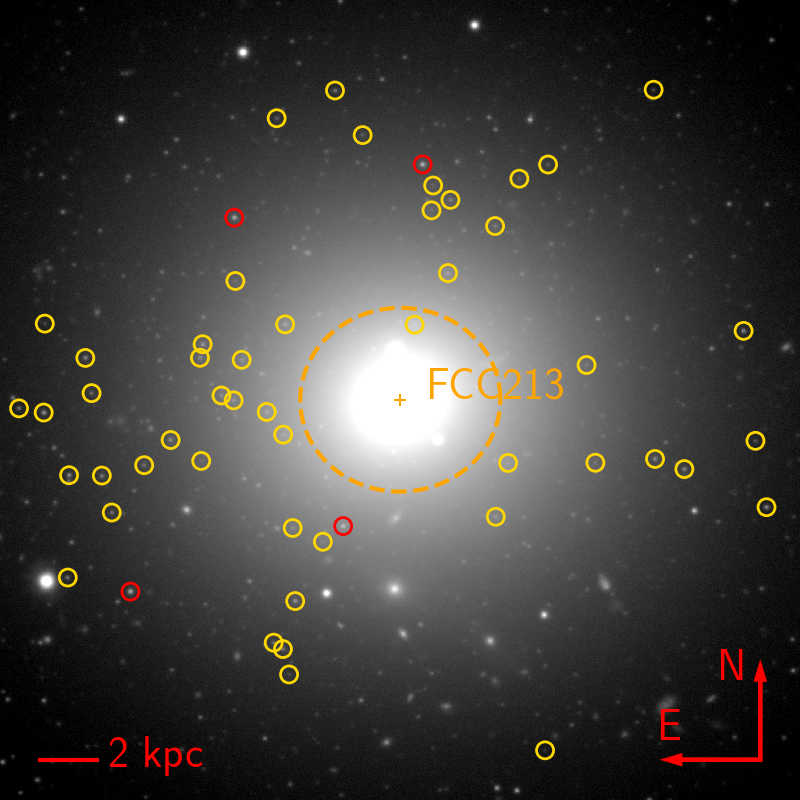}
        \includegraphics[width=0.32\linewidth]{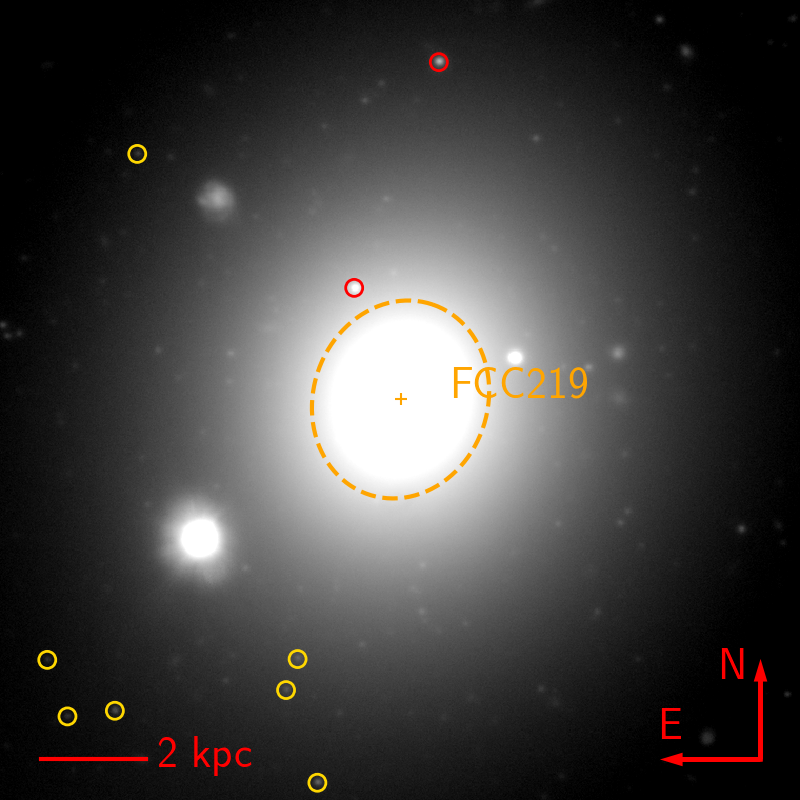}
        \includegraphics[width=0.32\linewidth]{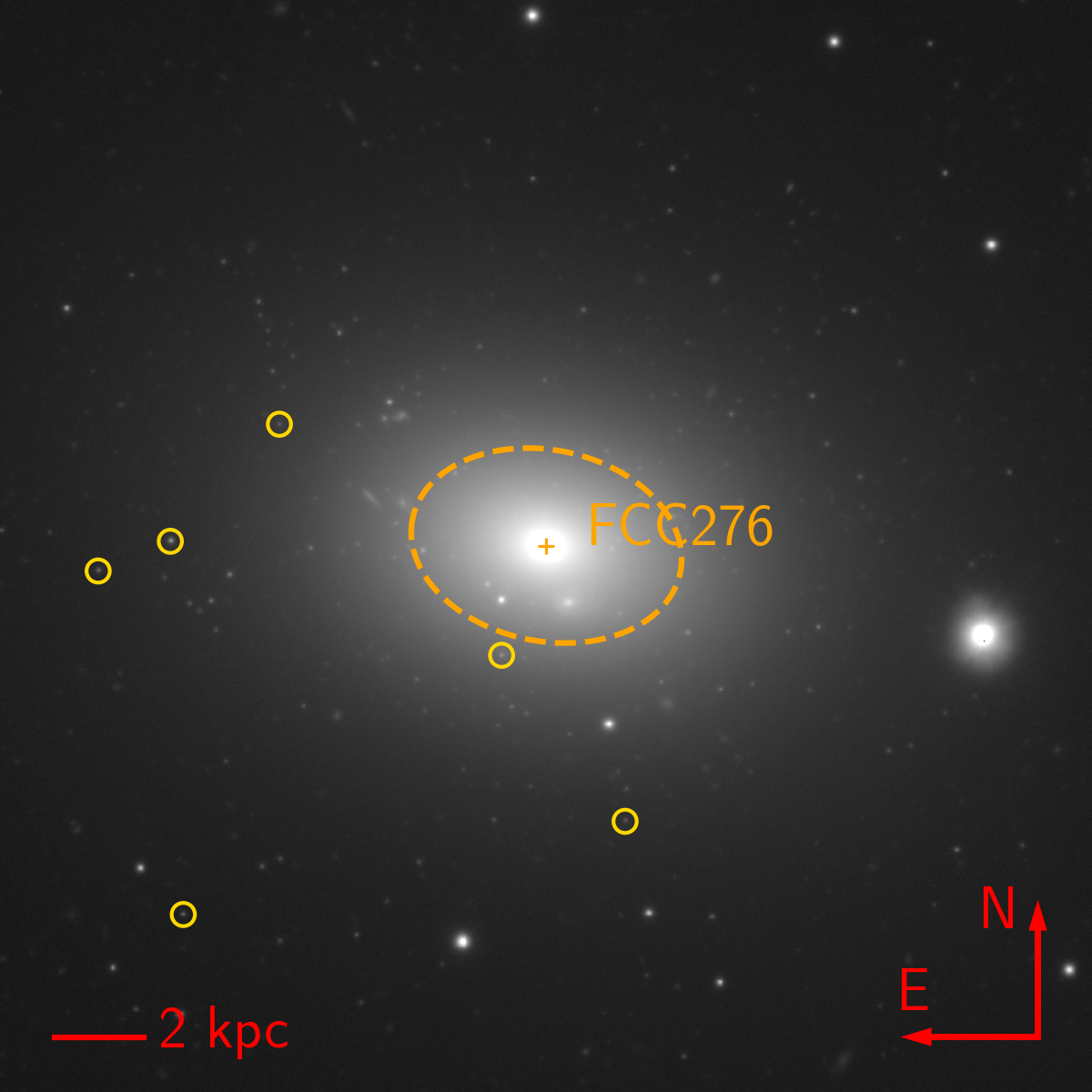}
        
\caption{UCD/GCs around the brightest galaxies in the Fornax cluster. These objects represent all the sources that are identified from the sources in the \textit{UNKNOWN} catalogue and the \textit{KNOWN} catalogue (recovered UCD/GCs from the \textit{KNOWN} catalogue or in other words, true positives in all the validation-sets). UCDs and GCs are indicated with red and yellow circles respectively. The dashed circles indicates one effective radius from the host galaxy (based on the values in the FCC catalogue, \citealp{fcc}). Fornax cluster galaxies with 2 or more identified UCD/GC candidates are shown here. Note that the inner 3 kpc were excluded from the detection of UCD/GC candidates.}
\label{gcsaround}
\end{figure*}

\begin{figure}
        \includegraphics[width=\linewidth]{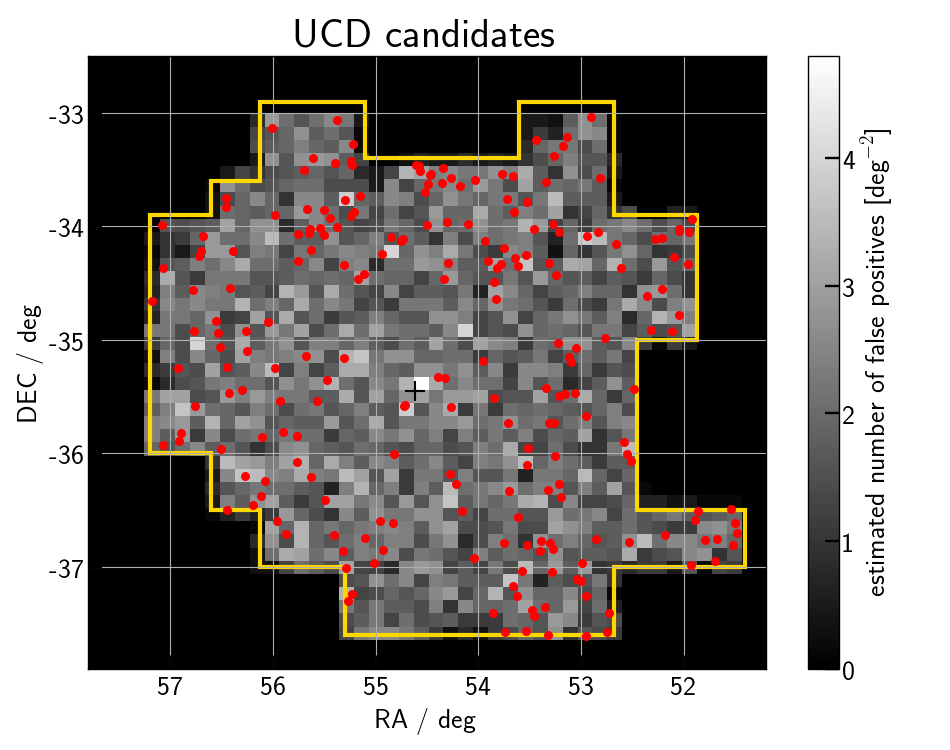}
        \includegraphics[width=\linewidth]{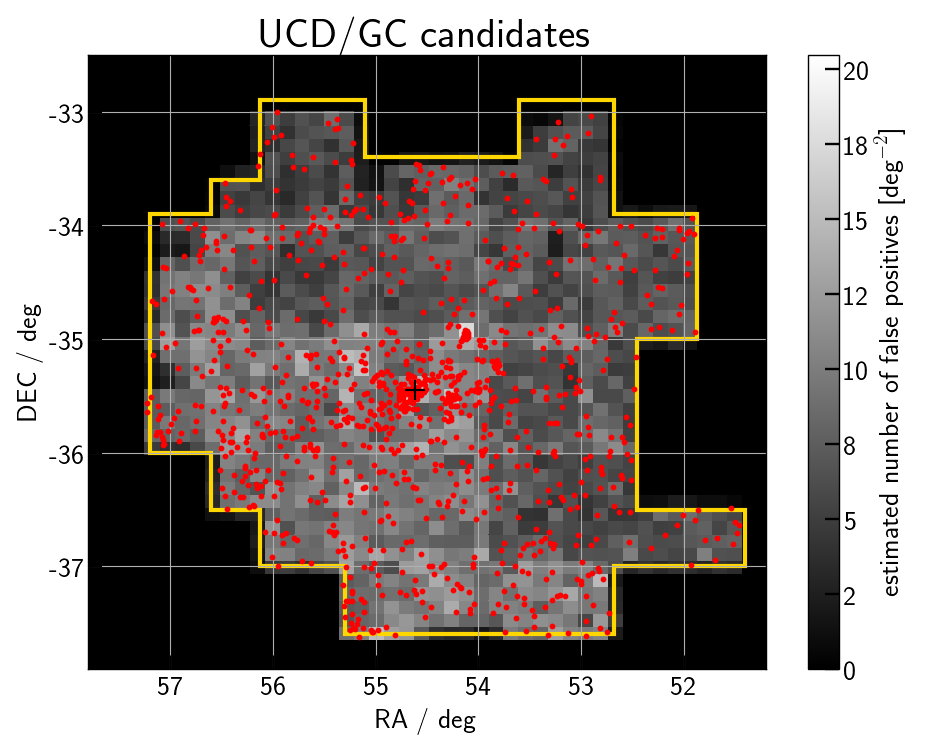}
\caption{Projected distribution of the candidates and the expected number of the false-positives across the observed area for UCD candidates (top) and all the UCD/GC candidates (bottom).}
\label{fieldsources}
\end{figure}

\begin{figure}
\includegraphics[width=\linewidth]{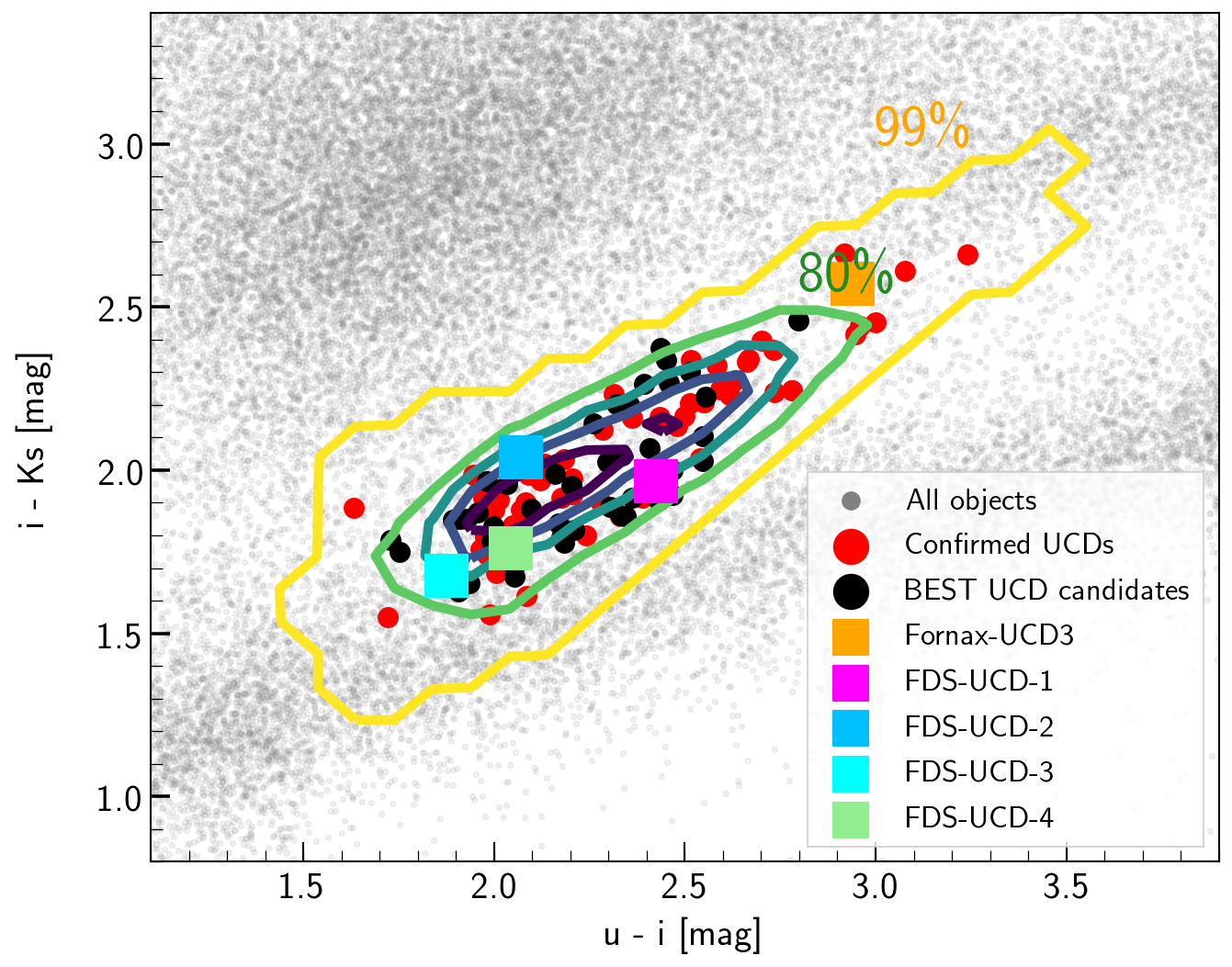}
\caption{$u-i$/$i-Ks$ colour-colour diagram of the confirmed UCDs (red) and the BEST UCD candidates (black). The 4 brightest BEST candidates brighter than m$_g$ = 19 mag (FDS-UCD-1, FDS-UCD-2, FDS-UCD-3, FDS-UCD-4) are indicated in the diagram.}
\label{uikbest}
\end{figure}

\begin{figure}
\includegraphics[width=0.9\linewidth]{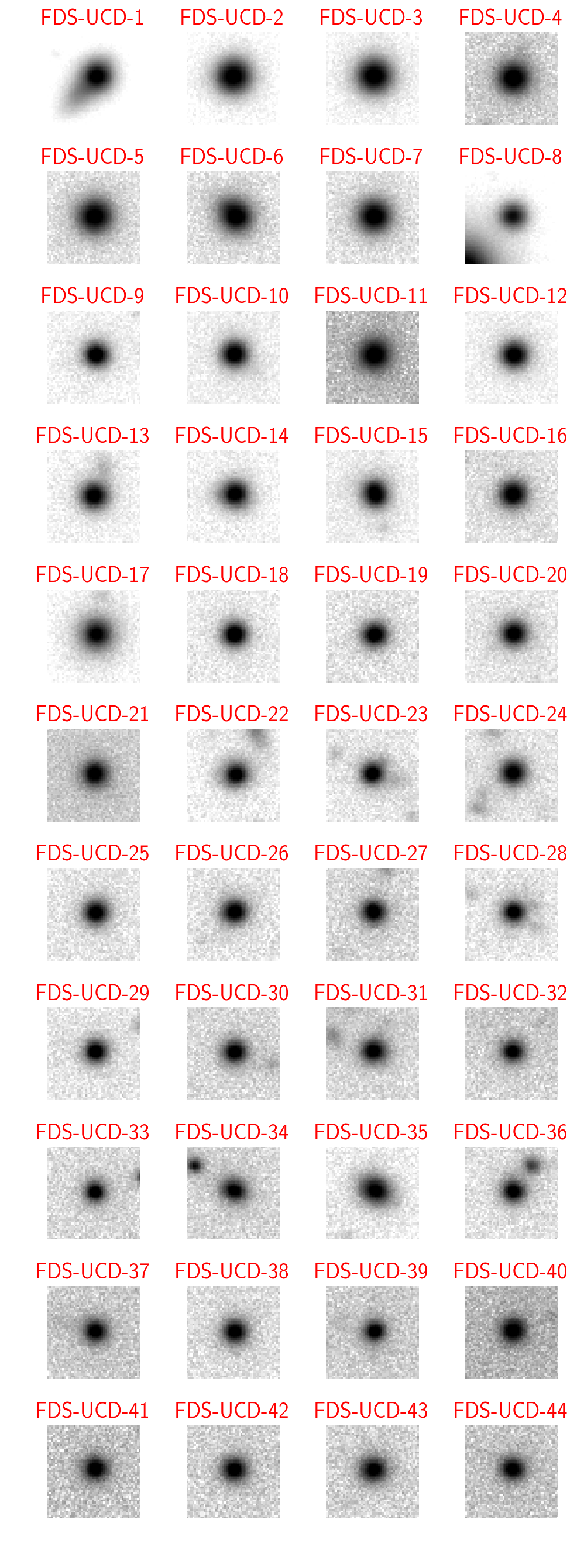}
\caption{The BEST UCD (candidates) identified in this work. Thumbnails have dimensions of 10 x 10 arcsec or 1 x 1 kpc at the distance of the Fornax cluster. This list is sorted based on m$_g$. The first 4 BEST candidates (first row) are the brightest candidates, brighter than m$_g$ = 19 mag. An elongated feature can be seen around FDS-UCD-1 which looks to be a blended with a background galaxy.}
\label{bestucds}
\end{figure}

\subsection{UCD/GC candidates}

Our method identifies 1155 \textit{UNKNOWN} UCD/GCs brighter than m$_g$ $=$ 24.5. By defining the UCDs as objects brighter than m$_g$ $=$ 21 mag (M$_g$ $=$ -10.5 mag at the distance of the Fornax cluster), we identified 222 UCD candidates and 933 GC candidates in total. The UCD/GC candidates are objects with UCD$\_$SCORE $>$ STAR$\_$SCORE and UCD$\_$SCORE $>$ GAL$\_$SCORE. Hence, all the UCD and GC candidates have UCD$\_$SCORE $\geq$ 50. A preview of the published catalogue of UCD and GC candidates in this paper is presented in Table \ref{ucdgccat} (in the appendix). The observed area of this work covers the Fornax cluster up to its viral radius and a bit beyond. Fig. \ref{gcsaround} shows the UCD/GC candidates and the recovered UCD/GCs from the \textit{KNOWN} catalogue (true positives) around some of the brightest Fornax cluster galaxies.

Among the UCD/GC candidates, 4 UCD candidates and 5 GC candidates (9 in total) have radial velocities in other references in the literature (different than our references in this paper) and are not included in the \textit{KNOWN} catalogue. Among these UCD candidates, 1 is a foreground star, 1 is a UCD (This means that 1 UCD is missed from the \textit{KNOWN} catalogue), 1 is a nucleated dwarf galaxy and 1 is a background galaxy. All these sources have UCD\_SCORE = 100. For the 5 GC candidates, 1 is a foreground star (UCD\_SCORE = 83), 2 are GCs (UCD\_SCORE = 100 for both), 1 is a nucleated dwarf galaxy (UCD\_SCORE = 60) and 1 is a background galaxy (UCD\_SCORE = 100). This implies that 55$\pm$25\% of the identified UCD/GCs are a true UCD/GC. Note that the nuclei of dwarf galaxies can also be considered as a true UCD since they share similar properties.

Out of about 3,000 size-magnitude selected foreground stars and background galaxies in the \textit{KNOWN} catalogue, our classifier ended up classifying about 30 as UCD/GCs (false positives). This implies an average false-positive rate of 1\%. In Fig. \ref{fieldsources} we show the projected distribution of the selected UCDs and GCs in comparison with the expected false positives derived from the total number of sources in the \textit{UNKNOWN} catalogue. We estimate $\sim$170 ($\sim$77\%) false positives and $\sim$50 ($\sim$23\%) true UCDs among the UCD candidates. For the GC candidates, we estimate $\sim$580 (63\%) false positives and $\sim$ 350 (37\%) true GCs. Since this is unsatisfactory, we now choose a subsample, with higher probabilities of being UCDs or, as we call it, the BEST UCD candidates.

\subsection{BEST UCD candidates}

Once the UCD candidates are identified, we select the most likely UCD candidates among them as follows. At first, among the candidates brighter than m$_g$ $=$ 21 mag (222 sources), \textit{UNKNOWN} objects inside the 80\% contour in the $u-i$/$i-Ks$ diagram and with high a UCD probability, estimated by the machine learning technique (UCD$\_$SCORE $>$ 80) were selected. Next, for this sample, objects with FWHM$^*$ = 0 are excluded. The criterion was applied because the majority of the confirmed UCDs (about 90\%) have a non-zero FWHM$^*$. Also, by visual inspection, we cleaned the data from blends and objects in the low-signal-to-noise regions of the $Ks$ images. Applying the criteria on the completeness, UCD\_SCORE, FWHM$^*$ and finally the visual inspection, we excluded 88, 27, 43 and 15 sources (173 in total) from the 222 sources respectively. These selection criteria leave us with 49 BEST UCD candidates. Applying the same criteria on the validation-sets leads to a UCD sample with completeness of 70\% (43 UCDs out of 61 UCDs). It also improves the rate of true positives by a factor of 2 from 23\% to 52\%.

Among these 49 selected candidates, 3 are nuclei of dwarf galaxies\footnote{Because of our methodology in source extraction, in addition to the size selection, we miss further nuclei of dwarf galaxies in the final catalogue.} in the cluster (FCC numbers FCC274, FCC188 and FCC 260), 1 is a UCD around FCC219 (Fornax 3D survey, \citealp{f3d,fahrion4}), 1 is a star (\citealp{schuberth2010}). In the following, all these sources were excluded from the BEST UCD sample and the final number of the BEST UCDs is 44. This number is within the expected range of the number of true positives (true UCDs) among all the UCD candidates ($\sim$50). The $u-i$/$i-Ks$ colour-colour diagram of the most likely UCD candidates or "BEST" UCD candidates is shown in Fig. \ref{uikbest}. 

\begin{figure}
\includegraphics[width=\linewidth]{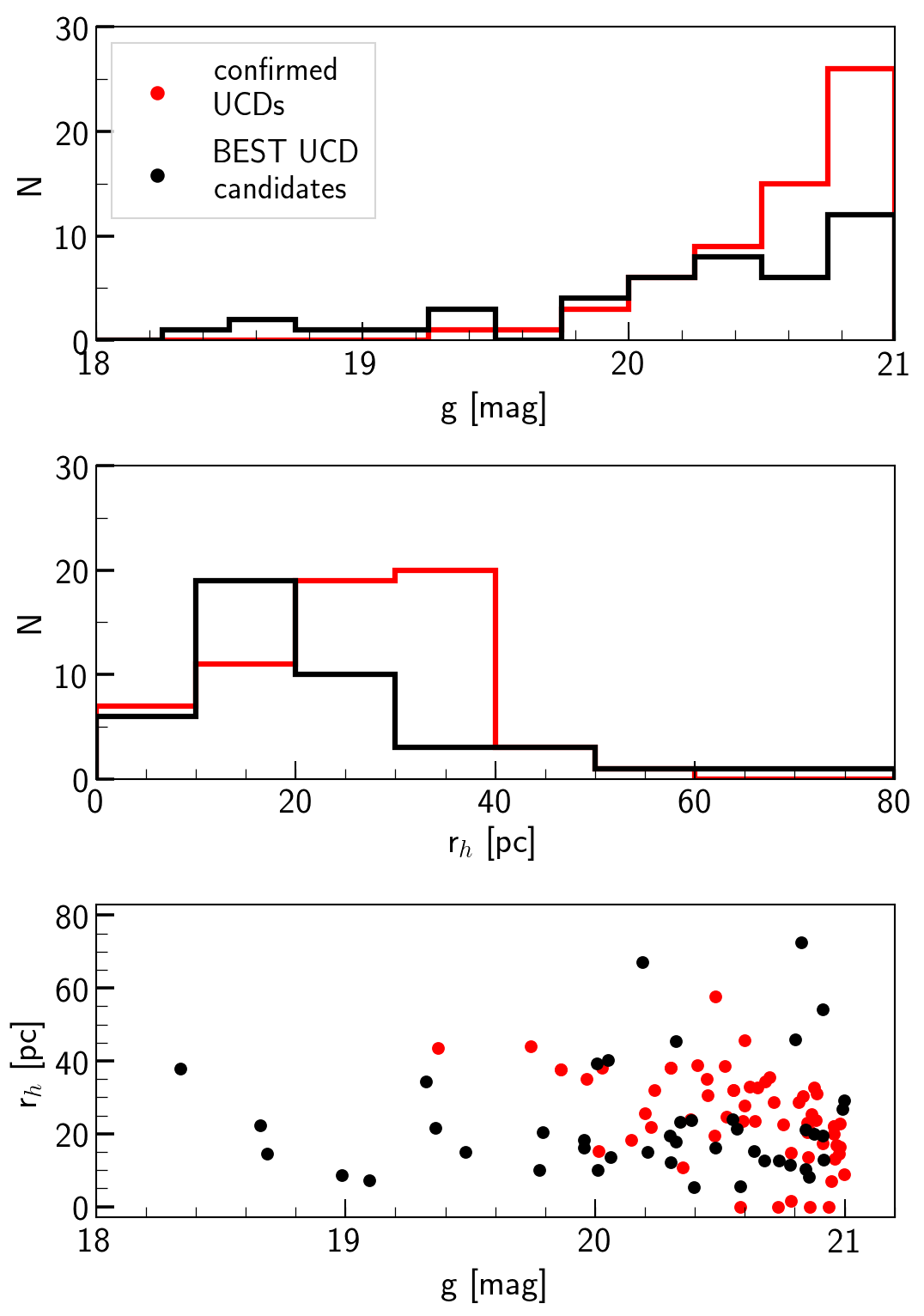}
\caption{Magnitude distribution, size distribution and size-magnitude relation of the BEST UCD candidates (in black) and the confirmed UCDs (in red). This figure does not show Fornax-UCD3 in the Fornax cluster. This object is located outside of the displayed size limits (r$_h$ = 139 pc based on our measurements).}
\label{magsizebest}
\end{figure}

\begin{figure}
\includegraphics[width=\linewidth]{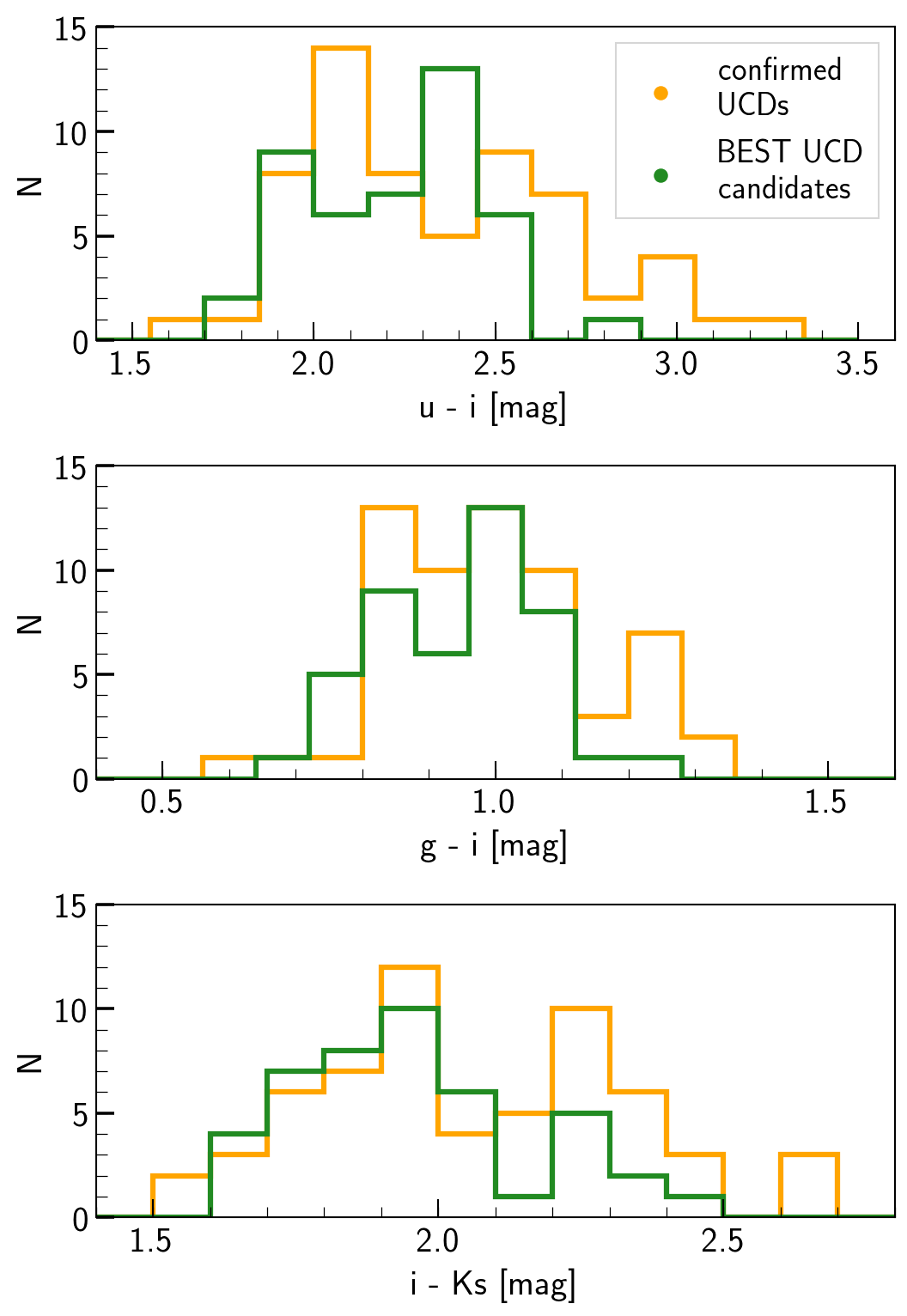}
\caption{$u-i$, $g-i$ and $i-Ks$ colours of the BEST UCD candidates (in green) and the confirmed UCDs (in yellow).}
\label{colorbest}
\end{figure}

Fig. \ref{bestucds} shows the thumbnails of the BEST UCDs, and the final catalogue is provided in Table \ref{ucdtable}. In Fig. \ref{magsizebest} and Fig. \ref{colorbest} we compare the observed properties of the BEST UCD candidates with the confirmed UCDs.
Since almost all of the BEST candidates are outside of the core while all the confirmed UCDs are inside the core of the cluster, the properties of these two samples represent the properties of UCDs in their respective environments. As is seen in Fig. \ref{magsizebest}, we have identified a few BEST UCD candidates brighter than the brightest confirmed UCD in our pre-selected sample. Among the brightest candidates, 4 are brighter than m$_g$ = 19 mag (FDS-UCD-1, FDS-UCD-2, FDS-UCD-3, FDS-UCD-4). These candidates in the $u-i$/$i-Ks$ colour-colour diagram are indicated in Fig. \ref{uikbest}. Note that, as was discussed earlier, for bright magnitudes, the rate of false positives is higher. 
The thumbnails of these objects can be seen in the first row of Fig. \ref{bestucds}. FDS-UCD-1 is brighter than Fornax-UCD3 (\citealp{drinkwater2000}). In Fig. \ref{bestucds}, FDS-UCD-1 seems to be blended with a background galaxy. This blend may had led to a larger measured size, brighter magnitude and redder colours for this object.

\subsection{ACSFCS UCD/GCs}

As an external check, we compared our UCD/GC candidates with the GC catalogue of the HST/ACS survey of the Fornax cluster (ACSFCS, \citealp{jordan2015}). For the comparison, we combined the selected UCD/GCs in the \textit{UNKNOWN} catalogue with the True positives. These true positives are objects in the 10 validation-sets that are recovered as UCD/GCs when they were not included in the training-set. Next, we selected sources in ACSFCS with GC probability\footnote{See \citet{jordan2009} for more details} p$_{GC}$ $>$ 0.75 (GC probability from \citealp{jordan2015}) and the projected distance to the host galaxy R$_{gal}$ $>$ 30 arcsec (3 kpc at the distance of the cluster). The limit on R$_{gal}$ was made since the source extraction efficiency (for optical data, especially in r) drops significantly at small projected distances to galaxies. The largest compact object in ACSFCS has a half-light radius r$_h$ $<$ 12 pc. The GC selection criteria of the ACSFCS catalogue do not capture possible UCD/GCs larger than this size. As a consequence, some UCDs can be missed from this catalogue. One example is the recently discovered UCD around the low-mass elliptical galaxy FCC 47 (\citealp{fahrion1}). The outcome of the comparison between our candidates and ACSFCS is presented in the following.

\begin{itemize}
\item The sample of UCDs in ACSFCS contains 13 objects of which 7 objects are within our coverage (6 objects are around FCC21 in Fornax A, which is not covered in $u$-band). \\

\item In our data, 6 of the 7 objects are pre-selected (magnitude, size and $u-i$/$i-Ks$ colours criteria). The missing object is in a low-signal-to-noise region in the $Ks$, and its $i-Ks$ colour is not reliable. \\

\item Among the 6 pre-selected sources, all 6 are classified as UCD (4 with UCD$\_$SCORE = 100, 1 with UCD$\_$SCORE = 98 and 1 with UCD$\_$SCORE = 61). Although the matched sample is small, hence might suffer for sampling effects, our source extraction and UCD selection when UCDs at R$_{gal}$ $>$ 30 arcsec are considered is 86$\pm$15\% complete. Moreover, based on this sample, the recovery rate of the selection is 100\%, however it is limited by the source extraction efficiency. \\

\item We also compared our classification outcome for objects with very low GC probability (p$_{GC}$ $<$ 25\%) which are very likely to be background galaxies. Our classifier did not pre-select these objects in the beginning and removed them from the list of possible UCDs. 
\end{itemize}

Next, we extend our comparison with ACSFCS to GCs (21 $<$ m$_g$ $<$ 21.5 mag) and investigate the accuracy of our classifier. 

\begin{itemize}
\item The ACSFCS GC catalogue includes 32 GCs that are overlapping with our dataset, within the magnitude range between 21 $<$ m$_g$ $<$ 21.5 mag and farther than R$_{gal}$ > 30 arcsec from the host galaxy. \\

\item 28 out of 32 GCs are extracted from the data and 4 are missing. This means that we were able to extract 88$\pm$6\% of the GCs in this magnitude range (source extraction efficiency). During the pre-selection step, 4 objects among 28 objects were excluded (not pre-selected). \\

\item At the end, among the 24 pre-selected sources, 23 are identified as GC with UCD$\_$SCORE > 95 (20 with UCD$\_$SCORE = 100) and 1 object is classified as a foreground star. This implies that the recovery rate of our selection is 72$\pm$10\%. The misclassified object is located in the blue or red tail of the UCD sequence where miss-classification is most likely.
\end{itemize}

\subsection{Optical vs. optical/near-infrared UCD selection}
Recently, \citet{Cantiello2020} published a catalogue of optically selected GCs in the Fornax cluster based on the FDS observations in $u$, $g$, $r$ and $i$. Their catalogue includes 8971 UCD/GCs (1649 UCDs, based on the definition in that paper) and has a larger coverage than the covered area of our work since our data is limited to the data in $J$ and $Ks$. Here, using the near-infrared photometry in $J$ and $Ks$ for the UCDs in \citet{Cantiello2020}, we investigate the optical/near-infrared UCD/GC selection vs. the optical selection (as in \citealp{Cantiello2020}).

\begin{itemize}
\item Among the 1649 GCs in \citet{Cantiello2020}, 1168 objects have $J$ and $Ks$ photometry in our source catalogue, 35 and 1153 in the \textit{KNOWN} and \textit{UNKNOWN} catalogues. \\

\item Among these objects, after pre-selection (\textit{step ii} in UCD/GC selection), 432 (around $\sim$40\% of the initial number) are left. The $u-i$/$i-Ks$ colours of the other 736 objects (not-selected, around $\sim$60\% of the initial number) show that, except for a few objects, the majority are foreground stars. This shows the well-known power of near-infrared photometry to exclude foreground stars. Among the 432 pre-selected sources, 35 are confirmed UCDs, 69 are foreground stars (spectroscopically confirmed), and 328 are unknown (no radial velocity available in our spectroscopic references). \\

\item Out of 328 \textit{UNKNOWN} objects, our classifier has selected 7 as background galaxies, 195 as foreground star and 126 as UCD (70\% with UCD$\_$SCORE $>$ 80). \\

\item In the end, 24 objects out of 126 ended up in our final BEST UCD catalogue. 26 of the BEST UCD candidates are not identified in \citet{Cantiello2020} of which 14 have FWHM$^*$ $>$ 0.6 and possibly are removed because of the adopted GC size criteria in \citet{Cantiello2020}. Another 5 BEST candidates are objects brighter than m$_g$ = 19.5 which is the upper magnitude limit of the UCD catalogue of \citet{Cantiello2020}. 
\end{itemize}

To summarize, out of 1168 sources in the UCD catalogue of \citet{Cantiello2020} with $J$ and $Ks$ data from our photometry, our classifier identified 126 as possible UCDs (not the BEST). Based on this result, while optical colours, including $u$-band and size criteria are enough to identify background galaxies, to distinguish stars and UCDs efficiently, near-infrared colours are needed. Based on the $u-i$/$i-Ks$ colour-colour diagram, optically selected UCD/GC samples are at least 70\% contaminated by foreground stars (the contamination from background galaxies is small). Additionally, machine learning on the 5D colour-colour space rejects another 60\% on the remaining 30\% which leads to a total contamination rate of 90\%. Hence, as in \citet{Cantiello2020}, a statistical decontamination from the background sources is required to characterize the properties of sources, if only optical data are used.

\subsection{BEST UCD candidates in Gaia DR2 and EDR3}

Once UCD candidates are identified, we checked if the sample is contaminated by Gaia stars. We selected Gaia sources with $\frac{\text{parallax}}{\text{parallax$\_$error}}$ $>$ 5.0 as foreground (Milky Way) stars. Out of $\sim$ 5,000 Gaia stars with full photometric data in 6 filters, none were selected as UCD/GC and our classifier successfully excluded all these objects using their $u-i$/$i-Ks$ colours in the pre-selection step. However, this test is biased toward the population of the nearby stars which may not be a representative of the stars that can be confused with UCDs. Among the BEST UCD candidates, 41 (out of 44) have been identified in Gaia DR2 and they all have $\frac{\text{parallax}}{\text{parallax$\_$error}}$ $<$ 4.0. However, in the most recent data release of Gaia (EDR3, \citealp{gaia3}), the two brightest candidates, FDS-UCD-1 and FDS-UCD-2 have $\frac{\text{parallax}}{\text{parallax$\_$error}}$ $\sim$ 6.0 ($>$ 5.0) which shows that they are very likely foreground stars. While this ratio is a good indicator of foreground (Milky Way) stars, but it is not definitive. Therefore, we do not exclude these two candidates from our catalogue. The rest of the objects have $\frac{\text{parallax}}{\text{parallax$\_$error}}$ $<$ 5.0 in Gaia EDR3.

Recently, \citet{voggel2020} showed that UCD/GCs in the NGC5128 group show larger $BP$/$RP$ Flux excess values in Gaia DR2 and used this parameter to identify UCD/GCs around NGC5128. As is described in the Gaia DR2 documentation\footnote{\url{https://gea.esac.esa.int/archive/documentation/GDR2/}}, the derived magnitudes in $BP$ and $RP$ filters are measured within a window size of 3.5 x 2.1 arcsec and can suffer from nearby bright sources while $G$ magnitudes are derived from accurate fitting of the objects within a small window and are less affected by the nearby sources. As a result, for blended objects or objects close to bright sources, the sum of the fluxes in $BP$ and $RP$ can be larger than the total flux in G. In the Gaia DR2, this ratio between two fluxes (I$_{BP}$+I$_{RP}$)/I$_{G}$ is reported as $BP$/$RP$ flux excess (\citealp{evans2018}). The values of $BP$/$RP$ flux excess closer to 1 indicate that the $BP$ and $RP$ magnitudes and colours of the corresponding object are reliable. For the unresolved foreground stars, the average $BP$/$RP$ flux excess value is 1.20 mag. The larger $BP$/$RP$ flux excess parameter can also happen when objects are resolved by Gaia (\citealp{voggel2020}). Later, \citet{Liu-2020} applied the same approach to investigate if their photometrically selected UCDs in the Virgo cluster show signs of being extended.

The Gaia DR2 data is very insensitive to the local background variations however the background estimation has been improved in Gaia EDR3. The $BP$/$RP$ flux excess is a complicated parameter and hard to interpret. Its behaviour depends on parameters such as colours of sources, magnitudes, variability, local background variations, nearby bright objects and processing problems. In case of the Fornax cluster, the confirmed UCDs are close to the bright galaxies and are located at the faint end of the Gaia detection limit. Therefore, while the approach discussed in \citet{voggel2020} based on the $BP$/$RP$ flux excess is useful for the nearby galaxy groups (such as NGC5128 group as is studied in their work), for the Fornax cluster, it is too uncertain. Therefore we did not apply this approach to our UCD selection.

\section{Discussion}
\label{sec6}

\begin{figure}
\includegraphics[width=\linewidth]{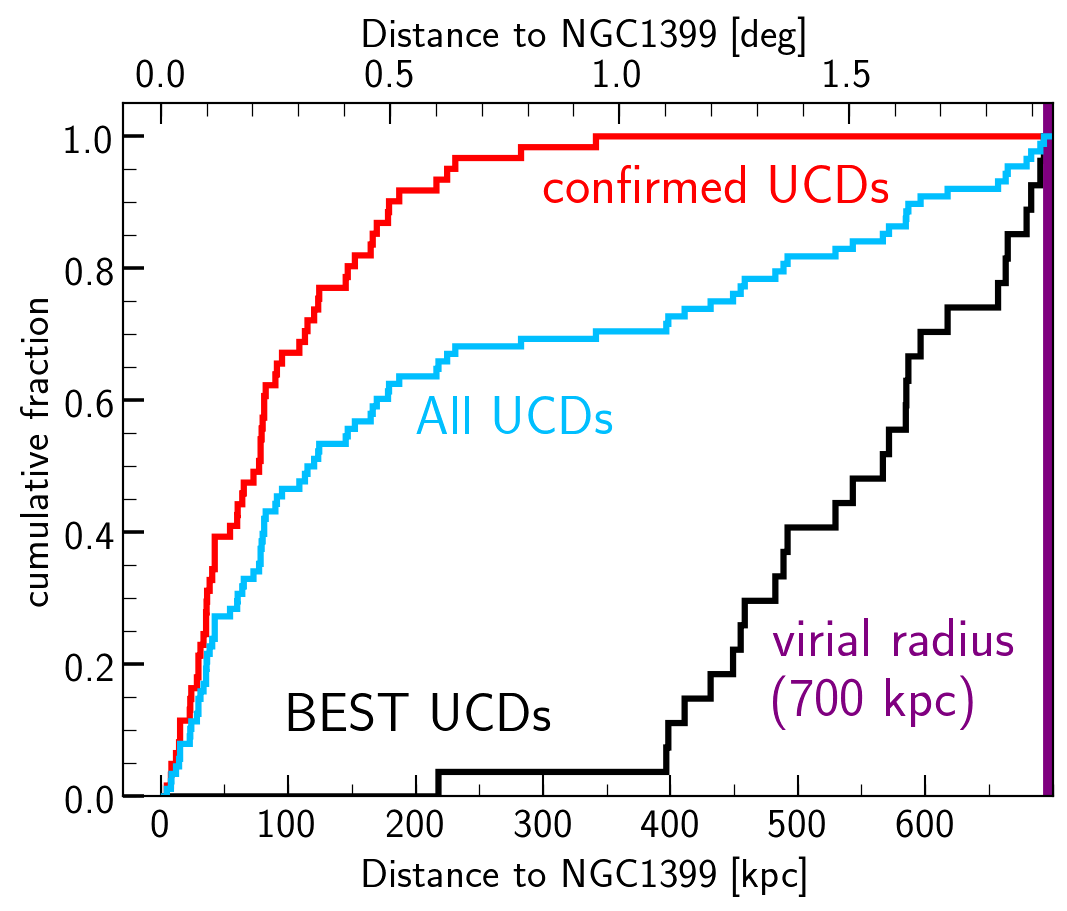}
\caption{Cumulative radial distribution of the BEST UCD candidates (black), confirmed UCDs (red) and the combination of both (blue). Almost all of the BEST candidates are found outside the 360 kpc radius from NGC1399. We assume that the spatial coverage of our dataset up to the cluster's virial radius is complete. Therefore, no completeness correction has been applied to the radial profile of the BEST UCD candidates.}
\label{radial}
\end{figure}

\begin{figure*}
\includegraphics[width=0.45\linewidth]{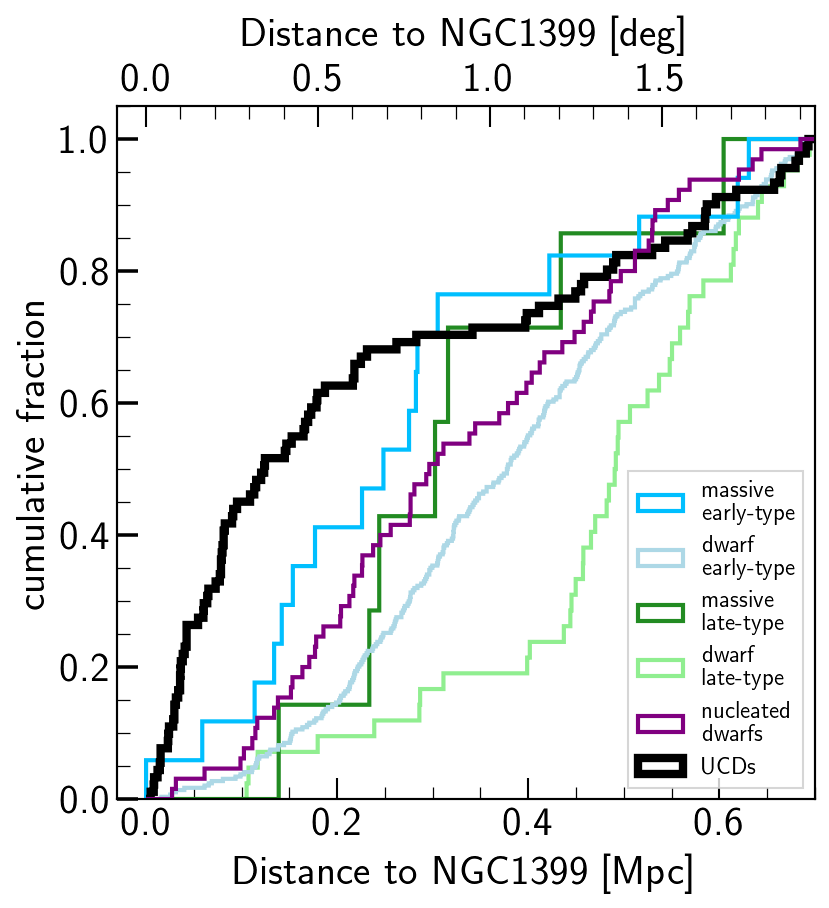}
\includegraphics[width=0.45\linewidth]{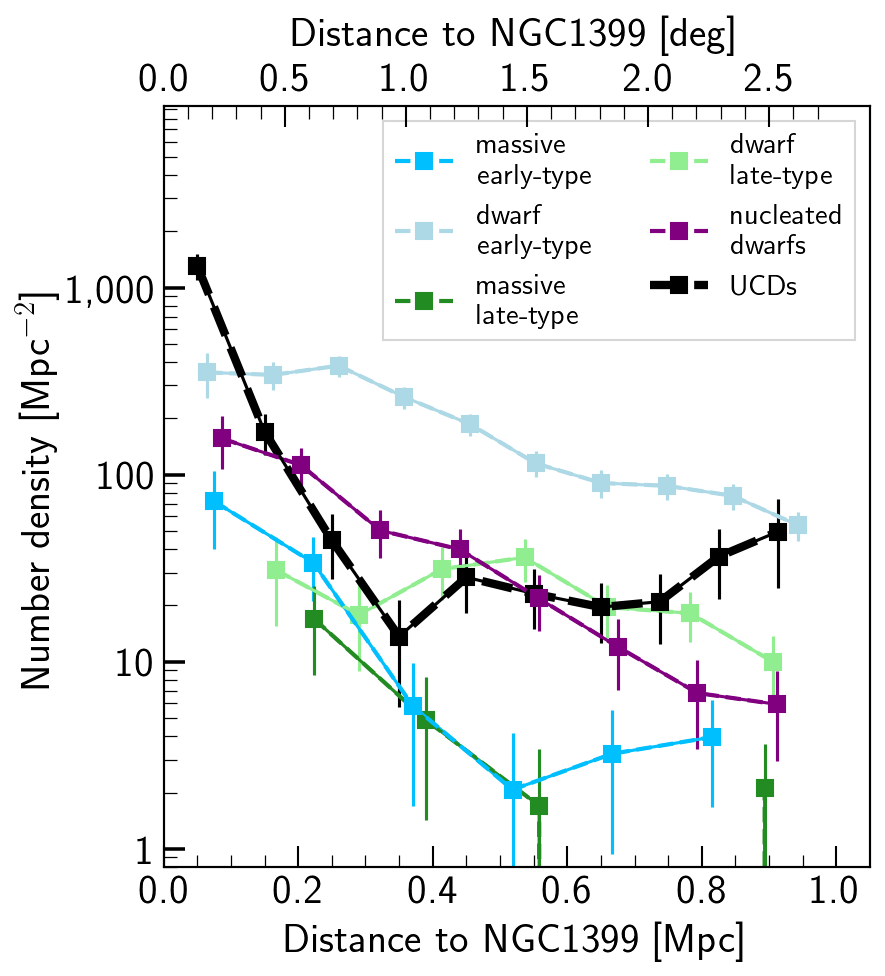}
\includegraphics[width=0.9\linewidth]{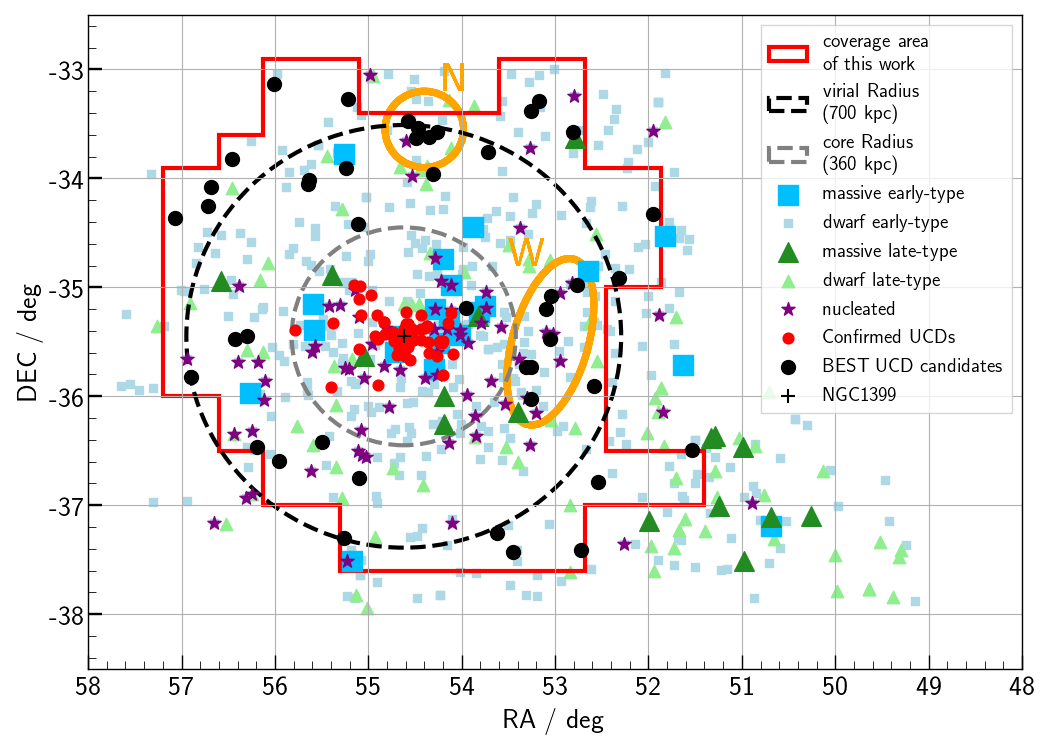}
\caption{Top: Radial distribution of the UCDs in the Fornax cluster (black) and Fornax cluster galaxies of different types. The cumulative distributions (on the left) are estimated up to the virial radius of the cluster (0.7 Mpc) where the data is spatially almost complete. The number density profiles are shown up to $~$1 Mpc, about 0.3 Mpc farther than the cluster's virial radius and are corrected for the spatial incompleteness.} Bottom: The projected distribution of the BEST UCDs and all the galaxies (massive, dwarf, nucleated) in the Fornax cluster. Massive/dwarf early-type and late-type galaxies are shown with larger/smaller blue squares and larger/smaller green triangles. Nucleated dwarf galaxies are indicated by purple stars. "N" and "W" indicate the two over-densities of UCDs in the northern side and western side of the cluster.
\label{radial2}
\end{figure*}

\begin{figure*}
\includegraphics[width=0.9\linewidth]{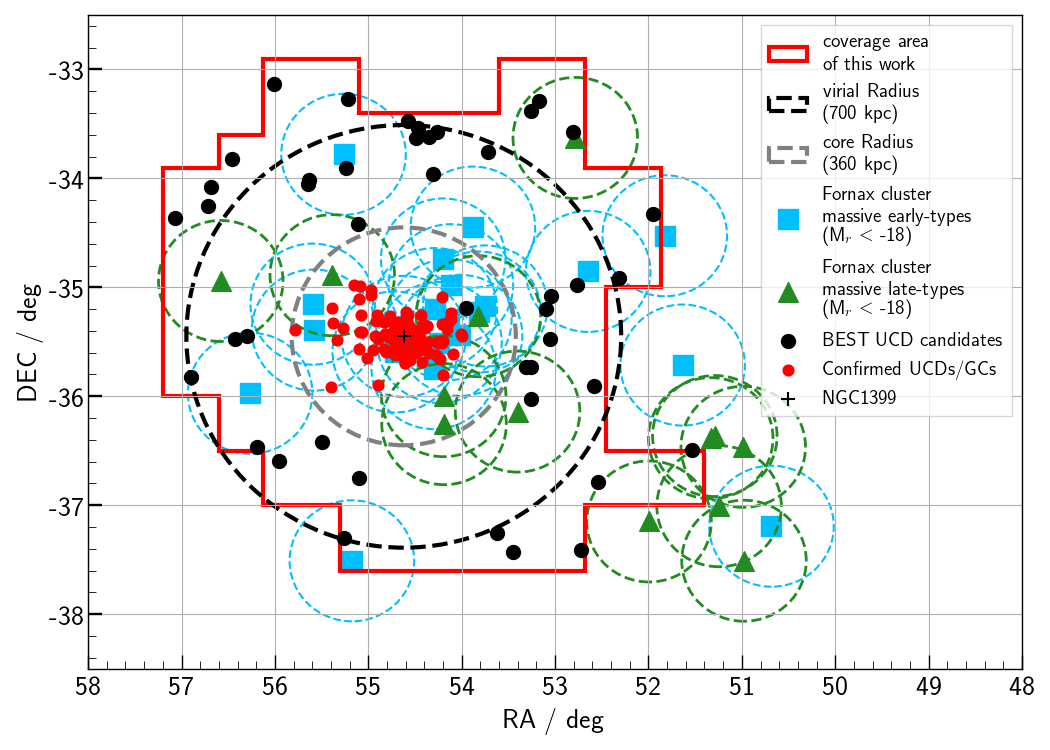}
\caption{The projected distribution of all the BEST UCD candidates (black circles). All the confirmed UCD/GCs (shown in red) are located in the core of the cluster, within 360 kpc from NGC1399 (grey dashed circle). Early-type and late-type Fornax cluster galaxies (\citealp{venhola2018}, \citealp{fcc}) brighter than M$_r$ $=$ -18 mag are shown with blue squares and green triangles. The circle around galaxies indicate a 200 kpc radius from them. Here we use these circles to investigate the association of UCD candidates to the massive galaxies in the cluster. Almost half (46\%) of the BEST UCD candidates are furthur than 200 kpc from any massive galaxies.}
\label{2dbestucdmap}
\end{figure*}

\subsection{The observed distribution of UCDs beyond the centre of the cluster}

Earlier in the paper, we defined the terms "core" and "outskirts" of the Fornax cluster as the area within the virial radius of the cluster, inside and outside the central 1 degree (360 kpc, $\sim$half the virial radius) from NGC1399. The majority of the identified BEST UCDs are located outside of the Fornax core, at a projected clusterocentric distance larger than 360 kpc in a region that has not been explored before to this depth. The radial distribution of the confirmed UCDs, BEST UCDs and the combined sample as representative of all the UCDs in the Fornax cluster is shown in Fig. \ref{radial}. The UCD radial and projected distribution along with the distribution of Fornax cluster galaxies (\citealp{venhola2018}) is investigated in Fig. \ref{radial2}. 70\% of the total UCDs within the virial radius are in the core of the cluster. UCDs in the outskirts are responsible for the remaining 30\% of the total UCD population. This fraction is in the same range as the fraction of massive galaxies in the outskirts (24\% and 29\% for massive early-type and late-type galaxies) and is smaller than the fraction of dwarfs (53\% and 81\% for early-type and late-type galaxies) and nucleated dwarfs (43\%). This implies that the UCDs are more likely to follow the massive galaxies and be connected to the high-density environments as it is expected. On the other hand, the projected distribution of the UCDs in Fig. \ref{radial2} in the bottom panel reveals over-densities of UCDs in the northern side and western side of the cluster, indicated by "N" and "W" in this figure. Both over-densities overlap with enhancements in population of dwarf galaxies in Fornax as shown in \citet{venhola-2019} (figure 3 in their paper) and \citet{yasna2018-2} (figure 3 in their paper). Thus, it suggests that a population of UCDs follow the dwarf galaxies and thus form in rather low-density, probably pre-processed group environments.

In the inner 360 kpc around NGC 1399, one object has been identified in a zone which is not covered in depth by previous surveys. This candidate (FDS-UCD-44) is not listed in any previous catalogue, it has a UCD$\_$SCORE = 92 and is located inside the 60\% completeness contour in the $u-i$/$i-Ks$ diagram. It is unlikely that the spectroscopic surveys in the core of Fornax have missed many UCDs, since they are bright and point-like, and are easy to observe with 4 or 8-meter class telescopes. However, not all UCD/GC candidates can be targeted in a multi-object approach with either slits or fibres and there will be always objects missed.
If we assume that current spectroscopic surveys have covered almost all UCDs in the centre of the Fornax cluster, we find that among the BEST UCDs there is one false positive. This fact can be used to calculate an upper limit to the number of false positives in the Fornax cluster, and therefore a lower limit to the total number of UCDs in it. Given an average of 1 source within 360 kpc from NGC1399, we estimate an average $\sim$6 false-positives within the virial radius of the cluster (700 kpc), 5 outside the core (360 kpc). For this estimate, we took into account the projected number density of \textit{UNKNOWN} sources in the core and in the outskirts. While the area in the outskirts is $\sim$3 times larger than in the core, the number of \textit{UNKNOWN} sources outside of the core is 10 times more than in the core. Considering the estimated background objects in the outskirts (10), the lower limit on the total number of the UCDs outside of the core is 33$\pm$10 (within uncertainties consistent with the expected rate of true positives of the BEST UCD sample). Among them, 18$\pm$7 are within the virial radius of the cluster. This lower limit implies that UCDs in the outskirts (within the virial radius of the cluster) are responsible for at least 23$\pm$6\ of the total UCDs within the virial radius.

While the majority of the BEST UCDs are outside the clusters' core, a fraction of them are still found close to the massive galaxies in the cluster. Fig. \ref{2dbestucdmap} shows the projected distribution of the BEST UCDs and the brightest early-type/late-type Fornax cluster galaxies (\citealp{venhola2018}, \citealp{fcc}). The circles around each galaxy indicate a projected radius of 200 kpc from each galaxy. This value is comparable to the virial radius of the Milky Way (R$_{vir}$ $\sim$ 200 kpc, \citealp{dehnen}). 20 BEST UCDs are located outside any of these radii. Some of the BEST UCD candidates, if confirmed spectroscopically, can improve the record of the most isolated compact object from a major galaxy. The most isolated compact stellar system in the Local Group known up to now, MGC1 (\citealp{mackey}) is located at de-projected (3D) distance of 200 kpc from M31. Outside of the Local Group, this record belongs to GC-2 in the M81 group. GC-2 has a de-projected distance of 400 kpc to M81 (\citealp{jang}). All these objects are fainter than M$_g$ = -10.5 (or M$_V$ = -11 given the average colour g-r $\sim$ 0.7 mag) and therefore fainter than the identified UCDs in the Fornax cluster.

If one assumes that the outskirts of galaxy clusters are formed from pre-processing in low-density environment such as groups, one can imagine that some of the nucleated group dEs/dIrrs (e.g. \citealp{Georgiev}) have been disrupted and left over the UCDs we see in the outskirts. In this case, a future follow-up spectroscopy and analysis of the radial velocities in a phase-space diagram can help to constrain the origin of these objects, whether they are in-falling systems or not.

\subsection{Rate of disruption of dwarf galaxies}

It is expected that the disruption of dwarf galaxies in galaxy clusters impacts on the faint-end of the galaxy luminosity function. The observed number of UCDs in such environments can constrain the number of disrupted nucleated dwarf galaxies and the rate of dwarf galaxy disruption can be derived. \citet{Ferrarese} used the ratio between the numbers of UCDs (N$_{UCD}$) and NSCs (N$_{NSC}$) in the Virgo cluster to estimate the disruption rate of dwarf galaxies and correct the galaxy luminosity function based on this estimate. The authors derived $\frac{N_{UCD}}{N_{NSC}}$ = 2.8 (N$_{UCD}$ = 92 and N$_{NSC}$ = 33) which implies that 70\% of all the infall halos to the cluster did not survive. In this calculation, two assumptions have been made: First, all the UCDs are stripped nuclei of dwarf galaxies, and second, the mass/luminosity of the UCD is the same as the mass/luminosity of the nucleus of the progenitor dwarf galaxy. The latter seems to be reasonable when the observed mass/luminosity of UCDs is compared to the predicated mass/luminosity of the nucleus derived from the observed mass of central SMBH (\citealp{alister}). However, the first assumption that all the observed UCDs are stripped nuclei is probably not correct. It is expected that the observed UCDs are a mixed bag of stripped nuclei and bright GCs, and that the ratio between stripped nuclei to GCs decreases with mass. Simulations of stripped nuclei (\citealp{Pfeffer2014,Pfeffer-2016}\footnote{Note that simulation of stripped nuclei in \citet{Pfeffer2014,Pfeffer-2016} have limitations in the known nucleation fraction and missed the low-surface brightness (LSB) population.}) show that, while stripped nuclei are responsible for around half of the UCDs more massive than 10$^7$ M$_{\odot}$, they are responsible for $\sim$10\% of the UCDs more massive than 4 $\times$ 10$^6$ M$_{\odot}$ (same as our mass limit for UCDs). The ratio between N$_{UCD}$ and N$_{NSC}$ for the Fornax cluster galaxies is calculated in \citet{voggel2019}, in which only UCDs with observed M$_{dyn}$/M$_*$ $>$ 1 are considered (which are very likely to be a stripped nuclei). The authors found $\frac{N_{UCD}}{N_{NSC}}$ $\sim$ 0.9. This ratio implies that around half of the nucleated dwarfs are disrupted. Note that they only took into account the known UCDs in the central regions and used an older UCD/GC catalogue (also see section 9.2 of \citet{nscreview} for a discussion on UCD/NSC ratio).

The combination of the BEST UCD candidates (44 UCDs) in our work and the confirmed UCDs (61) provides a sample of 105 UCDs in the Fornax cluster of which 88 are within its virial radius. On the other hand, The FDSDC (\citealp{venhola2018}) contains 62 nucleated dwarf galaxies (fainter than M$_r$ = -18 mag) within the virial radius of the cluster of which 13 have a nucleus brighter than M$_g$ = -10.5 mag (same magnitude as the limiting magnitude of UCDs). For our calculations here, first we make the same assumptions as \citet{Ferrarese} and find $\frac{N_{UCD}}{N_{NSC}}$ = 6.8 (lower limit 6.5$\pm$0.6) which means that 87\% (lower limit 81$\pm$8\% of the nucleated dwarfs (and dwarf galaxies in general) in the Fornax cluster have been disrupted to the present time. This rate is higher than the value estimated for the Virgo cluster (70\%, \citealp{Ferrarese}).

As was discussed above, not all the observed UCDs are stripped nuclei. Therefore, next, we assume that only 10\% of the UCDs are a stripped nucleus and the rest are GCs (\citealp{Pfeffer2014,Pfeffer-2016}). This assumption is not consistent with how \citet{Ferrarese} did this. We find a UCD to NSC (Nuclear star cluster) ratio of $ \frac{N_{UCD}}{N_{NSC}} $ = 0.68 for the Fornax cluster. This implies that about 40\% of the nucleated dwarfs have been disrupted. $\frac{N_{UCD}}{N_{NSC}}$ for the core and outskirts of the cluster is 0.88 and 0.43 which implies a rate of 47\% and 30\% of the nucleated dwarfs in the core and in the outskirts are stripped. Note that these values depend very strongly on our assumption for the ratio between the number of stripped nuclei to the total number of observed UCDs.

\begin{figure}
\includegraphics[width=0.9\linewidth]{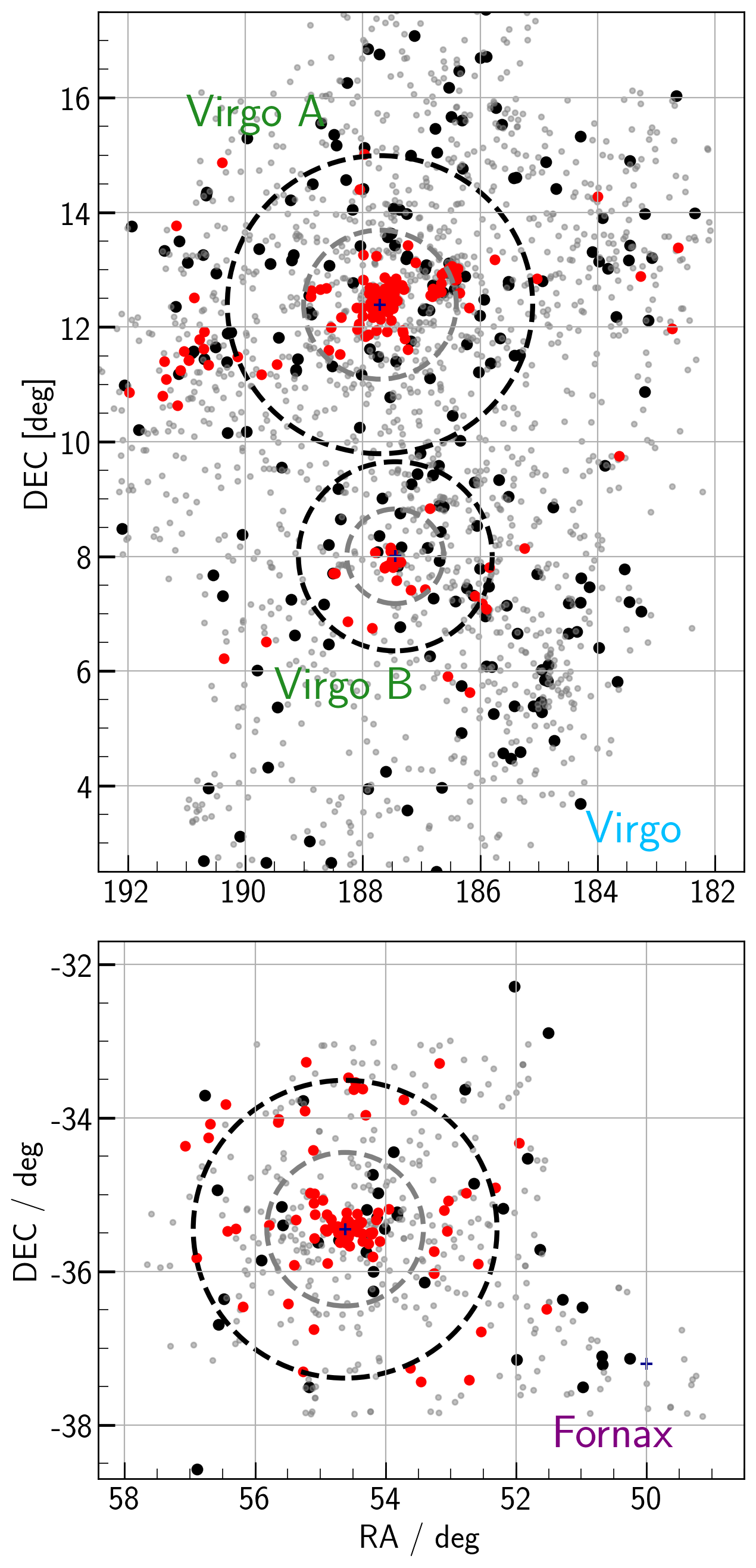}
\caption{UCDs in Virgo (top) vs. Fornax (bottom). UCDs are shown in red. Galaxies brighter than M$_B$ = -17 are shown in black and are from VCC (\citealp{vcc}) and FCC (\citealp{fcc}) catalogues. Grey points indicate galaxies, fainter than M$_B$ = -17 from VCC (\citealp{vcc}) and FDS (\citealp{venhola2018}) catalogues. The virial radius (black dashed circles) and inner region (half the virial radius, grey dashed circles) for Virgo A cluster (around M87) and Virgo B cluster (around M49) and Fornax cluster (around NGC1399) are indicated by black and grey circles.}
\label{fornaxvsvirgo}
\end{figure}

\subsection{Fornax vs. Virgo}

Recently \citet{Liu-2020} published a catalogue of UCD candidates in the Virgo cluster. The majority of these objects are selected using optical/near-infrared imaging data of the Next Generation Virgo Cluster Survey (NGVS) in $u$, $g$, $r$, $i$ and $Ks$ from the UKIRT Infrared Deep Sky Survey (UKIDSS). In their selection, they defined UCDs as objects brighter than m$_g$ = 21.5 (M$_g$ = -9.59 at the distance of the Virgo cluster). Therefore their magnitude limit for defining UCDs is $\sim$ 0.9 fainter than our adopted magnitude limit of UCDs. Additionally, the authors define a lower limit on sizes of objects and select UCDs as objects with a half-light radius (effective radius) larger than r$_h$ = 10 pc. In \citet{Liu-2020}, the authors published a catalogue of UCD candidates in the Virgo cluster. This catalogue provides an opportunity to compare the results of the Virgo cluster with our results of the Fornax cluster. Given the rich populations of massive and bright galaxies in the Virgo cluster, it is not possible to identify the possible free-floating/remote UCDs, far from the massive galaxies.

The distribution of the UCDs and the brightest galaxies (M$_B$ $>$ -17) in Virgo and Fornax are shown in Fig. \ref{fornaxvsvirgo}. For the UCDs in the Virgo cluster (top), we used the published catalogue of \citet{Liu-2020} and selected spectroscopic UCDs (confirmed) and photometrically (optical/near-infrared) identified UCDs brighter than m$_g$ = 20.6 mag (M$_g$ = -10.5 at the distance of the Virgo cluster). Of 612 UCDs in their main UCD sample, our selection criterion gives 109 and 24 UCDs within the virial radius of Virgo A and Virgo B clusters. For the UCDs in the Fornax cluster (bottom), we used the spectroscopically (confirmed) UCDs and the BEST UCD candidates. Additionally, for the BEST UCDs, we apply the size limit r$_h$ $>$ 10 pc to be consistent with the UCD sample of the Virgo cluster. This selection led to 87 UCDs within its virial radius. For Virgo A, Virgo B and Fornax, we counted 13, 7 and 26 UCDs outside of the cluster's core (grey dashed circle) corresponding to 12$\pm$4\%, 29$\pm$11\% and 30$\pm$6\% of the total UCDs within the virial radius. This implies that Fornax and Virgo B have more UCDs in their outskirts, almost twice more than Virgo A. Spectroscopic observations are needed to investigate the significance and the origin of this finding.

\section{Conclusion and Summary}
\label{sec7}
Our current knowledge of UCDs comes from limited and biased samples of confirmed UCDs. The known UCDs were found through spectroscopic surveys around massive galaxies or in the cores of galaxy clusters/groups. Therefore, there is not much known about UCDs (or similar objects) which might be located in the outskirts of galaxy clusters. In the Fornax cluster, all of the confirmed UCDs are located within half the virial radius of the cluster.

Using data from the Fornax Deep Survey (FDS), and the near-infrared observations with VISTA/VIRCAM (6 filters in total: $u$, $g$, $r$, $i$, $J$ and $Ks$), we identified, for the first time, UCD candidates in the outskirts of the Fornax cluster. However, follow-up spectroscopy is necessary to confirm the membership of the candidates. 

We prepared two catalogues containing the optical and near-infrared magnitudes, colours and sizes of the sources with and without spectroscopic data (radial velocities) based on our spectroscopic references (\citealp{wittmann-2016,Pota-2018,Maddox-2019}). Then, we pre-selected UCD/GCs based on their magnitude, size and $u-i$/$i-Ks$ colours and used the objects with available spectroscopic data (radial velocities) in the direction of the Fornax cluster as a reference-set (training-set) for a supervised machine learning method, evaluating a 5D colour space. As the result, all the pre-selected sources were classified into three classes: i. foreground stars, ii. UCD/GCs, and iii. background galaxies. Furthermore, we examined our classification with the HST/ACS observations of the Fornax cluster and Gaia DR2.

Among the $\sim$220 initially selected UCD candidates, we identified 44 BEST UCD candidates, which have a higher probability of being real UCDs. We also estimated that combining the near-infrared observations with the optical helps to remove 60-90\% of the contamination from foreground stars for the optically selected UCD. Based on our analysis on UCDs in the Fornax cluster in this paper we conclude:

\begin{itemize}

\item UCDs outside of the core and within the virial radius of the cluster are responsible for 30\% (lower limit 23$\pm$6\%) of the UCDs within the virial radius, more or less the same as the most massive galaxies in the cluster (early-type or late-type). \\

\item We identified two over-densities outside the core of the cluster in the northern and western sides, overlapping with the enhancements in the densities of the Fornax cluster dwarf galaxies (as is shown in the literature). This implies that a population of UCDs follow the dwarf galaxies in the cluster and may form in low-density, pre-processed group environments. \\

\item Almost half (46\%) of the identified UCDs in the outskirts of the Fornax cluster are further away than 200 kpc from any Fornax galaxy brighter than M$_r$ = -18 mag. The exact origin of this population is not clear since they can be UCDs associated to massive galaxies with an highly elongated orbit or UCDs formed in low-density environments (such as galaxy groups) and represent in-falling UCD populations. Future spectroscopic follow-ups to measure their radial velocities can constrain their origin. \\

\item We estimated that 87\% of the nucleated dwarfs (and dwarfs in general) in the Fornax cluster are stripped. This rate, compared to the estimated rate in the Virgo cluster (70\%, \citealp{Ferrarese} shows a higher disruption rate in Fornax versus Virgo. While these rates are strongly depend on the assumption on the fraction of stripped nuclei among the present day UCDs (100\% for the expressed values here), the relatively higher disruption rate in the Fornax cluster is independent of this assumption.\\

\item A comparison between the UCD populations in Virgo (\citealp{Liu-2020}) and Fornax (this work) shows that the Fornax cluster and Virgo B cluster (M49 sub-cluster) have relatively more UCDs in their outskirts (within their virial radius) than Virgo A (M87 sub-cluster).
\end{itemize}

Until now, most studies focused on compact sources around galaxies in the cores of galaxy clusters. This is mainly because of the lack of proper observations in the clusters’ outskirts. Recently, \citet{Liu-2020} published a catalogue of UCD candidates in the Virgo cluster, including intra-cluster UCD candidates. In the case of Fornax cluster, in this work we publish a catalogue of UCD candidates up to the virial radius of the cluster and a bit further. We investigated the properties of these candidates and their distribution which hints at a population of UCDs in low-density environments. However, the lack of realistic simulations in the outskirts of the clusters makes it very tricky to interpret our results. Upcoming cosmological simulation of GCs such as E-MOSAICS (\citealp{Pfeffer-2018, kuij}) can provide a framework to compare our results with the model predictions. Follow-up spectroscopy and radial velocity measurements in the future can confirm/reject the membership of the candidates and investigate the origin of the UCDs beyond the centre of the Fornax cluster.
\\

\textit{Acknowledgments:} We thank the referee for a number of excellent suggestions to improve the quality of the manuscript.
We want to thank Michael Biehl, Kerstin Bunte and Aleke Nolte from the Computational Intelligence Group at the Bernoulli Institute (University of Groningen) for useful discussions and their insightful comments. We thank Natasha Maddox, Nicola Napolitano and Katja Fahrion for sharing the latest spectroscopic datasets of the Fornax cluster. We thank Crescenzo Tortora for providing the latest tables of the massive compact galaxies in KiDS DR4. T.S. and R.F.P acknowledge financial support from the European Union's Horizon 2020 research and innovation programme under Marie Sk\l odowska-Curie grant agreement No 721463 to the SUNDIAL ITN network. AV would like to thank Eemil Aaltonen Foundation for funding during this project. MC acknowledges support from MIUR, PRIN 2017 (grant 20179ZF5KS). Based on observations collected at the European Organisation for Astronomical Research in the Southern Hemisphere under ESO programme(s) 092.B-0744(D), 094.B-0512(A), 096.B-0501(B), 098.B-0208(A), 098.B-0298(A), 099.B-0560(A), 0100.B-0148(A), 0100.B-0168(A), 090.B-0.477(A), 092.B-0.622(A), 094.B-0.536(A), 096.B-0.532(A). Based on observations obtained as part of the VISTA Hemisphere Survey, ESO Progamme, 179.A-2010 (PI: McMahon). This research has made use of the SIMBAD database (\citealp{simbad}), operated at CDS, Strasbourg, France. This research has made use of the VizieR catalogue access tool (\citealp{vizier}), CDS, Strasbourg, France (DOI : 10.26093/cds/vizier). This research has made use of Aladin sky atlas (\citealp{aladin1,aladin2}) developed at CDS, Strasbourg Observatory, France and SAOImageDS9 (\citealp{ds9}). This work has been done using the following software, packages and \textsc{python} libraries: Astro-WISE (\citealp{aw2,McFarland-2013}), \textsc{Numpy} (\citealp{numpy}), \textsc{Scipy} (\citealp{scipy}), \textsc{Astropy} (\citealp{astropy}), \textsc{Scikit-learn} (\citealp{scikit-learn}).

\section*{Data Availability}
The final mosaics of the optical data of Fornax Deep Survey (FDS), the final mosaics of the near-infrared data of VHS and the raw frames of the VISTA/VIRCAM observations are available trough ESO Science Archive Facility. The final mosaics of the Ks data are not available yet, however they will be available soon. The databases, software, packages and libraries that are used throughout this paper, \textsc{SIMBAD}, \textsc{VizieR}, \textsc{DS9}, \textsc{Aladin}, \textsc{SExtractor}, \textsc{SWarp}, \textsc{Astropy}, \textsc{Scikit-Learn} are publicly available. The catalogue of the UCD and GC candidates in the Fornax cluster generated in this research is available in the article and in its online supplementary material. 

\bibliographystyle{mnras}
\bibliography{main.bbl}

\begin{thebibliography}{}
\makeatletter
\relax
\def\mn@urlcharsother{\let\do\@makeother \do\$\do\&\do\#\do\^\do\_\do\%\do\~}
\def\mn@doi{\begingroup\mn@urlcharsother \@ifnextchar [ {\mn@doi@}
  {\mn@doi@[]}}
\def\mn@doi@[#1]#2{\def\@tempa{#1}\ifx\@tempa\@empty \href
  {http://dx.doi.org/#2} {doi:#2}\else \href {http://dx.doi.org/#2} {#1}\fi
  \endgroup}
\def\mn@eprint#1#2{\mn@eprint@#1:#2::\@nil}
\def\mn@eprint@arXiv#1{\href {http://arxiv.org/abs/#1} {{\tt arXiv:#1}}}
\def\mn@eprint@dblp#1{\href {http://dblp.uni-trier.de/rec/bibtex/#1.xml}
  {dblp:#1}}
\def\mn@eprint@#1:#2:#3:#4\@nil{\def\@tempa {#1}\def\@tempb {#2}\def\@tempc
  {#3}\ifx \@tempc \@empty \let \@tempc \@tempb \let \@tempb \@tempa \fi \ifx
  \@tempb \@empty \def\@tempb {arXiv}\fi \@ifundefined
  {mn@eprint@\@tempb}{\@tempb:\@tempc}{\expandafter \expandafter \csname
  mn@eprint@\@tempb\endcsname \expandafter{\@tempc}}}

\bibitem[\protect\citeauthoryear{{Afanasiev} et~al.,}{{Afanasiev}
  et~al.}{2018}]{afanasiev}
{Afanasiev} A.~V.,  et~al., 2018, \mn@doi [\mnras] {10.1093/mnras/sty913},
  \href {https://ui.adsabs.harvard.edu/abs/2018MNRAS.477.4856A} {477, 4856}

\bibitem[\protect\citeauthoryear{{Ahn} et~al.,}{{Ahn} et~al.}{2017}]{ahn2017}
{Ahn} C.~P.,  et~al., 2017, \mn@doi [\apj] {10.3847/1538-4357/aa6972}, \href
  {https://ui.adsabs.harvard.edu/abs/2017ApJ...839...72A} {839, 72}

\bibitem[\protect\citeauthoryear{{Ahn} et~al.,}{{Ahn} et~al.}{2018}]{ahn2018}
{Ahn} C.~P.,  et~al., 2018, \mn@doi [\apj] {10.3847/1538-4357/aabc57}, \href
  {https://ui.adsabs.harvard.edu/abs/2018ApJ...858..102A} {858, 102}

\bibitem[\protect\citeauthoryear{{Anders}, {Gieles}  \& {de Grijs}}{{Anders}
  et~al.}{2006}]{andres}
{Anders} P.,  {Gieles} M.,   {de Grijs} R.,  2006, \mn@doi [\aap]
  {10.1051/0004-6361:20054175}, \href
  {https://ui.adsabs.harvard.edu/abs/2006A&A...451..375A} {451, 375}

\bibitem[\protect\citeauthoryear{{Angora} et~al.,}{{Angora}
  et~al.}{2019}]{mlgc}
{Angora} G.,  et~al., 2019, \mn@doi [\mnras] {10.1093/mnras/stz2801}, \href
  {https://ui.adsabs.harvard.edu/abs/2019MNRAS.490.4080A} {490, 4080}

\bibitem[\protect\citeauthoryear{{Astropy Collaboration} et~al.,}{{Astropy
  Collaboration} et~al.}{2018}]{astropy}
{Astropy Collaboration} et~al., 2018, \mn@doi [\aj] {10.3847/1538-3881/aabc4f},
  \href {https://ui.adsabs.harvard.edu/abs/2018AJ....156..123A} {156, 123}

\bibitem[\protect\citeauthoryear{{Baumgardt} \& {Hilker}}{{Baumgardt} \&
  {Hilker}}{2018}]{Baumgardt2018}
{Baumgardt} H.,  {Hilker} M.,  2018, \mn@doi [\mnras] {10.1093/mnras/sty1057},
  \href {https://ui.adsabs.harvard.edu/abs/2018MNRAS.478.1520B} {478, 1520}

\bibitem[\protect\citeauthoryear{{Baumgardt}, {Makino}, {Hut}, {McMillan}  \&
  {Portegies Zwart}}{{Baumgardt} et~al.}{2003}]{Baum2003b}
{Baumgardt} H.,  {Makino} J.,  {Hut} P.,  {McMillan} S.,   {Portegies Zwart}
  S.,  2003, \mn@doi [\apjl] {10.1086/375802}, \href
  {https://ui.adsabs.harvard.edu/abs/2003ApJ...589L..25B} {589, L25}

\bibitem[\protect\citeauthoryear{{Begeman}, {Belikov}, {Boxhoorn}  \&
  {Valentijn}}{{Begeman} et~al.}{2013}]{aw2}
{Begeman} K.,  {Belikov} A.~N.,  {Boxhoorn} D.~R.,   {Valentijn} E.~A.,  2013,
  \mn@doi [Experimental Astronomy] {10.1007/s10686-012-9311-4}, \href
  {https://ui.adsabs.harvard.edu/abs/2013ExA....35....1B} {35, 1}

\bibitem[\protect\citeauthoryear{{Bekki}, {Couch}, {Drinkwater}  \&
  {Shioya}}{{Bekki} et~al.}{2003}]{Bekki-2003}
{Bekki} K.,  {Couch} W.~J.,  {Drinkwater} M.~J.,   {Shioya} Y.,  2003, \mn@doi
  [\mnras] {10.1046/j.1365-8711.2003.06916.x}, \href
  {https://ui.adsabs.harvard.edu/abs/2003MNRAS.344..399B} {344, 399}

\bibitem[\protect\citeauthoryear{{Bergond} et~al.,}{{Bergond}
  et~al.}{2007}]{bergond2007}
{Bergond} G.,  et~al., 2007, \mn@doi [\aap] {10.1051/0004-6361:20066963}, \href
  {https://ui.adsabs.harvard.edu/abs/2007A&A...464L..21B} {464, L21}

\bibitem[\protect\citeauthoryear{{Bertin} \& {Arnouts}}{{Bertin} \&
  {Arnouts}}{1996}]{sex}
{Bertin} E.,  {Arnouts} S.,  1996, \mn@doi [\aaps] {10.1051/aas:1996164}, \href
  {https://ui.adsabs.harvard.edu/abs/1996A&AS..117..393B} {117, 393}

\bibitem[\protect\citeauthoryear{{Bertin}, {Mellier}, {Radovich}, {Missonnier},
  {Didelon}  \& {Morin}}{{Bertin} et~al.}{2002}]{swarp}
{Bertin} E.,  {Mellier} Y.,  {Radovich} M.,  {Missonnier} G.,  {Didelon} P.,
  {Morin} B.,  2002, {The TERAPIX Pipeline}.
p.~228

\bibitem[\protect\citeauthoryear{{Binggeli}, {Sandage}  \&
  {Tammann}}{{Binggeli} et~al.}{1985}]{vcc}
{Binggeli} B.,  {Sandage} A.,   {Tammann} G.~A.,  1985, \mn@doi [\aj]
  {10.1086/113874}, \href
  {https://ui.adsabs.harvard.edu/abs/1985AJ.....90.1681B} {90, 1681}

\bibitem[\protect\citeauthoryear{{Blakeslee} et~al.,}{{Blakeslee}
  et~al.}{2009}]{Blakeslee2009}
{Blakeslee} J.~P.,  et~al., 2009, \mn@doi [\apj] {10.1088/0004-637X/694/1/556},
  \href {https://ui.adsabs.harvard.edu/abs/2009ApJ...694..556B} {694, 556}

\bibitem[\protect\citeauthoryear{{Boch} \& {Fernique}}{{Boch} \&
  {Fernique}}{2014}]{aladin2}
{Boch} T.,  {Fernique} P.,  2014, in {Manset} N.,  {Forshay} P.,  eds,
  Astronomical Society of the Pacific Conference Series Vol. 485, Astronomical
  Data Analysis Software and Systems XXIII. p.~277

\bibitem[\protect\citeauthoryear{{Bonnarel} et~al.,}{{Bonnarel}
  et~al.}{2000}]{aladin1}
{Bonnarel} F.,  et~al., 2000, \mn@doi [\aaps] {10.1051/aas:2000331}, \href
  {https://ui.adsabs.harvard.edu/abs/2000A&AS..143...33B} {143, 33}

\bibitem[\protect\citeauthoryear{{Brodie}, {Romanowsky}, {Strader}  \&
  {Forbes}}{{Brodie} et~al.}{2011}]{Brodie-2011}
{Brodie} J.~P.,  {Romanowsky} A.~J.,  {Strader} J.,   {Forbes} D.~A.,  2011,
  \mn@doi [\aj] {10.1088/0004-6256/142/6/199}, \href
  {https://ui.adsabs.harvard.edu/abs/2011AJ....142..199B} {142, 199}

\bibitem[\protect\citeauthoryear{{Cantiello}, {Grado}, {Rejkuba}, {Arnaboldi},
  {Capaccioli}, {Greggio}, {Iodice}  \& {Limatola}}{{Cantiello}
  et~al.}{2018}]{Cantiello2018}
{Cantiello} M.,  {Grado} A.,  {Rejkuba} M.,  {Arnaboldi} M.,  {Capaccioli} M.,
  {Greggio} L.,  {Iodice} E.,   {Limatola} L.,  2018, \mn@doi [\aap]
  {10.1051/0004-6361/201731325}, \href
  {https://ui.adsabs.harvard.edu/abs/2018A&A...611A..21C} {611, A21}

\bibitem[\protect\citeauthoryear{{Cantiello} et~al.,}{{Cantiello}
  et~al.}{2020}]{Cantiello2020}
{Cantiello} M.,  et~al., 2020, arXiv e-prints, \href
  {https://ui.adsabs.harvard.edu/abs/2020arXiv200512085C} {p. arXiv:2005.12085}

\bibitem[\protect\citeauthoryear{{Conroy}, {Loeb}  \& {Spergel}}{{Conroy}
  et~al.}{2011}]{conroy2011}
{Conroy} C.,  {Loeb} A.,   {Spergel} D.~N.,  2011, \mn@doi [\apj]
  {10.1088/0004-637X/741/2/72}, \href
  {https://ui.adsabs.harvard.edu/abs/2011ApJ...741...72C} {741, 72}

\bibitem[\protect\citeauthoryear{{Cover} \& {Hart}}{{Cover} \&
  {Hart}}{1967}]{knn}
{Cover} T.,  {Hart} P.,  1967, \mn@doi [IEEE Transactions on Information
  Theory] {10.1109/TIT.1967.1053964}, 13, 21

\bibitem[\protect\citeauthoryear{{Covey} et~al.,}{{Covey} et~al.}{2007}]{covey}
{Covey} K.~R.,  et~al., 2007, \mn@doi [\aj] {10.1086/522052}, \href
  {https://ui.adsabs.harvard.edu/abs/2007AJ....134.2398C} {134, 2398}

\bibitem[\protect\citeauthoryear{{D'Souza} \& {Rix}}{{D'Souza} \&
  {Rix}}{2013}]{dsouza}
{D'Souza} R.,  {Rix} H.-W.,  2013, \mn@doi [\mnras] {10.1093/mnras/sts426},
  \href {https://ui.adsabs.harvard.edu/abs/2013MNRAS.429.1887D} {429, 1887}

\bibitem[\protect\citeauthoryear{{Da Rocha}, {Mieske}, {Georgiev}, {Hilker},
  {Ziegler}  \& {Mendes de Oliveira}}{{Da Rocha} et~al.}{2011}]{DaRocha-2011}
{Da Rocha} C.,  {Mieske} S.,  {Georgiev} I.~Y.,  {Hilker} M.,  {Ziegler} B.~L.,
    {Mendes de Oliveira} C.,  2011, \mn@doi [\aap]
  {10.1051/0004-6361/201015353}, \href
  {https://ui.adsabs.harvard.edu/abs/2011A&A...525A..86D} {525, A86}

\bibitem[\protect\citeauthoryear{{Dehnen}, {McLaughlin}  \&
  {Sachania}}{{Dehnen} et~al.}{2006}]{dehnen}
{Dehnen} W.,  {McLaughlin} D.~E.,   {Sachania} J.,  2006, \mn@doi [\mnras]
  {10.1111/j.1365-2966.2006.10404.x}, \href
  {https://ui.adsabs.harvard.edu/abs/2006MNRAS.369.1688D} {369, 1688}

\bibitem[\protect\citeauthoryear{{Drinkwater}, {Jones}, {Gregg}  \&
  {Phillipps}}{{Drinkwater} et~al.}{2000a}]{Drinkwater-1999}
{Drinkwater} M.~J.,  {Jones} J.~B.,  {Gregg} M.~D.,   {Phillipps} S.,  2000a,
  \mn@doi [\pasa] {10.1071/AS00034}, \href
  {https://ui.adsabs.harvard.edu/abs/2000PASA...17..227D} {17, 227}

\bibitem[\protect\citeauthoryear{{Drinkwater} et~al.,}{{Drinkwater}
  et~al.}{2000b}]{drinkwater2000}
{Drinkwater} M.~J.,  et~al., 2000b, \aap, \href
  {https://ui.adsabs.harvard.edu/abs/2000A&A...355..900D} {355, 900}

\bibitem[\protect\citeauthoryear{{Drinkwater}, {Gregg}  \&
  {Colless}}{{Drinkwater} et~al.}{2001}]{drinkwater2001}
{Drinkwater} M.~J.,  {Gregg} M.~D.,   {Colless} M.,  2001, \mn@doi [\apjl]
  {10.1086/319113}, \href
  {https://ui.adsabs.harvard.edu/abs/2001ApJ...548L.139D} {548, L139}

\bibitem[\protect\citeauthoryear{{Evans} et~al.,}{{Evans}
  et~al.}{2018}]{evans2018}
{Evans} D.~W.,  et~al., 2018, \mn@doi [\aap] {10.1051/0004-6361/201832756},
  \href {https://ui.adsabs.harvard.edu/abs/2018A&A...616A...4E} {616, A4}

\bibitem[\protect\citeauthoryear{{Evstigneeva} et~al.,}{{Evstigneeva}
  et~al.}{2008}]{Evstigneeva-2008}
{Evstigneeva} E.~A.,  et~al., 2008, \mn@doi [\aj]
  {10.1088/0004-6256/136/1/461}, \href
  {https://ui.adsabs.harvard.edu/abs/2008AJ....136..461E} {136, 461}

\bibitem[\protect\citeauthoryear{{Fahrion} et~al.,}{{Fahrion}
  et~al.}{2019}]{fahrion1}
{Fahrion} K.,  et~al., 2019, \mn@doi [\aap] {10.1051/0004-6361/201834941},
  \href {https://ui.adsabs.harvard.edu/abs/2019A&A...625A..50F} {625, A50}

\bibitem[\protect\citeauthoryear{{Fahrion} et~al.,}{{Fahrion}
  et~al.}{2020a}]{fahrion3}
{Fahrion} K.,  et~al., 2020a, \mn@doi [\aap] {10.1051/0004-6361/201937120},
  \href {https://ui.adsabs.harvard.edu/abs/2020A&A...634A..53F} {634, A53}

\bibitem[\protect\citeauthoryear{{Fahrion} et~al.,}{{Fahrion}
  et~al.}{2020b}]{fahrion4}
{Fahrion} K.,  et~al., 2020b, \mn@doi [\aap] {10.1051/0004-6361/202037685},
  \href {https://ui.adsabs.harvard.edu/abs/2020A&A...637A..26F} {637, A26}

\bibitem[\protect\citeauthoryear{{Fellhauer} \& {Kroupa}}{{Fellhauer} \&
  {Kroupa}}{2002}]{Fellhauer-2002}
{Fellhauer} M.,  {Kroupa} P.,  2002, \mn@doi [\mnras]
  {10.1046/j.1365-8711.2002.05087.x}, \href
  {https://ui.adsabs.harvard.edu/abs/2002MNRAS.330..642F} {330, 642}

\bibitem[\protect\citeauthoryear{{Ferguson}}{{Ferguson}}{1989}]{fcc}
{Ferguson} H.~C.,  1989, \mn@doi [\aj] {10.1086/115152}, \href
  {https://ui.adsabs.harvard.edu/abs/1989AJ.....98..367F} {98, 367}

\bibitem[\protect\citeauthoryear{{Ferrarese} et~al.,}{{Ferrarese}
  et~al.}{2016}]{Ferrarese}
{Ferrarese} L.,  et~al., 2016, \mn@doi [\apj] {10.3847/0004-637X/824/1/10},
  \href {https://ui.adsabs.harvard.edu/abs/2016ApJ...824...10F} {824, 10}

\bibitem[\protect\citeauthoryear{{Firth}, {Drinkwater}, {Evstigneeva}, {Gregg},
  {Karick}, {Jones}  \& {Phillipps}}{{Firth} et~al.}{2007}]{firth2007}
{Firth} P.,  {Drinkwater} M.~J.,  {Evstigneeva} E.~A.,  {Gregg} M.~D.,
  {Karick} A.~M.,  {Jones} J.~B.,   {Phillipps} S.,  2007, \mn@doi [\mnras]
  {10.1111/j.1365-2966.2007.12474.x}, \href
  {https://ui.adsabs.harvard.edu/abs/2007MNRAS.382.1342F} {382, 1342}

\bibitem[\protect\citeauthoryear{{Firth}, {Drinkwater}  \& {Karick}}{{Firth}
  et~al.}{2008}]{firth2008}
{Firth} P.,  {Drinkwater} M.~J.,   {Karick} A.~M.,  2008, \mn@doi [\mnras]
  {10.1111/j.1365-2966.2008.13651.x}, \href
  {https://ui.adsabs.harvard.edu/abs/2008MNRAS.389.1539F} {389, 1539}

\bibitem[\protect\citeauthoryear{{Firth}, {Evstigneeva}  \&
  {Drinkwater}}{{Firth} et~al.}{2009}]{firth2009}
{Firth} P.,  {Evstigneeva} E.~A.,   {Drinkwater} M.~J.,  2009, \mn@doi [\mnras]
  {10.1111/j.1365-2966.2009.14479.x}, \href
  {https://ui.adsabs.harvard.edu/abs/2009MNRAS.394.1801F} {394, 1801}

\bibitem[\protect\citeauthoryear{{Forbes}, {Norris}, {Strader}, {Romanowsky},
  {Pota}, {Kannappan}, {Brodie}  \& {Huxor}}{{Forbes}
  et~al.}{2014}]{forbes2014}
{Forbes} D.~A.,  {Norris} M.~A.,  {Strader} J.,  {Romanowsky} A.~J.,  {Pota}
  V.,  {Kannappan} S.~J.,  {Brodie} J.~P.,   {Huxor} A.,  2014, \mn@doi
  [\mnras] {10.1093/mnras/stu1631}, \href
  {https://ui.adsabs.harvard.edu/abs/2014MNRAS.444.2993F} {444, 2993}

\bibitem[\protect\citeauthoryear{{Forbes}, {Ferr{\'e}-Mateu}, {Durr{\'e}},
  {Brodie}  \& {Romanowsky}}{{Forbes} et~al.}{2020}]{forbes2020}
{Forbes} D.~A.,  {Ferr{\'e}-Mateu} A.,  {Durr{\'e}} M.,  {Brodie} J.~P.,
  {Romanowsky} A.~J.,  2020, \mn@doi [\mnras] {10.1093/mnras/staa1924}, \href
  {https://ui.adsabs.harvard.edu/abs/2020MNRAS.497..765F} {497, 765}

\bibitem[\protect\citeauthoryear{{Francis}, {Drinkwater}, {Chilingarian},
  {Bolt}  \& {Firth}}{{Francis} et~al.}{2012}]{Francis-2012}
{Francis} K.~J.,  {Drinkwater} M.~J.,  {Chilingarian} I.~V.,  {Bolt} A.~M.,
  {Firth} P.,  2012, \mn@doi [\mnras] {10.1111/j.1365-2966.2012.21465.x}, \href
  {https://ui.adsabs.harvard.edu/abs/2012MNRAS.425..325F} {425, 325}

\bibitem[\protect\citeauthoryear{{Gaia Collaboration} et~al.,}{{Gaia
  Collaboration} et~al.}{2018}]{gaia}
{Gaia Collaboration} et~al., 2018, \mn@doi [\aap]
  {10.1051/0004-6361/201833051}, \href
  {https://ui.adsabs.harvard.edu/abs/2018A&A...616A...1G} {616, A1}

\bibitem[\protect\citeauthoryear{{Gaia Collaboration}, {Brown}, {Vallenari},
  {Prusti}, {de Bruijne}, {Babusiaux}  \& {Biermann}}{{Gaia Collaboration}
  et~al.}{2020}]{gaia3}
{Gaia Collaboration} {Brown} A.~G.~A.,  {Vallenari} A.,  {Prusti} T.,  {de
  Bruijne} J.~H.~J.,  {Babusiaux} C.,   {Biermann} M.,  2020, arXiv e-prints,
  \href {https://ui.adsabs.harvard.edu/abs/2020arXiv201201533G} {p.
  arXiv:2012.01533}

\bibitem[\protect\citeauthoryear{{Georgiev}, {Hilker}, {Puzia}, {Goudfrooij}
  \& {Baumgardt}}{{Georgiev} et~al.}{2009}]{Georgiev}
{Georgiev} I.~Y.,  {Hilker} M.,  {Puzia} T.~H.,  {Goudfrooij} P.,   {Baumgardt}
  H.,  2009, \mn@doi [\mnras] {10.1111/j.1365-2966.2009.14776.x}, \href
  {https://ui.adsabs.harvard.edu/abs/2009MNRAS.396.1075G} {396, 1075}

\bibitem[\protect\citeauthoryear{{Gonz{\'a}lez-L{\'o}pezlira}
  et~al.,}{{Gonz{\'a}lez-L{\'o}pezlira} et~al.}{2019}]{gonz2019}
{Gonz{\'a}lez-L{\'o}pezlira} R.~A.,  et~al., 2019, \mn@doi [\apj]
  {10.3847/1538-4357/ab113a}, \href
  {https://ui.adsabs.harvard.edu/abs/2019ApJ...876...39G} {876, 39}

\bibitem[\protect\citeauthoryear{{Graham}}{{Graham}}{2020}]{alister}
{Graham} A.~W.,  2020, \mn@doi [\mnras] {10.1093/mnras/stz3547}, \href
  {https://ui.adsabs.harvard.edu/abs/2020MNRAS.492.3263G} {492, 3263}

\bibitem[\protect\citeauthoryear{{Gregg} et~al.,}{{Gregg}
  et~al.}{2009}]{gregg2009}
{Gregg} M.~D.,  et~al., 2009, \mn@doi [\aj] {10.1088/0004-6256/137/1/498},
  \href {https://ui.adsabs.harvard.edu/abs/2009AJ....137..498G} {137, 498}

\bibitem[\protect\citeauthoryear{{Haghi}, {Safaei}, {Hasani Zonoozi}  \&
  {Kroupa}}{{Haghi} et~al.}{2020}]{haghi}
{Haghi} H.,  {Safaei} G.,  {Hasani Zonoozi} A.,   {Kroupa} P.,  2020, arXiv
  e-prints, \href {https://ui.adsabs.harvard.edu/abs/2020arXiv201207095H} {p.
  arXiv:2012.07095}

\bibitem[\protect\citeauthoryear{{Hilker}}{{Hilker}}{2006}]{Hilker2006}
{Hilker} M.,  2006, arXiv e-prints, \href
  {https://ui.adsabs.harvard.edu/abs/2006astro.ph..5447H} {pp
  astro--ph/0605447}

\bibitem[\protect\citeauthoryear{{Hilker}, {Infante}, {Vieira}, {Kissler-Patig}
   \& {Richtler}}{{Hilker} et~al.}{1999}]{Hilker-1999}
{Hilker} M.,  {Infante} L.,  {Vieira} G.,  {Kissler-Patig} M.,   {Richtler} T.,
   1999, \mn@doi [\aaps] {10.1051/aas:1999434}, \href
  {https://ui.adsabs.harvard.edu/abs/1999A&AS..134...75H} {134, 75}

\bibitem[\protect\citeauthoryear{{Ibata}, {Nipoti}, {Sollima}, {Bellazzini},
  {Chapman}  \& {Dalessandro}}{{Ibata} et~al.}{2013}]{iba2013}
{Ibata} R.,  {Nipoti} C.,  {Sollima} A.,  {Bellazzini} M.,  {Chapman} S.~C.,
  {Dalessandro} E.,  2013, \mn@doi [\mnras] {10.1093/mnras/sts302}, \href
  {https://ui.adsabs.harvard.edu/abs/2013MNRAS.428.3648I} {428, 3648}

\bibitem[\protect\citeauthoryear{{Iodice} et~al.,}{{Iodice}
  et~al.}{2016}]{Iodice-2016}
{Iodice} E.,  et~al., 2016, \mn@doi [\apj] {10.3847/0004-637X/820/1/42}, \href
  {https://ui.adsabs.harvard.edu/abs/2016ApJ...820...42I} {820, 42}

\bibitem[\protect\citeauthoryear{{Iodice} et~al.,}{{Iodice}
  et~al.}{2017}]{Iodice-2017}
{Iodice} E.,  et~al., 2017, \mn@doi [\apj] {10.3847/1538-4357/aa6846}, \href
  {https://ui.adsabs.harvard.edu/abs/2017ApJ...839...21I} {839, 21}

\bibitem[\protect\citeauthoryear{{Iodice} et~al.,}{{Iodice} et~al.}{2019}]{f3d}
{Iodice} E.,  et~al., 2019, \mn@doi [\aap] {10.1051/0004-6361/201935721}, \href
  {https://ui.adsabs.harvard.edu/abs/2019A&A...627A.136I} {627, A136}

\bibitem[\protect\citeauthoryear{{Ivezi{\'c}} et~al.,}{{Ivezi{\'c}}
  et~al.}{2004}]{ivezic}
{Ivezi{\'c}} {\v{Z}}.,  et~al., 2004, \mn@doi [Astronomische Nachrichten]
  {10.1002/asna.200410285}, \href
  {https://ui.adsabs.harvard.edu/abs/2004AN....325..583I} {325, 583}

\bibitem[\protect\citeauthoryear{{Jang}, {Lim}, {Park}  \& {Lee}}{{Jang}
  et~al.}{2012}]{jang}
{Jang} I.~S.,  {Lim} S.,  {Park} H.~S.,   {Lee} M.~G.,  2012, \mn@doi [\apjl]
  {10.1088/2041-8205/751/1/L19}, \href
  {https://ui.adsabs.harvard.edu/abs/2012ApJ...751L..19J} {751, L19}

\bibitem[\protect\citeauthoryear{{Janz}, {Forbes}, {Norris}, {Strader},
  {Penny}, {Fagioli}  \& {Romanowsky}}{{Janz} et~al.}{2015}]{janz2015}
{Janz} J.,  {Forbes} D.~A.,  {Norris} M.~A.,  {Strader} J.,  {Penny} S.~J.,
  {Fagioli} M.,   {Romanowsky} A.~J.,  2015, \mn@doi [\mnras]
  {10.1093/mnras/stv389}, \href
  {https://ui.adsabs.harvard.edu/abs/2015MNRAS.449.1716J} {449, 1716}

\bibitem[\protect\citeauthoryear{{Janz} et~al.,}{{Janz}
  et~al.}{2016}]{janz2016}
{Janz} J.,  et~al., 2016, \mn@doi [\mnras] {10.1093/mnras/stv2636}, \href
  {https://ui.adsabs.harvard.edu/abs/2016MNRAS.456..617J} {456, 617}

\bibitem[\protect\citeauthoryear{{Jerjen}}{{Jerjen}}{2003}]{jerjen2003}
{Jerjen} H.,  2003, \mn@doi [\aap] {10.1051/0004-6361:20021597}, \href
  {https://ui.adsabs.harvard.edu/abs/2003A&A...398...63J} {398, 63}

\bibitem[\protect\citeauthoryear{{Jester} et~al.,}{{Jester}
  et~al.}{2005}]{jester2005}
{Jester} S.,  et~al., 2005, \mn@doi [\aj] {10.1086/432466}, \href
  {https://ui.adsabs.harvard.edu/abs/2005AJ....130..873J} {130, 873}

\bibitem[\protect\citeauthoryear{{Johnston} et~al.,}{{Johnston}
  et~al.}{2020}]{johnston}
{Johnston} E.~J.,  et~al., 2020, \mn@doi [\mnras] {10.1093/mnras/staa1261},
  \href {https://ui.adsabs.harvard.edu/abs/2020MNRAS.495.2247J} {495, 2247}

\bibitem[\protect\citeauthoryear{{Jord{\'a}n} et~al.,}{{Jord{\'a}n}
  et~al.}{2004}]{jordan2004}
{Jord{\'a}n} A.,  et~al., 2004, \mn@doi [\apjs] {10.1086/422977}, \href
  {https://ui.adsabs.harvard.edu/abs/2004ApJS..154..509J} {154, 509}

\bibitem[\protect\citeauthoryear{{Jord{\'a}n} et~al.,}{{Jord{\'a}n}
  et~al.}{2007}]{jordan2007}
{Jord{\'a}n} A.,  et~al., 2007, \mn@doi [\apjs] {10.1086/512778}, \href
  {https://ui.adsabs.harvard.edu/abs/2007ApJS..169..213J} {169, 213}

\bibitem[\protect\citeauthoryear{{Jord{\'a}n} et~al.,}{{Jord{\'a}n}
  et~al.}{2009}]{jordan2009}
{Jord{\'a}n} A.,  et~al., 2009, \mn@doi [\apjs] {10.1088/0067-0049/180/1/54},
  \href {https://ui.adsabs.harvard.edu/abs/2009ApJS..180...54J} {180, 54}

\bibitem[\protect\citeauthoryear{{Jord{\'a}n}, {Peng}, {Blakeslee},
  {C{\^o}t{\'e}}, {Eyheramendy}  \& {Ferrarese}}{{Jord{\'a}n}
  et~al.}{2015}]{jordan2015}
{Jord{\'a}n} A.,  {Peng} E.~W.,  {Blakeslee} J.~P.,  {C{\^o}t{\'e}} P.,
  {Eyheramendy} S.,   {Ferrarese} L.,  2015, \mn@doi [\apjs]
  {10.1088/0067-0049/221/1/13}, \href
  {https://ui.adsabs.harvard.edu/abs/2015ApJS..221...13J} {221, 13}

\bibitem[\protect\citeauthoryear{{Joye} \& {Mandel}}{{Joye} \&
  {Mandel}}{2003}]{ds9}
{Joye} W.~A.,  {Mandel} E.,  2003, in {Payne} H.~E.,  {Jedrzejewski} R.~I.,
  {Hook} R.~N.,  eds,  Astronomical Society of the Pacific Conference Series
  Vol. 295, Astronomical Data Analysis Software and Systems XII. p.~489

\bibitem[\protect\citeauthoryear{{King}}{{King}}{1966}]{king}
{King} I.~R.,  1966, \mn@doi [\aj] {10.1086/109918}, \href
  {https://ui.adsabs.harvard.edu/abs/1966AJ.....71..276K} {71, 276}

\bibitem[\protect\citeauthoryear{{Kron}}{{Kron}}{1980}]{kron}
{Kron} R.~G.,  1980, \mn@doi [\apjs] {10.1086/190669}, \href
  {https://ui.adsabs.harvard.edu/abs/1980ApJS...43..305K} {43, 305}

\bibitem[\protect\citeauthoryear{{Kroupa}}{{Kroupa}}{2020}]{kroupa}
{Kroupa} P.,  2020, in {Bragaglia} A.,  {Davies} M.,  {Sills} A.,   {Vesperini}
  E.,  eds,  IAU Symposium Vol. 351, IAU Symposium. pp 117--121 (\mn@eprint
  {arXiv} {1910.06971}), \mn@doi{10.1017/S1743921319007749}

\bibitem[\protect\citeauthoryear{{Kruijssen}}{{Kruijssen}}{2008}]{kruij2008}
{Kruijssen} J.~M.~D.,  2008, \mn@doi [\aap] {10.1051/0004-6361:200810237},
  \href {https://ui.adsabs.harvard.edu/abs/2008A&A...486L..21K} {486, L21}

\bibitem[\protect\citeauthoryear{{Kruijssen}, {Pfeffer}, {Crain}  \&
  {Bastian}}{{Kruijssen} et~al.}{2019}]{kuij}
{Kruijssen} J.~M.~D.,  {Pfeffer} J.~L.,  {Crain} R.~A.,   {Bastian} N.,  2019,
  \mn@doi [\mnras] {10.1093/mnras/stz968}, \href
  {https://ui.adsabs.harvard.edu/abs/2019MNRAS.486.3134K} {486, 3134}

\bibitem[\protect\citeauthoryear{{Kuijken} et~al.,}{{Kuijken}
  et~al.}{2019}]{kids-dr4}
{Kuijken} K.,  et~al., 2019, \mn@doi [\aap] {10.1051/0004-6361/201834918},
  \href {https://ui.adsabs.harvard.edu/abs/2019A&A...625A...2K} {625, A2}

\bibitem[\protect\citeauthoryear{{Liu} et~al.,}{{Liu} et~al.}{2015}]{Liu-2015}
{Liu} C.,  et~al., 2015, \mn@doi [\apj] {10.1088/0004-637X/812/1/34}, \href
  {https://ui.adsabs.harvard.edu/abs/2015ApJ...812...34L} {812, 34}

\bibitem[\protect\citeauthoryear{{Liu} et~al.,}{{Liu} et~al.}{2020}]{Liu-2020}
{Liu} C.,  et~al., 2020, arXiv e-prints, \href
  {https://ui.adsabs.harvard.edu/abs/2020arXiv200715275L} {p. arXiv:2007.15275}

\bibitem[\protect\citeauthoryear{{Mackey} et~al.,}{{Mackey}
  et~al.}{2010}]{mackey}
{Mackey} A.~D.,  et~al., 2010, \mn@doi [\mnras]
  {10.1111/j.1365-2966.2009.15678.x}, \href
  {https://ui.adsabs.harvard.edu/abs/2010MNRAS.401..533M} {401, 533}

\bibitem[\protect\citeauthoryear{{Maddox}, {Serra}, {Venhola}, {Peletier},
  {Loubser}  \& {Iodice}}{{Maddox} et~al.}{2019}]{Maddox-2019}
{Maddox} N.,  {Serra} P.,  {Venhola} A.,  {Peletier} R.,  {Loubser} I.,
  {Iodice} E.,  2019, \mn@doi [\mnras] {10.1093/mnras/stz2530}, \href
  {https://ui.adsabs.harvard.edu/abs/2019MNRAS.490.1666M} {490, 1666}

\bibitem[\protect\citeauthoryear{{Mahani}, {Zonoozi}, {Haghi},
  {Je{\v{r}}{\'a}bkov{\'a}}, {Kroupa}  \& {Mieske}}{{Mahani}
  et~al.}{2021}]{mahani}
{Mahani} H.,  {Zonoozi} A.~H.,  {Haghi} H.,  {Je{\v{r}}{\'a}bkov{\'a}} T.,
  {Kroupa} P.,   {Mieske} S.,  2021, \mn@doi [\mnras] {10.1093/mnras/stab330},
  \href {https://ui.adsabs.harvard.edu/abs/2021MNRAS.502.5185M} {502, 5185}

\bibitem[\protect\citeauthoryear{{Mashchenko} \& {Sills}}{{Mashchenko} \&
  {Sills}}{2005}]{mash2005}
{Mashchenko} S.,  {Sills} A.,  2005, \mn@doi [\apj] {10.1086/426132}, \href
  {https://ui.adsabs.harvard.edu/abs/2005ApJ...619..243M} {619, 243}

\bibitem[\protect\citeauthoryear{{Mayes}, {Drinkwater}, {Pfeffer}, {Baumgardt},
  {Liu}, {Ferrarese}, {C{\^o}t{\'e}}  \& {Peng}}{{Mayes} et~al.}{2020}]{mayes}
{Mayes} R.~J.,  {Drinkwater} M.~J.,  {Pfeffer} J.,  {Baumgardt} H.,  {Liu} C.,
  {Ferrarese} L.,  {C{\^o}t{\'e}} P.,   {Peng} E.~W.,  2020, \mn@doi [\mnras]
  {10.1093/mnras/staa3731}, \href
  {https://ui.adsabs.harvard.edu/abs/2020MNRAS.tmp.3534M} {}

\bibitem[\protect\citeauthoryear{{McFarland}, {Verdoes-Kleijn}, {Sikkema},
  {Helmich}, {Boxhoorn}  \& {Valentijn}}{{McFarland}
  et~al.}{2013}]{McFarland-2013}
{McFarland} J.~P.,  {Verdoes-Kleijn} G.,  {Sikkema} G.,  {Helmich} E.~M.,
  {Boxhoorn} D.~R.,   {Valentijn} E.~A.,  2013, \mn@doi [Experimental
  Astronomy] {10.1007/s10686-011-9266-x}, \href
  {https://ui.adsabs.harvard.edu/abs/2013ExA....35...45M} {35, 45}

\bibitem[\protect\citeauthoryear{{McMahon}, {Banerji}, {Gonzalez}, {Koposov},
  {Bejar}, {Lodieu}, {Rebolo}  \& {VHS Collaboration}}{{McMahon}
  et~al.}{2013}]{vhs}
{McMahon} R.~G.,  {Banerji} M.,  {Gonzalez} E.,  {Koposov} S.~E.,  {Bejar}
  V.~J.,  {Lodieu} N.,  {Rebolo} R.,   {VHS Collaboration} 2013, The Messenger,
  \href {https://ui.adsabs.harvard.edu/abs/2013Msngr.154...35M} {154, 35}

\bibitem[\protect\citeauthoryear{{Mieske}, {Hilker}  \& {Infante}}{{Mieske}
  et~al.}{2002}]{mieske2002}
{Mieske} S.,  {Hilker} M.,   {Infante} L.,  2002, \mn@doi [\aap]
  {10.1051/0004-6361:20011833}, \href
  {https://ui.adsabs.harvard.edu/abs/2002A&A...383..823M} {383, 823}

\bibitem[\protect\citeauthoryear{{Mieske}, {Hilker}  \& {Infante}}{{Mieske}
  et~al.}{2004}]{Mieske2004}
{Mieske} S.,  {Hilker} M.,   {Infante} L.,  2004, \mn@doi [\aap]
  {10.1051/0004-6361:20035723}, \href
  {https://ui.adsabs.harvard.edu/abs/2004A&A...418..445M} {418, 445}

\bibitem[\protect\citeauthoryear{{Mieske} et~al.,}{{Mieske}
  et~al.}{2008}]{Mieske-2008}
{Mieske} S.,  et~al., 2008, \mn@doi [\aap] {10.1051/0004-6361:200810077}, \href
  {https://ui.adsabs.harvard.edu/abs/2008A&A...487..921M} {487, 921}

\bibitem[\protect\citeauthoryear{{Mieske}, {Frank}, {Baumgardt},
  {L{\"u}tzgendorf}, {Neumayer}  \& {Hilker}}{{Mieske}
  et~al.}{2013}]{Mieske-2013}
{Mieske} S.,  {Frank} M.~J.,  {Baumgardt} H.,  {L{\"u}tzgendorf} N.,
  {Neumayer} N.,   {Hilker} M.,  2013, \mn@doi [\aap]
  {10.1051/0004-6361/201322167}, \href
  {https://ui.adsabs.harvard.edu/abs/2013A&A...558A..14M} {558, A14}

\bibitem[\protect\citeauthoryear{{Misgeld} \& {Hilker}}{{Misgeld} \&
  {Hilker}}{2011}]{Misgeld-2011}
{Misgeld} I.,  {Hilker} M.,  2011, \mn@doi [\mnras]
  {10.1111/j.1365-2966.2011.18669.x}, \href
  {https://ui.adsabs.harvard.edu/abs/2011MNRAS.414.3699M} {414, 3699}

\bibitem[\protect\citeauthoryear{{Mu{\~n}oz} et~al.,}{{Mu{\~n}oz}
  et~al.}{2014}]{Munoz-2014}
{Mu{\~n}oz} R.~P.,  et~al., 2014, \mn@doi [\apjs] {10.1088/0067-0049/210/1/4},
  \href {https://ui.adsabs.harvard.edu/abs/2014ApJS..210....4M} {210, 4}

\bibitem[\protect\citeauthoryear{{Neumayer}, {Seth}  \& {B{\"o}ker}}{{Neumayer}
  et~al.}{2020}]{nscreview}
{Neumayer} N.,  {Seth} A.,   {B{\"o}ker} T.,  2020, \mn@doi [\aapr]
  {10.1007/s00159-020-00125-0}, \href
  {https://ui.adsabs.harvard.edu/abs/2020A&ARv..28....4N} {28, 4}

\bibitem[\protect\citeauthoryear{{Norris} et~al.,}{{Norris}
  et~al.}{2014}]{Norris-2014}
{Norris} M.~A.,  et~al., 2014, \mn@doi [\mnras] {10.1093/mnras/stu1186}, \href
  {https://ui.adsabs.harvard.edu/abs/2014MNRAS.443.1151N} {443, 1151}

\bibitem[\protect\citeauthoryear{{Norris}, {Escudero}, {Faifer}, {Kannappan},
  {Forte}  \& {van den Bosch}}{{Norris} et~al.}{2015}]{Norris-2015}
{Norris} M.~A.,  {Escudero} C.~G.,  {Faifer} F.~R.,  {Kannappan} S.~J.,
  {Forte} J.~C.,   {van den Bosch} R. C.~E.,  2015, \mn@doi [\mnras]
  {10.1093/mnras/stv1221}, \href
  {https://ui.adsabs.harvard.edu/abs/2015MNRAS.451.3615N} {451, 3615}

\bibitem[\protect\citeauthoryear{{Ordenes-Brice{\~n}o}
  et~al.,}{{Ordenes-Brice{\~n}o} et~al.}{2018a}]{yasna2018-2}
{Ordenes-Brice{\~n}o} Y.,  et~al., 2018a, \mn@doi [\apj]
  {10.3847/1538-4357/aaba70}, \href
  {https://ui.adsabs.harvard.edu/abs/2018ApJ...859...52O} {859, 52}

\bibitem[\protect\citeauthoryear{{Ordenes-Brice{\~n}o}
  et~al.,}{{Ordenes-Brice{\~n}o} et~al.}{2018b}]{yasna2018}
{Ordenes-Brice{\~n}o} Y.,  et~al., 2018b, \mn@doi [\apj]
  {10.3847/1538-4357/aac1b8}, \href
  {https://ui.adsabs.harvard.edu/abs/2018ApJ...860....4O} {860, 4}

\bibitem[\protect\citeauthoryear{Pedregosa et~al.,}{Pedregosa
  et~al.}{2011}]{scikit-learn}
Pedregosa F.,  et~al., 2011, Journal of Machine Learning Research, 12, 2825

\bibitem[\protect\citeauthoryear{{Peletier} et~al.,}{{Peletier}
  et~al.}{2020}]{fds}
{Peletier} R.,  et~al., 2020, arXiv e-prints, \href
  {https://ui.adsabs.harvard.edu/abs/2020arXiv200812633P} {p. arXiv:2008.12633}

\bibitem[\protect\citeauthoryear{{Pfeffer}, {Griffen}, {Baumgardt}  \&
  {Hilker}}{{Pfeffer} et~al.}{2014}]{Pfeffer2014}
{Pfeffer} J.,  {Griffen} B.~F.,  {Baumgardt} H.,   {Hilker} M.,  2014, \mn@doi
  [\mnras] {10.1093/mnras/stu1705}, \href
  {https://ui.adsabs.harvard.edu/abs/2014MNRAS.444.3670P} {444, 3670}

\bibitem[\protect\citeauthoryear{{Pfeffer}, {Hilker}, {Baumgardt}  \&
  {Griffen}}{{Pfeffer} et~al.}{2016}]{Pfeffer-2016}
{Pfeffer} J.,  {Hilker} M.,  {Baumgardt} H.,   {Griffen} B.~F.,  2016, \mn@doi
  [\mnras] {10.1093/mnras/stw498}, \href
  {https://ui.adsabs.harvard.edu/abs/2016MNRAS.458.2492P} {458, 2492}

\bibitem[\protect\citeauthoryear{{Pfeffer}, {Kruijssen}, {Crain}  \&
  {Bastian}}{{Pfeffer} et~al.}{2018}]{Pfeffer-2018}
{Pfeffer} J.,  {Kruijssen} J.~M.~D.,  {Crain} R.~A.,   {Bastian} N.,  2018,
  \mn@doi [\mnras] {10.1093/mnras/stx3124}, \href
  {https://ui.adsabs.harvard.edu/abs/2018MNRAS.475.4309P} {475, 4309}

\bibitem[\protect\citeauthoryear{{Phillipps}, {Drinkwater}, {Gregg}  \&
  {Jones}}{{Phillipps} et~al.}{2001}]{phi2001}
{Phillipps} S.,  {Drinkwater} M.~J.,  {Gregg} M.~D.,   {Jones} J.~B.,  2001,
  \mn@doi [\apj] {10.1086/322517}, \href
  {https://ui.adsabs.harvard.edu/abs/2001ApJ...560..201P} {560, 201}

\bibitem[\protect\citeauthoryear{{Pota} et~al.,}{{Pota}
  et~al.}{2018}]{Pota-2018}
{Pota} V.,  et~al., 2018, \mn@doi [\mnras] {10.1093/mnras/sty2149}, \href
  {https://ui.adsabs.harvard.edu/abs/2018MNRAS.481.1744P} {481, 1744}

\bibitem[\protect\citeauthoryear{{Powalka} et~al.,}{{Powalka}
  et~al.}{2017}]{Powalka-2017}
{Powalka} M.,  et~al., 2017, \mn@doi [\apj] {10.3847/1538-4357/aa77b1}, \href
  {https://ui.adsabs.harvard.edu/abs/2017ApJ...844..104P} {844, 104}

\bibitem[\protect\citeauthoryear{{Prole} et~al.,}{{Prole}
  et~al.}{2019}]{prole2019}
{Prole} D.~J.,  et~al., 2019, \mn@doi [\mnras] {10.1093/mnras/stz326}, \href
  {https://ui.adsabs.harvard.edu/abs/2019MNRAS.484.4865P} {484, 4865}

\bibitem[\protect\citeauthoryear{{Raj} et~al.,}{{Raj} et~al.}{2020}]{raj}
{Raj} M.~A.,  et~al., 2020, \mn@doi [\aap] {10.1051/0004-6361/202038043}, \href
  {https://ui.adsabs.harvard.edu/abs/2020A&A...640A.137R} {640, A137}

\bibitem[\protect\citeauthoryear{{Richtler}, {Dirsch}, {Larsen}, {Hilker}  \&
  {Infante}}{{Richtler} et~al.}{2005}]{richtler2005}
{Richtler} T.,  {Dirsch} B.,  {Larsen} S.,  {Hilker} M.,   {Infante} L.,  2005,
  \mn@doi [\aap] {10.1051/0004-6361:20052705}, \href
  {https://ui.adsabs.harvard.edu/abs/2005A&A...439..533R} {439, 533}

\bibitem[\protect\citeauthoryear{Rijsbergen}{Rijsbergen}{1979}]{f1}
Rijsbergen C. J.~V.,  1979, Information Retrieval, 2nd edn.
Butterworth-Heinemann, USA

\bibitem[\protect\citeauthoryear{{Schuberth}, {Richtler}, {Hilker}, {Dirsch},
  {Bassino}, {Romanowsky}  \& {Infante}}{{Schuberth}
  et~al.}{2010}]{schuberth2010}
{Schuberth} Y.,  {Richtler} T.,  {Hilker} M.,  {Dirsch} B.,  {Bassino} L.~P.,
  {Romanowsky} A.~J.,   {Infante} L.,  2010, \mn@doi [\aap]
  {10.1051/0004-6361/200912482}, \href
  {https://ui.adsabs.harvard.edu/abs/2010A&A...513A..52S} {513, A52}

\bibitem[\protect\citeauthoryear{{Schweizer}, {Seitzer}, {Whitmore}, {Kelson}
  \& {Villanueva}}{{Schweizer} et~al.}{2018}]{Schweizer-2018}
{Schweizer} F.,  {Seitzer} P.,  {Whitmore} B.~C.,  {Kelson} D.~D.,
  {Villanueva} E.~V.,  2018, \mn@doi [\apj] {10.3847/1538-4357/aaa424}, \href
  {https://ui.adsabs.harvard.edu/abs/2018ApJ...853...54S} {853, 54}

\bibitem[\protect\citeauthoryear{{Seth} et~al.,}{{Seth}
  et~al.}{2014}]{Seth-2014}
{Seth} A.~C.,  et~al., 2014, \mn@doi [\nat] {10.1038/nature13762}, \href
  {https://ui.adsabs.harvard.edu/abs/2014Natur.513..398S} {513, 398}

\bibitem[\protect\citeauthoryear{{Sharina}, {Puzia}  \& {Makarov}}{{Sharina}
  et~al.}{2005}]{sharina2005}
{Sharina} M.~E.,  {Puzia} T.~H.,   {Makarov} D.~I.,  2005, \mn@doi [\aap]
  {10.1051/0004-6361:20052921}, \href
  {https://ui.adsabs.harvard.edu/abs/2005A&A...442...85S} {442, 85}

\bibitem[\protect\citeauthoryear{{Skrutskie} et~al.,}{{Skrutskie}
  et~al.}{2006}]{2mass}
{Skrutskie} M.~F.,  et~al., 2006, \mn@doi [\aj] {10.1086/498708}, \href
  {https://ui.adsabs.harvard.edu/abs/2006AJ....131.1163S} {131, 1163}

\bibitem[\protect\citeauthoryear{{Spavone} et~al.,}{{Spavone}
  et~al.}{2020}]{spavone}
{Spavone} M.,  et~al., 2020, \mn@doi [\aap] {10.1051/0004-6361/202038015},
  \href {https://ui.adsabs.harvard.edu/abs/2020A&A...639A..14S} {639, A14}

\bibitem[\protect\citeauthoryear{{Taylor}, {Mu{\~n}oz}, {Puzia}, {Mieske},
  {Eigenthaler}  \& {Bovill}}{{Taylor} et~al.}{2016}]{taylor2016}
{Taylor} M.~A.,  {Mu{\~n}oz} R.~P.,  {Puzia} T.~H.,  {Mieske} S.,
  {Eigenthaler} P.,   {Bovill} M.~S.,  2016, arXiv e-prints, \href
  {https://ui.adsabs.harvard.edu/abs/2016arXiv160807285T} {p. arXiv:1608.07285}

\bibitem[\protect\citeauthoryear{{Vazdekis}, {S{\'a}nchez-Bl{\'a}zquez},
  {Falc{\'o}n-Barroso}, {Cenarro}, {Beasley}, {Cardiel}, {Gorgas}  \&
  {Peletier}}{{Vazdekis} et~al.}{2010}]{vazdekis}
{Vazdekis} A.,  {S{\'a}nchez-Bl{\'a}zquez} P.,  {Falc{\'o}n-Barroso} J.,
  {Cenarro} A.~J.,  {Beasley} M.~A.,  {Cardiel} N.,  {Gorgas} J.,   {Peletier}
  R.~F.,  2010, \mn@doi [\mnras] {10.1111/j.1365-2966.2010.16407.x}, \href
  {https://ui.adsabs.harvard.edu/abs/2010MNRAS.404.1639V} {404, 1639}

\bibitem[\protect\citeauthoryear{{Vazdekis}, {Koleva}, {Ricciardelli},
  {R{\"o}ck}  \& {Falc{\'o}n-Barroso}}{{Vazdekis} et~al.}{2016}]{vazdekis2016}
{Vazdekis} A.,  {Koleva} M.,  {Ricciardelli} E.,  {R{\"o}ck} B.,
  {Falc{\'o}n-Barroso} J.,  2016, \mn@doi [\mnras] {10.1093/mnras/stw2231},
  \href {https://ui.adsabs.harvard.edu/abs/2016MNRAS.463.3409V} {463, 3409}

\bibitem[\protect\citeauthoryear{{Venhola} et~al.,}{{Venhola}
  et~al.}{2017}]{Venhola-2017}
{Venhola} A.,  et~al., 2017, \mn@doi [\aap] {10.1051/0004-6361/201730696},
  \href {https://ui.adsabs.harvard.edu/abs/2017A&A...608A.142V} {608, A142}

\bibitem[\protect\citeauthoryear{{Venhola} et~al.,}{{Venhola}
  et~al.}{2018}]{venhola2018}
{Venhola} A.,  et~al., 2018, \mn@doi [\aap] {10.1051/0004-6361/201833933},
  \href {https://ui.adsabs.harvard.edu/abs/2018A&A...620A.165V} {620, A165}

\bibitem[\protect\citeauthoryear{{Venhola} et~al.,}{{Venhola}
  et~al.}{2019}]{venhola-2019}
{Venhola} A.,  et~al., 2019, \mn@doi [\aap] {10.1051/0004-6361/201935231},
  \href {https://ui.adsabs.harvard.edu/abs/2019A&A...625A.143V} {625, A143}

\bibitem[\protect\citeauthoryear{{Villaume}, {Brodie}, {Conroy}, {Romanowsky}
  \& {van Dokkum}}{{Villaume} et~al.}{2017}]{alexa}
{Villaume} A.,  {Brodie} J.,  {Conroy} C.,  {Romanowsky} A.~J.,   {van Dokkum}
  P.,  2017, \mn@doi [\apjl] {10.3847/2041-8213/aa970f}, \href
  {https://ui.adsabs.harvard.edu/abs/2017ApJ...850L..14V} {850, L14}

\bibitem[\protect\citeauthoryear{Virtanen et~al.,}{Virtanen
  et~al.}{2020}]{scipy}
Virtanen P.,  et~al., 2020, \mn@doi [Nature Methods]
  {10.1038/s41592-019-0686-2}, \href {https://rdcu.be/b08Wh} {17, 261}

\bibitem[\protect\citeauthoryear{{Voggel}, {Hilker}, {Baumgardt}, {Collins},
  {Grebel}, {Husemann}, {Richtler}  \& {Frank}}{{Voggel}
  et~al.}{2016a}]{Voggel2016}
{Voggel} K.,  {Hilker} M.,  {Baumgardt} H.,  {Collins} M. L.~M.,  {Grebel}
  E.~K.,  {Husemann} B.,  {Richtler} T.,   {Frank} M.~J.,  2016a, \mn@doi
  [\mnras] {10.1093/mnras/stw1132}, \href
  {https://ui.adsabs.harvard.edu/abs/2016MNRAS.460.3384V} {460, 3384}

\bibitem[\protect\citeauthoryear{{Voggel}, {Hilker}  \& {Richtler}}{{Voggel}
  et~al.}{2016b}]{voggel-2016}
{Voggel} K.,  {Hilker} M.,   {Richtler} T.,  2016b, \mn@doi [\aap]
  {10.1051/0004-6361/201527070}, \href
  {https://ui.adsabs.harvard.edu/abs/2016A&A...586A.102V} {586, A102}

\bibitem[\protect\citeauthoryear{{Voggel}, {Seth}, {Baumgardt}, {Mieske},
  {Pfeffer}  \& {Rasskazov}}{{Voggel} et~al.}{2019}]{voggel2019}
{Voggel} K.~T.,  {Seth} A.~C.,  {Baumgardt} H.,  {Mieske} S.,  {Pfeffer} J.,
  {Rasskazov} A.,  2019, \mn@doi [\apj] {10.3847/1538-4357/aaf735}, \href
  {https://ui.adsabs.harvard.edu/abs/2019ApJ...871..159V} {871, 159}

\bibitem[\protect\citeauthoryear{{Voggel}, {Seth}, {Sand}, {Hughes}, {Strader},
  {Crnojevic}  \& {Caldwell}}{{Voggel} et~al.}{2020}]{voggel2020}
{Voggel} K.~T.,  {Seth} A.~C.,  {Sand} D.~J.,  {Hughes} A.,  {Strader} J.,
  {Crnojevic} D.,   {Caldwell} N.,  2020, \mn@doi [\apj]
  {10.3847/1538-4357/ab6f69}, \href
  {https://ui.adsabs.harvard.edu/abs/2020ApJ...899..140V} {899, 140}

\bibitem[\protect\citeauthoryear{{Wenger} et~al.,}{{Wenger}
  et~al.}{2000}]{simbad}
{Wenger} M.,  et~al., 2000, \mn@doi [\aaps] {10.1051/aas:2000332}, \href
  {https://ui.adsabs.harvard.edu/abs/2000A&AS..143....9W} {143, 9}

\bibitem[\protect\citeauthoryear{{Wittmann}, {Lisker}, {Pasquali}, {Hilker}  \&
  {Grebel}}{{Wittmann} et~al.}{2016}]{wittmann-2016}
{Wittmann} C.,  {Lisker} T.,  {Pasquali} A.,  {Hilker} M.,   {Grebel} E.~K.,
  2016, \mn@doi [\mnras] {10.1093/mnras/stw827}, \href
  {https://ui.adsabs.harvard.edu/abs/2016MNRAS.459.4450W} {459, 4450}

\bibitem[\protect\citeauthoryear{{Zhang} et~al.,}{{Zhang}
  et~al.}{2018}]{Zhang-2018}
{Zhang} H.-X.,  et~al., 2018, \mn@doi [\apj] {10.3847/1538-4357/aab88a}, \href
  {https://ui.adsabs.harvard.edu/abs/2018ApJ...858...37Z} {858, 37}

\bibitem[\protect\citeauthoryear{{den Brok} et~al.,}{{den Brok}
  et~al.}{2014}]{mark}
{den Brok} M.,  et~al., 2014, \mn@doi [\mnras] {10.1093/mnras/stu1906}, \href
  {https://ui.adsabs.harvard.edu/abs/2014MNRAS.445.2385D} {445, 2385}

\bibitem[\protect\citeauthoryear{et. al}{et. al}{}]{vizier}
et. al O.~F., , { The VizieR database of astronomical catalogues },
  \mn@doi{10.26093/cds/vizier}

\bibitem[\protect\citeauthoryear{{van der Walt}, {Colbert}  \&
  {Varoquaux}}{{van der Walt} et~al.}{2011}]{numpy}
{van der Walt} S.,  {Colbert} S.~C.,   {Varoquaux} G.,  2011, \mn@doi
  [Computing in Science and Engineering] {10.1109/MCSE.2011.37}, \href
  {https://ui.adsabs.harvard.edu/abs/2011CSE....13b..22V} {13, 22}

\makeatother
\end{thebibliography}











\bsp    
\appendix


\section{Catalogues}
\begin{table*}
\caption {Catalogue of the spectroscopically confirmed UCD/GC (sample table, The full table is available online). Columns from left to right represent R.A. (a), Declination (b), magnitude in $u$, $g$, $r$, $i$, $J$ and $Ks$ (c to h), measured half-light (effective) radius (i). Quoted errors are statistical ones, systematic errors are not included. }
\begin{tabular}{ ccccccccccccccc } \hline  
RA & DEC & $u$ & $g$ & $r$ & $i$ & $J$ & $Ks$ & r$_h$ \\
$hms$ & $dms$ & AB mag & AB mag & AB mag & AB mag & Vega mag & Vega mag & pc \\
(a) & (b) &  (c) &  (d) &  (e) &  (f) &  (g)  &  (h) & (i) \\
\hline
03h 38m 54.0s & -35d 33m 33.01s & 20.78$\pm$0.014 & 19.28$\pm$0.010 & 18.28$\pm$0.016 & 17.84$\pm$0.029 & 16.19$\pm$0.010 & 15.27$\pm$0.006 & 139.0 \\
03h 38m 5.04s & -35d 24m 09.48s & 20.87$\pm$0.013 & 19.37$\pm$0.015 & 18.59$\pm$0.006 & 18.28$\pm$0.005 & 16.74$\pm$0.014 & 16.03$\pm$0.012 & 43.6 \\
03h 39m 35.9s & -35d 28m 24.57s & 21.14$\pm$0.017 & 19.74$\pm$0.010 & 19.16$\pm$0.016 & 18.71$\pm$0.029 & 17.30$\pm$0.018 & 16.55$\pm$0.019 & 43.9 \\
03h 38m 6.29s & -35d 28m 58.65s & 21.29$\pm$0.019 & 19.86$\pm$0.016 & 19.14$\pm$0.006 & 18.74$\pm$0.006 & 17.21$\pm$0.018 & 16.53$\pm$0.019 & 37.5 \\
03h 38m 6.29s & -35d 28m 58.65s & 21.29$\pm$0.019 & 19.86$\pm$0.016 & 19.14$\pm$0.006 & 18.74$\pm$0.006 & 17.21$\pm$0.018 & 16.53$\pm$0.019 & 37.5 \\
03h 37m 3.22s & -35d 38m 04.54s & 21.46$\pm$0.024 & 19.96$\pm$0.039 & 19.19$\pm$0.035 & 18.80$\pm$0.016 & 17.22$\pm$0.018 & 16.47$\pm$0.017 & 35.0 \\
03h 38m 10.3s & -35d 24m 05.79s & 21.45$\pm$0.022 & 20.01$\pm$0.016 & 19.26$\pm$0.006 & 18.97$\pm$0.006 & 17.63$\pm$0.022 & 16.84$\pm$0.025 & 15.3 \\
03h 39m 52.5s & -35d 04m 24.00s & 21.23$\pm$0.022 & 20.02$\pm$0.030 & 19.45$\pm$0.088 & 19.17$\pm$0.040 & 17.71$\pm$0.110 & 17.13$\pm$0.032 & 38.0 \\
03h 38m 23.7s & -35d 13m 49.48s & 21.31$\pm$0.020 & 20.14$\pm$0.002 & 19.56$\pm$0.015 & 19.33$\pm$0.004 & 18.08$\pm$0.028 & 17.42$\pm$0.042 & 18.3 \\
03h 37m 43.5s & -35d 22m 51.38s & 21.40$\pm$0.022 & 20.20$\pm$0.006 & 19.60$\pm$0.023 & 19.31$\pm$0.006 & 18.10$\pm$0.030 & 17.32$\pm$0.039 & 25.5 \\

... & ... & ... & ... & ... & ... & ... & ... & ...  \\
\hline
\label{specucdgccat}
\end{tabular}
\end{table*}


\begin{table*}
\caption{Catalogue of the spectroscopic foreground stars (sample table, The full table is available online). Columns from left to right represent R.A. (a), Declination (b), magnitude in $u$, $g$, $r$, $i$, $J$ and $Ks$ (c to h). Quoted errors are statistical ones, systematic errors are not included.}
\begin{tabular}{ cccccccccccccc } \hline  
RA & DEC & $u$ & $g$ & $r$ & $i$ & $J$ & $Ks$ \\
$hms$ & $dms$ & AB mag & AB mag & AB mag & AB mag & Vega mag & Vega mag \\
(a) & (b) &  (c) &  (d) &  (e) &  (f) &  (g)  &  (h)\\
\hline
03h 31m 38.5s & -35d 53m 13.4s & 16.95$\pm$0.004 & 15.89$\pm$0.027 & 15.90$\pm$0.027 & 16.10$\pm$0.024 & 15.31$\pm$0.007 & 15.54$\pm$0.017  \\
03h 32m 43.2s & -35d 48m 24.1s & 17.29$\pm$0.011 & 16.30$\pm$0.032 & 16.25$\pm$0.028 & 16.22$\pm$0.027 & 15.54$\pm$0.008 & 15.45$\pm$0.012  \\
03h 36m 18.8s & -35d 14m 58.6s & 16.92$\pm$0.010 & 16.34$\pm$0.020 & 16.15$\pm$0.033 & 16.15$\pm$0.023 & 15.33$\pm$0.008 & 15.17$\pm$0.006  \\
03h 38m 31.0s & -35d 16m 38.9s & 17.31$\pm$0.003 & 16.37$\pm$0.001 & 16.45$\pm$0.015 & 16.59$\pm$0.002 & 16.00$\pm$0.010 & 16.15$\pm$0.013  \\
03h 37m 14.3s & -35d 17m 41.0s & 17.23$\pm$0.007 & 16.38$\pm$0.005 & 15.96$\pm$0.023 & 15.94$\pm$0.004 & 14.78$\pm$0.010 & 14.51$\pm$0.003  \\
03h 38m 18.0s & -34d 44m 38.8s & 17.48$\pm$0.022 & 16.42$\pm$0.008 & 16.06$\pm$0.007 & 15.96$\pm$0.007 & 15.11$\pm$0.022 & 14.78$\pm$0.009  \\
03h 40m 21.8s & -36d 10m 40.6s & 17.72$\pm$0.030 & 16.42$\pm$0.031 & 16.05$\pm$0.025 & 15.96$\pm$0.033 & 15.01$\pm$0.012 & 14.72$\pm$0.006  \\
03h 28m 54.3s & -34d 55m 21.1s & 17.14$\pm$0.026 & 16.45$\pm$0.048 & 16.04$\pm$0.032 & 15.96$\pm$0.042 & 15.00$\pm$0.015 & 14.65$\pm$0.008  \\
03h 37m 42.4s & -35d 29m 54.7s & 17.33$\pm$0.006 & 16.46$\pm$0.020 & 16.07$\pm$0.017 & 15.99$\pm$0.004 & 15.19$\pm$0.011 & 14.91$\pm$0.004  \\
03h 30m 00.3s & -35d 29m 56.8s & 17.22$\pm$0.005 & 16.47$\pm$0.004 & 16.18$\pm$0.027 & 16.00$\pm$0.004 & 15.13$\pm$0.011 & 14.79$\pm$0.008  \\

... & ... & ... & ... & ... & ... & ... & ... \\
\hline
\label{fore}
\end{tabular}
\end{table*}

\begin{table*}
\caption{Catalogue of the spectroscopic background galaxies (sample table, The full table is available online). Columns from left to right represent R.A. (a), Declination (b), magnitude in $u$, $g$, $r$, $i$, $J$ and $Ks$ (c to h). Quoted errors are statistical ones, systematic errors are not included.}
\begin{tabular}{ cccccccccccccc } \hline  
RA & DEC & $u$ & $g$ & $r$ & $i$ & $J$ & $Ks$ \\
$hms$ & $dms$ & AB mag & AB mag & AB mag & AB mag & Vega mag & Vega mag \\
(a) & (b) &  (c) &  (d) &  (e) &  (f) &  (g)  &  (h)\\
\hline
03h 38m 46.8s & -35d 47m 17.4s & 17.47$\pm$0.002 & 17.53$\pm$0.030 & 17.37$\pm$0.013 & 17.13$\pm$0.014 & 16.06$\pm$0.012 & 15.11$\pm$0.005  \\
03h 35m 40.1s & -34d 07m 49.3s & 17.77$\pm$0.021 & 17.56$\pm$0.009 & 17.78$\pm$0.031 & 17.59$\pm$0.013 & 16.76$\pm$0.023 & 15.39$\pm$0.018  \\
03h 35m 18.5s & -34d 44m 03.5s & 19.94$\pm$0.013 & 18.16$\pm$0.008 & 17.41$\pm$0.022 & 16.85$\pm$0.014 & 15.12$\pm$0.008 & 14.19$\pm$0.009  \\
03h 30m 59.3s & -34d 14m 57.0s & 18.93$\pm$0.009 & 18.17$\pm$0.007 & 17.56$\pm$0.003 & 17.23$\pm$0.019 & 15.50$\pm$0.010 & 14.64$\pm$0.007  \\
03h 34m 13.4s & -33d 36m 26.2s & 18.77$\pm$0.006 & 18.25$\pm$0.002 & 17.72$\pm$0.004 & 17.26$\pm$0.009 & 15.57$\pm$0.012 & 14.43$\pm$0.011  \\
03h 37m 37.3s & -36d 06m 04.7s & 18.24$\pm$0.008 & 18.26$\pm$0.030 & 18.29$\pm$0.027 & 18.08$\pm$0.025 & 16.91$\pm$0.017 & 15.90$\pm$0.017  \\
03h 38m 47.5s & -34d 28m 20.2s & 18.67$\pm$0.016 & 18.27$\pm$0.005 & 18.04$\pm$0.002 & 17.65$\pm$0.002 & 15.71$\pm$0.030 & 14.21$\pm$0.013  \\
03h 37m 54.2s & -33d 39m 50.5s & 19.09$\pm$0.012 & 18.31$\pm$0.004 & 17.72$\pm$0.006 & 17.29$\pm$0.014 & 15.87$\pm$0.038 & 14.86$\pm$0.010  \\
03h 38m 00.5s & -33d 34m 17.4s & 19.76$\pm$0.016 & 18.33$\pm$0.004 & 17.55$\pm$0.003 & 17.19$\pm$0.004 & 15.40$\pm$0.089 & 14.34$\pm$0.015  \\
03h 31m 50.9s & -34d 35m 55.6s & 19.35$\pm$0.004 & 18.37$\pm$0.005 & 17.85$\pm$0.001 & 17.55$\pm$0.021 & 15.87$\pm$0.010 & 14.88$\pm$0.010  \\

... & ... & ... & ... & ... & ... & ... & ... \\
\hline
\label{back}
\end{tabular}
\end{table*}

\begin{table*}
\caption{The \textit{UNKNOWN} catalogue (sample table, The full table is available online). This catalogue contains sources without radial velocity measurements in the literature. Columns from left to right represent R.A. (a), Declination (b), magnitude in $u$, $g$, $r$, $i$, $J$ and $Ks$ (c to h). Quoted errors are statistical ones, systematic errors are not included.}
\begin{tabular}{ ccccccccccccccc } \hline  
RA & DEC & $u$ & $g$ & $r$ & $i$ & $J$ & $Ks$ \\
$hms$ & $dms$ & AB mag & AB mag & AB mag & AB mag & Vega mag & Vega mag \\
(a) & (b) &  (c) &  (d) &  (e) &  (f) &  (g)  &  (h)\\
\hline
03h 31m 51.2s & -37d 02m 56.7s & 17.00$\pm$0.021 & 16.07$\pm$0.031 & 16.14$\pm$0.052 & 16.17$\pm$0.037 & 15.57$\pm$0.009 & 15.65$\pm$0.014  \\
03h 45m 22.4s & -35d 01m 03.4s & 17.20$\pm$0.028 & 16.28$\pm$0.040 & 16.18$\pm$0.053 & 16.02$\pm$0.076 & 15.11$\pm$0.039 & 14.95$\pm$0.009  \\
03h 43m 25.3s & -36d 30m 07.6s & 17.12$\pm$0.025 & 16.31$\pm$0.010 & 16.07$\pm$0.015 & 15.95$\pm$0.027 & 15.07$\pm$0.021 & 14.89$\pm$0.012  \\
03h 41m 57.3s & -34d 38m 23.9s & 17.35$\pm$0.031 & 16.31$\pm$0.026 & 16.24$\pm$0.003 & 16.29$\pm$0.010 & 15.75$\pm$0.059 & 15.60$\pm$0.018  \\
03h 39m 56.3s & -37d 08m 13.6s & 17.26$\pm$0.016 & 16.34$\pm$0.010 & 16.12$\pm$0.005 & 16.11$\pm$0.039 & 15.41$\pm$0.009 & 15.23$\pm$0.009  \\
03h 43m 46.6s & -33d 13m 52.5s & 17.16$\pm$0.003 & 16.35$\pm$0.006 & 16.16$\pm$0.004 & 15.98$\pm$0.002 & 15.13$\pm$0.009 & 15.16$\pm$0.016  \\
03h 42m 04.4s & -34d 10m 55.0s & 17.31$\pm$0.029 & 16.35$\pm$0.032 & 16.01$\pm$0.036 & 16.01$\pm$0.063 & 15.12$\pm$0.009 & 14.79$\pm$0.009  \\
03h 33m 32.2s & -33d 35m 01.8s & 16.96$\pm$0.009 & 16.36$\pm$0.005 & 16.11$\pm$0.004 & 16.03$\pm$0.002 & 15.38$\pm$0.156 & 15.00$\pm$0.014  \\
03h 41m 27.2s & -36d 15m 13.2s & 17.19$\pm$0.015 & 16.36$\pm$0.003 & 16.13$\pm$0.015 & 15.92$\pm$0.016 & 15.09$\pm$0.006 & 14.94$\pm$0.012  \\
03h 45m 13.2s & -35d 07m 22.7s & 17.34$\pm$0.028 & 16.36$\pm$0.040 & 16.22$\pm$0.053 & 16.02$\pm$0.076 & 14.98$\pm$0.039 & 14.73$\pm$0.007  \\

... & ... & ... & ... & ... & ... & ... & ... \\
\hline
\label{unk}
\end{tabular}
\end{table*}


\begin{table*}
\caption {Catalogue of the UCD/GC candidates (sample table, the full table is available online). Columns from left to right represent Candidate ID (a), R.A. (b), Declination (c), magnitude in $u$, $g$, $r$, $i$, $J$ and $Ks$ (d to i). Quoted errors are statistical ones, systematic errors are not included.}
\begin{tabular}{ cccccccccccccc } \hline  
ID & RA & DEC & $u$ & $g$ & $r$ & $i$ & $J$ & $Ks$ \\
- & $hms$ & $dms$ & AB mag & AB mag & AB mag & AB mag & Vega mag & Vega mag \\
(a) & (b) &  (c) &  (d) &  (e) &  (f) &  (g)  &  (h) & (i) \\
\hline
UCD-CAND-1 & 03h 37m 23.7s & -33d 37m 13.2s & 19.79$\pm$0.013 & 18.33$\pm$0.004 & 17.66$\pm$0.006 & 17.36$\pm$0.014 & 16.06$\pm$0.038 & 15.40$\pm$0.014  \\ 
UCD-CAND-2 & 03h 33m 24.8s & -37d 21m 40.4s & 19.75$\pm$0.008 & 18.58$\pm$0.002 & 17.96$\pm$0.004 & 17.65$\pm$0.002 & 16.48$\pm$0.016 & 16.05$\pm$0.021  \\ 
UCD-CAND-3 & 03h 41m 35.2s & -33d 26m 41.8s & 19.79$\pm$0.023 & 18.61$\pm$0.021 & 18.02$\pm$0.032 & 17.69$\pm$0.021 & 16.50$\pm$0.042 & 16.10$\pm$0.035  \\ 
UCD-CAND-4 & 03h 43m 54.4s & -33d 54m 10.4s & 19.83$\pm$0.059 & 18.64$\pm$0.027 & 18.09$\pm$0.081 & 17.87$\pm$0.081 & 16.61$\pm$0.017 & 15.84$\pm$0.027  \\ 
UCD-CAND-5 & 03h 33m 08.3s & -37d 03m 02.7s & 19.81$\pm$0.038 & 18.65$\pm$0.038 & 18.12$\pm$0.059 & 17.90$\pm$0.012 & 16.76$\pm$0.015 & 16.18$\pm$0.023  \\ 
UCD-CAND-6 & 03h 46m 52.4s & -34d 15m 24.1s & 19.92$\pm$0.006 & 18.65$\pm$0.009 & 18.17$\pm$0.012 & 17.85$\pm$0.012 & 16.58$\pm$0.025 & 15.82$\pm$0.014  \\ 
UCD-CAND-7 & 03h 41m 30.2s & -34d 00m 20.4s & 20.23$\pm$0.038 & 18.68$\pm$0.031 & 18.09$\pm$0.023 & 17.84$\pm$0.050 & 16.45$\pm$0.074 & 15.62$\pm$0.024  \\ 
UCD-CAND-8 & 03h 48m 16.3s & -34d 22m 01.9s & 19.83$\pm$0.007 & 18.68$\pm$0.008 & 18.27$\pm$0.008 & 17.95$\pm$0.002 & 16.89$\pm$0.018 & 16.27$\pm$0.032  \\ 
UCD-CAND-9 & 03h 38m 52.6s & -35d 35m 17.7s & 19.95$\pm$0.007 & 18.70$\pm$0.010 & 18.25$\pm$0.016 & 17.87$\pm$0.029 & 16.99$\pm$0.016 & 16.24$\pm$0.014  \\ 
UCD-CAND-10 & 03h 32m 51.7s & -34d 03m 01.6s & 19.74$\pm$0.016 & 18.72$\pm$0.010 & 18.13$\pm$0.007 & 17.93$\pm$0.012 & 16.69$\pm$0.044 & 16.22$\pm$0.037  \\ 

... ... & ... & ... & ... & ... & ... & ... & ... & ... \\

GC-CAND-1 & 03h 43m 02.5s & -34d 05m 12.8s & 21.99$\pm$0.047 & 21.00$\pm$0.038 & 20.50$\pm$0.046 & 20.31$\pm$0.064 & 19.10$\pm$0.068 & 18.31$\pm$0.229  \\ 
GC-CAND-2 & 03h 42m 26.8s & -33d 59m 44.1s & 22.44$\pm$0.087 & 21.00$\pm$0.044 & 20.91$\pm$0.085 & 20.26$\pm$0.081 & 19.13$\pm$0.069 & 18.23$\pm$0.216  \\ 
GC-CAND-3 & 03h 39m 47.6s & -36d 35m 32.7s & 22.03$\pm$0.025 & 21.01$\pm$0.003 & 20.61$\pm$0.004 & 20.31$\pm$0.024 & 19.39$\pm$0.070 & 18.65$\pm$0.217  \\ 
GC-CAND-4 & 03h 37m 32.5s & -36d 27m 12.9s & 22.26$\pm$0.032 & 21.01$\pm$0.006 & 20.50$\pm$0.020 & 20.14$\pm$0.025 & 19.01$\pm$0.053 & 18.62$\pm$0.211  \\ 
GC-CAND-5 & 03h 31m 13.4s & -36d 07m 21.6s & 22.31$\pm$0.053 & 21.02$\pm$0.059 & 20.26$\pm$0.041 & 20.03$\pm$0.054 & 18.93$\pm$0.050 & 18.11$\pm$0.164  \\ 
GC-CAND-6 & 03h 35m 46.3s & -35d 59m 21.4s & 22.27$\pm$0.045 & 21.02$\pm$0.047 & 20.42$\pm$0.052 & 20.18$\pm$0.031 & 18.94$\pm$0.047 & 18.39$\pm$0.203  \\ 
GC-CAND-7 & 03h 42m 03.6s & -34d 38m 18.3s & 22.66$\pm$0.071 & 21.02$\pm$0.032 & 20.15$\pm$0.004 & 19.82$\pm$0.009 & 17.89$\pm$0.033 & 17.36$\pm$0.094  \\ 
GC-CAND-8 & 03h 41m 09.1s & -36d 14m 44.1s & 22.05$\pm$0.031 & 21.02$\pm$0.007 & 20.56$\pm$0.011 & 20.28$\pm$0.018 & 19.12$\pm$0.051 & 18.31$\pm$0.159  \\ 
GC-CAND-9 & 03h 35m 22.0s & -35d 14m 00.0s & 22.36$\pm$0.046 & 21.03$\pm$0.009 & 20.46$\pm$0.022 & 20.17$\pm$0.015 & 18.74$\pm$0.042 & 18.24$\pm$0.136  \\ 
GC-CAND-10 & 03h 35m 21.5s & -35d 14m 42.1s & 22.23$\pm$0.041 & 21.03$\pm$0.009 & 20.53$\pm$0.022 & 20.29$\pm$0.016 & 18.95$\pm$0.048 & 18.35$\pm$0.151  \\

... ... & ... & ... & ... & ... & ... & ... & ... & ... \\
\hline
\label{ucdgccat}
\end{tabular}
\end{table*}

\begin{table*}
\caption {Catalogue of the BEST UCD candidates. Columns from left to right represent candidate ID (a), R.A. (b), Declination (c), magnitude in $u$, $g$, $r$, $i$, $J$ and $Ks$ (d to i), measured half-light (effective) radius (j). Quoted errors are statistical ones, systematic errors are not included.}
\begin{tabular}{ ccccccccccccccc } \hline  
ID & RA & DEC & $u$ & $g$ & $r$ & $i$ & $J$ & $Ks$ & r$_h$ \\
- & $hms$ & $dms$ & AB mag & AB mag & AB mag & AB mag & Vega mag & Vega mag & pc \\
(a) & (b) &  (c) &  (d) &  (e) &  (f) &  (g)  &  (h) & (i) & (j)\\
\hline

FDS-UCD-1 & 03h 37m 23.7s & -33d 37m 13.2s & 19.79$\pm$0.013 & 18.33$\pm$0.004 & 17.66$\pm$0.006 & 17.36$\pm$0.014 & 16.06$\pm$0.038 & 15.40$\pm$0.014 & 37.8 \\ 
FDS-UCD-2 & 03h 46m 52.4s & -34d 15m 24.1s & 19.92$\pm$0.006 & 18.65$\pm$0.009 & 18.17$\pm$0.012 & 17.85$\pm$0.012 & 16.58$\pm$0.025 & 15.82$\pm$0.014 & 22.3 \\ 
FDS-UCD-3 & 03h 48m 16.3s & -34d 22m 01.9s & 19.83$\pm$0.007 & 18.68$\pm$0.008 & 18.27$\pm$0.008 & 17.95$\pm$0.002 & 16.89$\pm$0.018 & 16.27$\pm$0.032 & 14.6 \\ 
FDS-UCD-4 & 03h 31m 14.0s & -33d 34m 37.3s & 20.18$\pm$0.017 & 18.98$\pm$0.019 & 18.38$\pm$0.004 & 18.14$\pm$0.002 & 17.24$\pm$0.184 & 16.37$\pm$0.043 & 8.5 \\ 
FDS-UCD-5 & 03h 44m 01.8s & -33d 08m 11.0s & 20.22$\pm$0.007 & 19.09$\pm$0.001 & 18.50$\pm$0.001 & 18.16$\pm$0.003 & 17.06$\pm$0.018 & 16.49$\pm$0.050 & 7.1 \\ 
FDS-UCD-6 & 03h 26m 09.0s & -36d 29m 20.8s & 20.64$\pm$0.012 & 19.32$\pm$0.005 & 18.64$\pm$0.002 & 18.32$\pm$0.006 & 16.93$\pm$0.015 & 16.12$\pm$0.024 & 34.3 \\ 
FDS-UCD-7 & 03h 42m 34.9s & -34d 03m 14.4s & 20.36$\pm$0.030 & 19.35$\pm$0.032 & 18.88$\pm$0.036 & 18.63$\pm$0.063 & 17.77$\pm$0.030 & 16.84$\pm$0.059 & 21.5 \\ 
FDS-UCD-8 & 03h 40m 53.3s & -33d 16m 24.6s & 20.64$\pm$0.083 & 19.47$\pm$0.037 & 18.90$\pm$0.053 & 18.59$\pm$0.020 & 17.45$\pm$0.022 & 16.88$\pm$0.056 & 15.1 \\ 
FDS-UCD-9 & 03h 33m 49.2s & -37d 26m 02.4s & 21.06$\pm$0.012 & 19.77$\pm$0.004 & 19.03$\pm$0.004 & 18.71$\pm$0.002 & 17.50$\pm$0.023 & 16.85$\pm$0.045 & 10.0 \\ 
FDS-UCD-10 & 03h 38m 17.6s & -33d 28m 12.0s & 21.25$\pm$0.025 & 19.78$\pm$0.004 & 18.97$\pm$0.003 & 18.74$\pm$0.004 & 17.36$\pm$0.092 & 16.44$\pm$0.038 & 20.5 \\ 
FDS-UCD-11 & 03h 37m 58.3s & -33d 37m 34.4s & 21.14$\pm$0.021 & 19.95$\pm$0.004 & 19.26$\pm$0.006 & 18.96$\pm$0.014 & 17.90$\pm$0.046 & 17.18$\pm$0.071 & 16.1 \\ 
FDS-UCD-12 & 03h 42m 33.1s & -34d 01m 08.6s & 21.07$\pm$0.032 & 19.95$\pm$0.032 & 19.46$\pm$0.036 & 19.13$\pm$0.063 & 18.25$\pm$0.039 & 17.48$\pm$0.107 & 18.2 \\ 
FDS-UCD-13 & 03h 46m 44.4s & -34d 04m 53.8s & 21.50$\pm$0.020 & 20.00$\pm$0.016 & 19.53$\pm$0.077 & 19.13$\pm$0.016 & 17.73$\pm$0.125 & 17.22$\pm$0.054 & 39.2 \\ 
FDS-UCD-14 & 03h 33m 14.3s & -35d 44m 05.3s & 21.36$\pm$0.020 & 20.01$\pm$0.005 & 19.22$\pm$0.020 & 19.03$\pm$0.023 & 17.79$\pm$0.025 & 17.17$\pm$0.051 & 9.9 \\ 
FDS-UCD-15 & 03h 37m 13.4s & -33d 57m 43.5s & 21.23$\pm$0.022 & 20.05$\pm$0.005 & 19.79$\pm$0.022 & 19.25$\pm$0.042 & 18.04$\pm$0.033 & 17.29$\pm$0.085 & 40.2 \\ 
FDS-UCD-16 & 03h 45m 49.1s & -33d 49m 32.8s & 21.46$\pm$0.059 & 20.06$\pm$0.010 & 19.54$\pm$0.062 & 19.17$\pm$0.036 & 17.73$\pm$0.031 & 17.14$\pm$0.091 & 13.5 \\ 
FDS-UCD-17 & 03h 30m 08.3s & -36d 47m 01.4s & 21.87$\pm$0.034 & 20.18$\pm$0.015 & 19.46$\pm$0.012 & 19.07$\pm$0.023 & 17.43$\pm$0.021 & 16.61$\pm$0.042 & 67.0 \\ 
FDS-UCD-18 & 03h 44m 45.8s & -36d 27m 41.2s & 21.50$\pm$0.039 & 20.20$\pm$0.014 & 19.60$\pm$0.024 & 19.33$\pm$0.025 & 18.26$\pm$0.036 & 17.53$\pm$0.083 & 15.0 \\ 
FDS-UCD-19 & 03h 32m 41.4s & -33d 17m 22.6s & 21.71$\pm$0.037 & 20.30$\pm$0.004 & 19.50$\pm$0.005 & 19.24$\pm$0.003 & 17.93$\pm$0.025 & 17.24$\pm$0.094 & 19.3 \\ 
FDS-UCD-20 & 03h 37m 51.1s & -33d 32m 08.1s & 21.37$\pm$0.026 & 20.30$\pm$0.002 & 19.67$\pm$0.003 & 19.44$\pm$0.005 & 18.23$\pm$0.087 & 17.59$\pm$0.103 & 12.1 \\ 
FDS-UCD-21 & 03h 32m 13.3s & -35d 28m 14.3s & 21.43$\pm$0.021 & 20.32$\pm$0.003 & 19.57$\pm$0.006 & 19.52$\pm$0.016 & 18.29$\pm$0.033 & 17.67$\pm$0.081 & 17.9 \\ 
FDS-UCD-22 & 03h 45m 42.3s & -35d 28m 19.2s & 21.55$\pm$0.018 & 20.32$\pm$0.051 & 19.48$\pm$0.099 & 19.11$\pm$0.136 & 17.97$\pm$0.028 & 17.19$\pm$0.062 & 45.4 \\ 
FDS-UCD-23 & 03h 27m 49.1s & -34d 19m 45.6s & 21.60$\pm$0.025 & 20.34$\pm$0.014 & 19.55$\pm$0.003 & 19.29$\pm$0.003 & 17.92$\pm$0.034 & 17.40$\pm$0.091 & 23.3 \\ 
FDS-UCD-24 & 03h 31m 02.9s & -34d 58m 49.0s & 21.55$\pm$0.026 & 20.38$\pm$0.006 & 19.74$\pm$0.029 & 19.51$\pm$0.009 & 18.23$\pm$0.037 & 17.55$\pm$0.105 & 23.6 \\ 
FDS-UCD-25 & 03h 33m 02.6s & -33d 22m 54.1s & 21.72$\pm$0.037 & 20.39$\pm$0.004 & 19.60$\pm$0.005 & 19.32$\pm$0.003 & 18.05$\pm$0.027 & 17.33$\pm$0.103 & 5.4 \\ 
FDS-UCD-26 & 03h 34m 53.3s & -33d 45m 23.1s & 21.85$\pm$0.037 & 20.48$\pm$0.003 & 19.86$\pm$0.011 & 19.59$\pm$0.010 & 18.28$\pm$0.035 & 17.45$\pm$0.117 & 16.1 \\ 
FDS-UCD-27 & 03h 29m 15.7s & -34d 54m 36.5s & 21.86$\pm$0.039 & 20.54$\pm$0.038 & 19.94$\pm$0.027 & 19.65$\pm$0.040 & 18.58$\pm$0.046 & 17.83$\pm$0.133 & 23.9 \\ 
FDS-UCD-28 & 03h 34m 28.8s & -37d 15m 26.9s & 21.55$\pm$0.017 & 20.56$\pm$0.006 & 20.01$\pm$0.003 & 19.80$\pm$0.004 & 18.75$\pm$0.048 & 18.05$\pm$0.131 & 21.4 \\ 
FDS-UCD-29 & 03h 43m 49.6s & -36d 35m 41.8s & 21.99$\pm$0.027 & 20.58$\pm$0.011 & 19.95$\pm$0.017 & 19.60$\pm$0.024 & 18.45$\pm$0.043 & 17.34$\pm$0.070 & 5.5 \\ 
FDS-UCD-30 & 03h 40m 26.4s & -34d 25m 05.9s & 22.10$\pm$0.035 & 20.63$\pm$0.010 & 19.92$\pm$0.005 & 19.63$\pm$0.010 & 18.37$\pm$0.043 & 17.71$\pm$0.125 & 15.3 \\ 
FDS-UCD-31 & 03h 30m 19.8s & -35d 54m 13.5s & 21.96$\pm$0.053 & 20.67$\pm$0.023 & 20.22$\pm$0.070 & 19.96$\pm$0.052 & 18.91$\pm$0.049 & 18.14$\pm$0.176 & 12.7 \\ 
FDS-UCD-32 & 03h 30m 53.2s & -37d 24m 46.1s & 22.08$\pm$0.027 & 20.73$\pm$0.014 & 19.96$\pm$0.040 & 19.67$\pm$0.036 & 18.60$\pm$0.038 & 17.60$\pm$0.102 & 12.5 \\ 
FDS-UCD-33 & 03h 41m 59.0s & -36d 25m 00.4s & 21.93$\pm$0.031 & 20.78$\pm$0.003 & 20.21$\pm$0.013 & 19.94$\pm$0.010 & 18.91$\pm$0.045 & 18.16$\pm$0.149 & 11.4 \\ 
FDS-UCD-34 & 03h 41m 03.5s & -37d 18m 16.8s & 22.20$\pm$0.027 & 20.80$\pm$0.013 & 20.21$\pm$0.008 & 19.85$\pm$0.028 & 18.46$\pm$0.037 & 17.65$\pm$0.087 & 45.9 \\ 
FDS-UCD-35 & 03h 40m 23.7s & -36d 45m 07.3s & 22.13$\pm$0.033 & 20.82$\pm$0.017 & 19.99$\pm$0.009 & 19.68$\pm$0.025 & 18.19$\pm$0.032 & 17.34$\pm$0.065 & 72.4 \\ 
FDS-UCD-36 & 03h 32m 24.0s & -35d 11m 56.7s & 22.00$\pm$0.035 & 20.84$\pm$0.054 & 20.00$\pm$0.012 & 19.84$\pm$0.034 & 18.54$\pm$0.087 & 17.85$\pm$0.099 & 21.1 \\ 
FDS-UCD-37 & 03h 32m 10.2s & -35d 04m 32.6s & 21.90$\pm$0.032 & 20.84$\pm$0.053 & 20.04$\pm$0.010 & 19.82$\pm$0.022 & 18.59$\pm$0.100 & 17.76$\pm$0.091 & 10.2 \\ 
FDS-UCD-38 & 03h 37m 02.4s & -33d 34m 29.9s & 22.12$\pm$0.039 & 20.85$\pm$0.010 & 20.14$\pm$0.004 & 19.78$\pm$0.003 & 18.65$\pm$0.085 & 17.74$\pm$0.120 & 8.1 \\ 
FDS-UCD-39 & 03h 45m 11.7s & -35d 26m 44.4s & 21.90$\pm$0.025 & 20.87$\pm$0.019 & 20.36$\pm$0.013 & 19.99$\pm$0.037 & 19.06$\pm$0.053 & 18.37$\pm$0.186 & 19.9 \\ 
FDS-UCD-40 & 03h 47m 35.3s & -35d 49m 30.5s & 22.33$\pm$0.042 & 20.91$\pm$0.012 & 20.34$\pm$0.024 & 19.90$\pm$0.108 & 18.57$\pm$0.038 & 17.52$\pm$0.062 & 54.0 \\ 
FDS-UCD-41 & 03h 33m 01.4s & -36d 01m 34.6s & 22.13$\pm$0.037 & 20.91$\pm$0.018 & 20.16$\pm$0.022 & 19.92$\pm$0.030 & 18.64$\pm$0.041 & 17.97$\pm$0.143 & 19.4 \\ 
FDS-UCD-42 & 03h 40m 56.4s & -33d 54m 15.8s & 22.43$\pm$0.058 & 20.91$\pm$0.019 & 20.64$\pm$0.052 & 19.89$\pm$0.041 & 18.83$\pm$0.053 & 17.86$\pm$0.143 & 12.8 \\ 
FDS-UCD-43 & 03h 33m 02.2s & -35d 44m 06.1s & 22.44$\pm$0.049 & 20.99$\pm$0.005 & 20.28$\pm$0.013 & 19.99$\pm$0.022 & 18.47$\pm$0.037 & 17.72$\pm$0.085 & 26.8 \\ 
FDS-UCD-44 & 03h 35m 48.1s & -35d 11m 10.9s & 21.92$\pm$0.034 & 20.99$\pm$0.028 & 20.20$\pm$0.028 & 20.03$\pm$0.015 & 18.63$\pm$0.054 & 18.18$\pm$0.128 & 29.0 \\ 

\hline 
\label{ucdtable}
\end{tabular}
\end{table*}

\label{lastpage}
\end{document}